\documentclass[aps,amssymb,amsmath, twocolumn, prx, superscriptaddress]{revtex4-2}

\usepackage{graphicx}
\usepackage{latexsym}
\usepackage[english]{babel}
\usepackage{dsfont}
\usepackage{epsfig}
\usepackage{amsmath}
\usepackage{lipsum}
\usepackage{bm}
\usepackage{braket}
\usepackage{mathtools}
\usepackage{natbib}
\usepackage{enumitem}
\usepackage{longtable}
\usepackage[T1]{fontenc}
\usepackage{bbold}
\usepackage[table]{xcolor}
\usepackage[colorlinks=true, citecolor=blue, urlcolor=blue]{hyperref}
\usepackage{float}
\usepackage{amsfonts}
\usepackage{soul}
\usepackage{cleveref}
\usepackage{multirow}
\usepackage{arydshln} 

\usepackage[up]{subfigure}
\DeclareMathOperator{\Tr}{Tr}
\usepackage{epstopdf}
\newcommand\norm[1]{\left\lVert#1\right\rVert}

\usepackage{booktabs}

\usepackage{amsthm}

\newcounter{rtaskno}

\def\<{\langle}
\def\>{\rangle}

\def\braket#1{\langle#1\rangle}
\def\ketbra#1#2{|#1\rangle\!\langle#2|}

\DeclareMathAlphabet\mathbfcal{OMS}{cmsy}{b}{n}

\mathchardef\mhyphen="2D 

\newcommand{\eq}{\mathrm{eq}}

\DeclareMathOperator{\ii}{\mathrm{i}}
\newcommand{\ts}{\raisebox{0.4ex}{\scriptsize$\intercal$}}

\begin{document}

\title{Intrinsic Hamiltonian of Mean Force and Strong-Coupling Quantum Thermodynamics}

\author{Ignacio Gonz\'alez}
\email{ignago10@ucm.es}
\affiliation{Departamento de F\'{\i}sica Te\'orica, Facultad de Ciencias F\'isicas, Universidad Complutense, 28040 Madrid, Spain.}
	
\author{Sagnik Chakraborty}
\email{sagnikch@ucm.es}
\affiliation{Departamento de F\'{\i}sica Te\'orica, Facultad de Ciencias F\'isicas, Universidad Complutense, 28040 Madrid, Spain.}

\author{\'Angel Rivas}
\email{anrivas@ucm.es}
\affiliation{Departamento de F\'{\i}sica Te\'orica, Facultad de Ciencias F\'isicas, Universidad Complutense, 28040 Madrid, Spain.}
\affiliation{CCS-Center for Computational Simulation, Campus de Montegancedo UPM, 28660 Boadilla del Monte, Madrid, Spain.}

\begin{abstract}
We present a universal thermodynamic framework for quantum systems strongly coupled to thermal reservoirs, identifying the class of initial conditions required for consistent thermodynamic behavior. Unlike other approaches, our framework provides a unified description of both thermodynamic processes and thermostatic properties, while preserving the same gauge freedoms as in the standard weak-coupling regime and retaining the von Neumann expression for the thermodynamic entropy. Furthermore, the formulation is entirely in terms of the system’s reduced state, relying only on variables accessible through local microscopic control. It is therefore directly connected to experimentally accessible quantities, making it possible to address operational questions concerning the controlled system—such as determining the minimum work required to drive it between two prescribed states in the presence of a strongly coupled reservoir. We validate our approach using a paradigmatic model of a structured bosonic reservoir.
\end{abstract}

\maketitle

\section{Introduction}\label{Sec:I}
The thermodynamics of systems strongly coupled to thermal reservoirs is a challenging problem that has puzzled the scientific community for decades. Fundamental concepts such as internal energy become ambiguous when the interaction energy cannot be neglected, making it difficult to establish general thermodynamic laws \cite{Jarzynski2004,Gelin2009,Seifert2016,Talkner2016,Jarzynski2017}. 

This question has gained renewed interest with the advent of quantum thermodynamics \cite{Janet2016,Binder2018,Deffner2019,StrasbergBook,Potts2024review}, which focuses on applying thermodynamic principles to quantum systems coupled to thermal sources. Although most theoretical developments and experimental work~\cite{Pekola2015,An2015,Rossnagel2016,Zou2017,Ronzani2018,vonLindenfels2019,Peterson2019,Klatzow2019,Bouton2021,Zhang2022,Sundelin2024,Uusnakki2025} in quantum thermodynamics assume weak system-reservoir coupling, the contribution of the interaction energy to the energetic balance is often non-negligible for microscopic systems. In fact, many modern experimental platforms, such as superconducting circuits, nanoscale solid-state devices, and complex systems in chemical physics (e.g., light-harvesting molecules and polariton chemistry), are inherently subject to strong environmental interactions. Consequently, recent technological advances are rapidly bringing us into a regime where such interactions must be explicitly accounted for \cite{VanHorne2020,Colla2025,Pekola2024}. Thus, the development of a strong-coupling quantum thermodynamic framework is no longer merely a foundational problem, but also a practical imperative for understanding the energetic costs and ultimate performance bounds of quantum-controlled systems across multiple disciplines.

In order to include strong-coupling corrections to quantum thermodynamic laws several approaches have been suggested \cite{Hanggi2009, Esposito2010, Strasberg2016, Strasberg2017,Hu2018,Perarnau-Llobet2018,Miller2018,Dou2018,Strasberg2019,Rivas2019,Strasberg2019b,Strasberg2021,Schaller2020,Rivas2020,Strasberg2021,Colla2022,Savoudi2025}, (see also \cite{Binder2015,Esposito2015,Bruch2016,Alipour2016,Tanimura2016,Ludovico2016,Newman2017,Hsiang2018,Thomas2018,Hanggi2020,Anto-Sztrikacs2021,Seshadri2021,Alipour2022,Chakraborty2022,Huang2022,Ahmadi2023,Lacerda2023,Elouard2023,Kaneyasu2023,Latune2023,Rolandi2023,Xu2023,Davoudi2024,Yao2024,Aguilar2024} for further developments, including multi-bath cases or other kinds of reservoirs). Not all of them allow for the formulation of general thermodynamic laws. Among the most general ones is the approach of \cite{Esposito2010,Strasberg2017,Strasberg2019b,Tanimura2016}, where the interaction energy is counted as part of the system's internal energy. The main feature of this framework is that general first and second laws can be derived using the von Neumann entropy. However, it fails to satisfy the usual relations between thermodynamic variables at equilibrium. More importantly, it defines certain thermodynamic variables of the system—namely, its internal and free energies—in such a way that they are no longer functions of the system’s reduced state, but instead depend on the global system–reservoir state. Since, according to the quantum postulates, the density matrix contains all the available physical information about a system, if internal energy, free energy, and thermodynamic entropy are genuine physical properties of a quantum system, they must ultimately be expressible in terms of its quantum state. Moreover, from a practical point of view, this global dependence implies that determining these thermodynamic variables requires microscopic control over the reservoir’s degrees of freedom. This is problematic, as the reservoir is typically an exceedingly large system for which we can control, at most, only macroscopic properties such as its temperature.

Another general approach, somewhat aimed at overcoming these difficulties, is based on the concept of the Hamiltonian of mean force \cite{Hanggi2009,Hu2018,Miller2018,Strasberg2019}. This formulation recovers the usual relations between thermodynamic variables at equilibrium and yields thermodynamic laws using only system mean values. However, the entropy in this framework deviates from the von Neumann formula, making it difficult to interpret thermodynamics from an informational perspective. Additionally, although in this approach all thermodynamic variables depend solely on system operators, determining the Hamiltonian of mean force still requires control over the reservoir's microscopic degrees of freedom. This requirement holds unless the set of thermodynamic gauge freedoms in the strong-coupling regime is expanded beyond that of the weak-coupling case, resulting in a theory agnostic to thermostatic properties \cite{Strasberg2020}.

To avoid this issue, a different Hamiltonian of mean force formulation was suggested in \cite{Rivas2020} based exclusively on the reduced dynamics of the system. While this improves the measurability and practical applicability of the framework, it renders it insensitive to any strong-coupling effect at equilibrium (e.g., level-dressing). This is so because in the approach of \cite{Rivas2020} the thermodynamic variables take the same values at equilibrium regardless of the strength of the system-reservoir coupling. Furthermore, the connection between information and thermodynamics at strong coupling remains unclear since, similar to previous approaches, the entropy function used becomes equivalent to the von Neumann function only in the weak-coupling limit.

The purpose of this work is to overcome most of these difficulties by presenting a thermodynamic framework for the strong-coupling regime that:

i) Formulates general thermodynamic laws in terms of system variables that depend strictly on the reduced quantum state, thereby complying with quantum principles and guaranteeing experimental accessibility through purely local microscopic control.

ii) Preserves the same fundamental gauge freedom as in the weak-coupling regime, establishing a rigorously consistent and well-defined theory of strong-coupling thermostatics.

iii) Retains the von Neumann expression for the entropy, extending the foundational link between information theory and physical thermodynamics to arbitrarily strong system-environment couplings.

Before presenting this new framework, in Sec. \ref{Sec:II} we establish the specific dynamical setting under consideration. Then, we introduce the concept of the ``intrinsic Hamiltonian of mean force'' to describe strong-coupling equilibrium thermodynamics in Sec. \ref{Sec:III}. We proceed to formulate general nonequilibrium thermodynamic laws within this setting in Sec. \ref{Sec:IV}. In Section \ref{sec:example}, we apply the approach to a simple yet illustrative model of strong coupling with a composite bosonic reservoir. We compute thermodynamic variables, solve the dynamics, and verify the validity of the proposed thermodynamic laws. Detailed steps of the model's solution, as well as additional technical results, are provided in the appendices. The final section is devoted to conclusions and perspectives.

\section{Quantum Dynamics of a System Coupled to a Thermal Reservoir} \label{Sec:II} 
As already hinted at in the introduction, we choose a system, with Hamiltonian $H_{\rm S}$, interacting with a thermal reservoir or bath with Hamiltonian $H_{\rm R}$ at a temperature $T=1/k_B\beta$. The interaction Hamiltonian is denoted by $V_{\rm SR}$. As a result, the total Hamiltonian is given by
\begin{equation}\label{Hamiltonian}
    H=H_{\rm S}+H_{\rm R}+V_{\rm SR}.
\end{equation}
Assuming that at $t=0$ the system is put into contact with the reservoir, the system-reservoir state at a time $t$ is formally given by 
\begin{align}\label{rhoSRt}
    \rho_{\rm SR}(t)=U_{\rm SR}(t)\big(\rho_{\rm S}(0)\otimes\rho_{\rm R,\beta}\big)\, U^{\dagger}_{\rm SR}(t),
\end{align}
where $U_{\rm SR}(t)=\exp\big(-\frac{\ii}{\hbar} H t \big)$, and 
\begin{equation}
    \rho_{\rm R,\beta}=\frac{e^{-\beta H_{\rm R}}}{Z_{\rm R}},\quad \text{with } Z_{\rm R}=\Tr \big(e^{-\beta H_{\rm R}}\big). 
\end{equation}
Strictly speaking, $\rho_{\rm R,\beta}$ must be rigorously defined as a positive functional in the algebraic formulation of quantum mechanics \cite{Bratteli}. Indeed, for a reservoir with an infinitely large, potentially continuous, number of degrees of freedom, the continuous spectrum of $H_{\rm R}$ prevents $\exp(-\beta H_{\rm R})$ from being a trace-class operator. However, we shall continue using the density matrix notation $\rho_{\rm R,\beta}$ in a formal sense.

For a total time-independent Hamiltonian $H$, it has been shown \cite{Bach2000,Derezinski2001,Merkli2001,Derezinski2003,Frohlich2004,Merkli2008,Linden2009,Reimann2010,Short2012,Subasi2012,Iles-Smith2014,Konenberg2016,Gogolin2016} that, under certain conditions—such as the interaction term $V_{\rm SR}$ being quasilocal (i.e., a local observable or the limit of a sequence of local observables)—the joint system-reservoir state thermalizes in the following way:
\begin{align}\label{TotalGibbs}
    \rho_{\rm SR}(t)\xrightarrow{t\rightarrow \infty}\rho_{\rm SR,\beta}= \frac{e^{-\beta H}}{Z_{\rm SR}},
\end{align}
where $Z_{\rm SR}=\Tr\big(e^{-\beta H}\big)$. More specifically, if $\langle X \rangle_t$ and $\langle X\rangle_\beta$ denote the mean value of a quasilocal observable $X$ of the total system-reservoir composite, in the states $\rho_{\rm SR}(t)$ and $\rho_{\rm SR,\beta}$, respectively, the convergence \eqref{TotalGibbs} means
\begin{align}\label{TotalGibbs2}
    \lim_{t\to\infty}\langle X\rangle_{t}=\langle X\rangle_{\beta}\,.
\end{align} 
As a result, the system state
\begin{align}\label{Lambdat}
    \rho_{\rm S}(t)=\text{Tr}_{\rm R} \big[ U_{\rm SR}(t)\big(\rho_{\rm S}(0)\otimes\rho_{\rm R,\beta}\big)\, U^{\dagger}_{\rm SR}(t) \big],
\end{align}
relaxes in the long-time limit to the so-called ``mean force Gibbs state'' \cite{revHMF}, namely, $\lim_{t\to\infty}\rho_{\rm S}(t)=\rho_{\rm S,\eq},$ where
\begin{equation}
    \rho_{\rm S, \eq}=\Tr_{\rm R}\bigg(\frac{e^{-\beta H}}{Z_{\rm SR}}\bigg).\label{system-equilibrium}
\end{equation}

\section{Strong-coupling thermodynamics at equilibrium}\label{Sec:III}

\subsection{Standard Hamiltonian of mean force thermodynamics} \label{Sec:IIA}
Once the equilibrium state has been reached, one way to construct a strong-coupling thermodynamic framework is to express the mean-force Gibbs state \eqref{system-equilibrium} in the form
\begin{align}\label{rhoH*}
    \rho_{\rm S,\eq}=\frac{e^{-\beta H_{\rm S}^*}}{Z_{\rm S}^*},
\end{align}
where 
\begin{align}
    H_{\rm S}^*(\beta):= -\beta^{-1} \log \bigg[\frac{Z_{\rm SR}}{Z_{\rm R}}\Tr_{\rm R} \big(\rho_{\rm SR,\beta}\big)\bigg],
\end{align}
is the so-called ``Hamiltonian of mean force,'' and the associated system partition function is given by 
\[
Z_{\rm S}^*=\frac{Z_{\rm SR}}{Z_{\rm R}}=\mathrm{Tr}\big(e^{-\beta H_{\rm S}^*}\big)
\]
\cite{Hanggi2009,Hu2018,Miller2018,Strasberg2019,revHMF}. This partition function reduces to the usual one if the coupling is very weak so that $Z_{\rm SR}\simeq Z_{\rm S} Z_{\rm R}$, with $Z_{\rm S}=\Tr(e^{-\beta H_{\rm S}})$ the usual partition function of the system. 

Similarly to the weak coupling situation where the equilibrium thermodynamic variables are specified by the partition function, analogous equations can be introduced by using $Z_{\rm S}^*(\beta)$:
\begin{align}
    &F_{\rm S}^*(\beta):=-\beta^{-1} \log Z_{\rm S}^*(\beta), \label{F*}\\
    &E_{\rm S}^*(\beta):=-\partial_{\beta} \log Z_{\rm S}^*(\beta)=F_{\rm S}^*(\beta)+\beta\partial_\beta F_{\rm S}^*(\beta),\label{E*_Z}\\
    &S_{\rm S}^*(\beta):=T^{-1}[E_{\rm S}^*(\beta)-F_{\rm S}^*(\beta)]=k_B \beta^2 \partial_\beta F_{\rm S}^*(\beta) \label{S*},
\end{align}
where $F_{\rm S}^*(\beta)$, $E_{\rm S}^*(\beta)$ and $S_{\rm S}^*(\beta)$ play the role of free energy, internal energy and thermodynamic entropy of the system at strong coupling. These definitions can also be motivated by considering an isothermal process connecting two system and reservoir equilibrium states \eqref{TotalGibbs} with different system Hamiltonians, $H_{\rm S}(0)$ and $H_{\rm S}(1)$, respectively. The free energy change is given by
\begin{equation}\label{DeltaF}
    \Delta F_{\rm SR}=-\beta^{-1}\log\frac{Z_{\rm SR}(\beta,1)}{Z_{\rm SR}(\beta,0)}=-\beta^{-1}\log\frac{Z_{\rm S}^*(\beta,1)}{Z_{\rm S}^*(\beta,0)},
\end{equation}
where $Z_{\rm SR}(\beta,j)=\Tr[e^{-\beta H(j)}]$, with $H(j)=H_{\rm S}(j)+H_{\rm R}+V_{\rm SR}$ for $j=0,1$, and we have used that $Z_{\rm SR}(\beta,j)=Z_{\rm S}^*(\beta,j)Z_{\rm R}(\beta)$, as $H_{\rm R}$ does not change in the process. If we assume that, regardless of the nature of the isothermal process, this free energy change $\Delta F_{\rm SR}=\Delta F_{\rm S}^*$ should be entirely understood as a system free energy change, because only the system part of the Hamiltonian has been changed, then one would identify \eqref{F*}-\eqref{S*} with system thermodynamic variables. 

Equations \eqref{F*}-\eqref{S*} can be rewritten in terms of the Hamiltonian of mean force by using $Z_{\rm S}^*=\mathrm{Tr}(e^{-\beta H_{\rm S}^*})$:
\begin{align}
    &F_{\rm S}^*(\beta)=\langle H_{\rm S}^*\rangle_{\eq}-\beta^{-1} S_{\rm vN}(\rho_{\rm S,\eq}), \label{FH*}\\
    &E_{\rm S}^*(\beta)=\langle H_{\rm S}^*\rangle_{\eq} +\beta\langle \partial_\beta H_{\rm S}^*\rangle_{\eq},\label{EH*} \\
    &S_{\rm S}^*(\beta)=k_B S_{\rm vN}(\rho_{\rm S,\eq}) +k_B \beta^2 \langle \partial_\beta H_{\rm S}^*\rangle_{\eq}\label{SH*},
\end{align}
with $S_{\rm vN}(\rho)=-\Tr(\rho\log\rho)$ the von Neumann entropy, that has a well-known meaning from the information theoretic point of view, i.e., it quantifies the optimal compression rate of a state $\rho$ \cite{Schumacher}. We see that because $H_{\rm S}^*$ is a function of $\beta$,  the thermodynamic entropy departs from the von Neumann expression in this approach, except in the weak-coupling limit where $\langle \partial_\beta H_{\rm S}^*\rangle_{\eq}\to 0$.

\subsection{The intrinsic Hamiltonian of mean force}
Despite the motivation to introduce the previous definition of Hamiltonian of mean force $H^*_{\rm S}$ and its associated partition function $Z^*_{\rm S}$, Eqs. such as \eqref{F*}-\eqref{S*} and \eqref{FH*}-\eqref{SH*} present certain inconveniences. From a practical point of view, $Z^*_{\rm S} = Z_{\rm SR} / Z_{\rm R}$ is generally a function of the reservoir degrees of freedom, which makes $F_{\rm S}^*$ (and so $E_{\rm S}^*$ and $S_{\rm S}^*$) inaccessible through system measurements alone. In addition, the extra term in the entropy \eqref{SH*} due to $\partial_\beta H_{\rm S}^*$ makes it difficult to establish a relationship between informational and thermodynamic properties in the strong-coupling regime.

Nevertheless, all these expressions are subject to a large degree of ambiguity due to the freedom in the choice of $H_{\rm S}^*$. Indeed, Eq. \eqref{rhoH*} remains invariant under the transformation 
\begin{equation}\label{GaugeT}
\begin{cases}
    H_{\rm S}^*\to H_{\rm S}^*+f(\beta)\mathds{1},\\
    Z^*_{\rm S}\to Z^*_{\rm S}e^{-\beta f(\beta)},
    \end{cases}
\end{equation}
with $f(\beta)$ an arbitrary function of $\beta$ (and potentially other system or environmental parameters). 

Actually, this apparent ``gauge'' transformation for the equilibrium quantum state can also be considered in the weak-coupling case. That is, the equilibrium state $\rho_{\rm S,\beta}=e^{-\beta H_{\rm S}}/Z_{\rm S}$ is also invariant under \eqref{GaugeT}, with $H_{\rm S}$ and $Z_{\rm S}$ playing the roles of $H_{\rm S}^*$ and $Z_{\rm S}^*$, respectively. Therefore, from this point of view, there is also a large degree of ambiguity in the choice of the exponent for the standard Gibbs state. 
However, this ambiguity is strongly reduced by the weak-coupling relation $E_{\rm S}=\langle H_{\rm S} \rangle_{\mathrm{eq}}$. Namely, the exponent of the weak-coupling equilibrium Gibbs state is chosen such that it also represents the internal energy operator (up to the $-\beta$ factor, which ensures the correct physical dimensions). This is not the case in the Hamiltonian of mean force approach, where the internal energy is given in \eqref{EH*} by the expectation value of the operator $H_{\rm S}^* + \beta \partial_\beta H_{\rm S}^*$. This operator does not coincide with the exponent in \eqref{rhoH*} due to the presence of the $\beta$-derivative term.

In order to correct this mismatch between the internal energy operator and the exponent of the Hamiltonian of mean force state, we now rewrite \eqref{rhoH*} as
\begin{align}\label{rhoHash}
    \rho_{\rm S,\eq}=\frac{e^{-\beta H_{\rm S}^\sharp}}{Z_{\rm S}^\sharp},
\end{align}
with $H_{\rm S}^\sharp$ a new Hamiltonian of mean force that satisfies the property 
\begin{equation}\label{GaugeFixing}
\langle \partial_\beta H_{\rm S}^\sharp\rangle_{\eq}=0. 
\end{equation}
Notably, the partition function associated with this Hamiltonian of mean force, 
\begin{equation}
    Z_{\rm S}^\sharp=\Tr\big(e^{-\beta H_{\rm S}^\sharp}\big),
\end{equation}
is just a function of the system equilibrium state. To see this, we use \eqref{rhoHash} to write
\begin{equation}\label{Hash}
    H_{\rm S}^\sharp=-\beta^{-1}(\log Z_{\rm S}^\sharp +\log \rho_{\rm S,\eq}),
\end{equation}
and then
\begin{align}\label{aux1}
    \langle \partial_\beta H_{\rm S}^\sharp\rangle_{\eq}=&\beta^{-2}\left(\log Z_{\rm S}^\sharp +\langle\log \rho_{\rm S,\eq}\rangle_{\eq}\right)\nonumber\\
    &-\beta^{-1}\left(\partial_\beta \log Z_{\rm S}^\sharp+\langle\partial_\beta\log \rho_{\rm S,\eq}\rangle_{\eq}\right).
\end{align}
The last term vanishes $\langle\partial_\beta\log \rho_{\rm S,\eq}\rangle_{\eq}=\Tr\left(\rho_{\rm S,\eq}\partial_\beta\log \rho_{\rm S,\eq}\right)=0$. Indeed, if $\rho_{\rm S,\eq}$ is full rank, this can be proven using the spectral theorem $\rho_{\rm S,\eq}=\sum_{j=1}^d p_j(\beta) \Pi_{j} (\beta)$, so that $\log(\rho_{\rm S,\eq})=\sum_{j=1}^d \log[p_j(\beta)] \Pi_{j} (\beta)$ and
\begin{align}  \Tr\big(&\rho_{\rm S,\eq}\partial_{\beta}\log\rho_{\rm S,\eq}\big)\nonumber\\
&=\sum_{j,k}p_j \Tr\left[\frac{(\partial_\beta p_k)}{p_k}\Pi_j\Pi_k+\log(p_k)\Pi_j\partial_\beta \Pi_k\right]\nonumber \\
&=\Tr\sum_{j}(\partial_\beta p_j)\Pi_j=\partial_\beta \Tr[\rho_{\rm S,\eq}]=0.
\end{align}
Here, we have used that $\Pi_j=\Pi_j^2$ inside the derivative to obtain $\Tr(\Pi_j\partial_\beta \Pi_k)=\Tr(\Pi_j\partial_\beta \Pi_k^2)=2\Tr(\Pi_j\partial_\beta \Pi_j)\delta_{jk}$, which implies $\Tr(\Pi_j\partial_\beta \Pi_k)=0$ $\forall j,k$, and, jointly with the orthogonality of the projections, the second equality. Similarly, $\Tr[\partial_{\beta}\Pi_j]=\Tr[\partial_{\beta}\Pi_j^2]=2\Tr[\Pi_j\partial_{\beta}\Pi_j]=0$ and the third equality follows. If $\rho_{\rm S,\eq}$ is not full rank, or its rank depends on $\beta$, the proof follows using a more refined argument \cite{Wilde_entropy}.

Going back to \eqref{aux1}, since $\langle\log \rho_{\rm S,\eq}\rangle_{\eq}=-S_{\rm vN}(\rho_{\rm S,\eq})$, the condition $\langle \partial_\beta H_{\rm S}^\sharp\rangle_{\eq}=0$ implies
\begin{align}
    \log Z^{\sharp}_{\rm S}-\beta\partial_{\beta}\log Z^{\sharp}_{\rm S} - S_{\rm vN}(\rho_{\rm S,\eq})=0.
\end{align}
Rewriting this result in terms of the free energy
\begin{equation}
    F_{\rm S}^{\sharp}(\beta):=-\beta^{-1}\log Z^{\sharp}_{\rm S}(\beta),
\end{equation} 
and using the shorthand notation $S_{\rm vN}(\rho_{\rm S,\eq})\equiv S_{\rm vN,S}(\beta)$, we obtain
\begin{align}\label{aux2}
     \beta^2\partial_{\beta}F_{\rm S}^{\sharp}(\beta) - S_{\rm vN,S}(\beta)=0.
\end{align}
This implies that the thermodynamic entropy associated with the Hamiltonian of mean force $H^{\sharp}_{\rm S}$ is the von Neumann entropy (up to the Boltzmann factor $k_{B}$). This may be expected from the condition $\langle \partial_\beta H_{\rm S}^\sharp\rangle_{\eq}=0$ and Eq. \eqref{SH*}. Therefore
\begin{equation}\label{Shash}
    S_{\rm S}(\beta):=k_B\beta^2\partial_{\beta}F_{\rm S}^{\sharp}(\beta)=k_B S_{\rm vN,S}(\beta).
\end{equation}
On solving \eqref{aux2}, we get
\begin{align}
  F_{\rm S}^{\sharp}(\beta) = F_{\rm S}^{\sharp}(\beta_0)+\int_{\beta_0}^{\beta} \frac{S_{\rm vN,S}(\alpha)}{\alpha^2}d\alpha .
\end{align}
Taking the limit of zero temperature $\beta_0\to\infty$, 
\begin{align}\label{FHashS}
  F_{\rm S}^{\sharp}(\beta) = F_{\rm S}^{\sharp}(\infty)-\int_{\beta}^{\infty} \frac{S_{\rm vN,S}(\alpha)}{\alpha^2}d\alpha,
\end{align}
and so the partition function reads
\begin{align}\label{Zhash}
Z^{\sharp}_{\rm S}(\beta) = \exp\Bigg\{-\beta\bigg( F_{\rm S}^\sharp(\infty) -\int_{\beta}^{\infty}\frac{S_{\rm vN,S}(\alpha)}{\alpha^2}d\alpha\bigg)\Bigg\}.
\end{align}
These last results in \eqref{Hash} yield
\begin{align}\label{HashS}
    H_{\rm S}^\sharp&=F_{\rm S}^{\sharp}(\beta)-\beta^{-1}\log \rho_{\rm S,\eq}\nonumber\\
    &=F_{\rm S}^{\sharp}(\infty)-\beta^{-1}\log \rho_{\rm S,\eq}-\int_{\beta}^{\infty} \frac{S_{\rm vN,S}(\alpha)}{\alpha^2}d\alpha,
\end{align}
which is just a function of the equilibrium state of the system, up to the additive constant $F_{\rm S}^{\sharp}(\infty)$ that shifts the zero point energy. This is the only freedom that remains, matching the situation in standard weak coupling thermodynamics. The internal energy is
\begin{align}\label{EUeq}
  E_{\rm S}^{\sharp}(\beta) &:= F_{\rm S}^\sharp(\beta)+\beta \partial_\beta F_{\rm S}^\sharp(\beta)\nonumber\\
  &= F_{\rm S}^{\sharp}(\infty)-\int_{\beta}^{\infty} \frac{S_{\rm vN,S}(\alpha)}{\alpha^2}d\alpha+\frac{S_{\rm vN,S}(\beta)}{\beta}\nonumber\\
  &=\Tr(\rho_{\rm S,\eq} H_{\rm S}^\sharp),
\end{align}
as desired. Notice that the second equality can also be read as
\begin{equation}
    F_{\rm S}^{\sharp}(\beta)=E_{\rm S}^{\sharp}(\beta)-T S_{\rm S}(\beta),
\end{equation} 
which was anticipated on consistency grounds.

Therefore, the condition $E_{\rm S}^{\sharp}(\beta)=\langle H_{\rm S}^\sharp\rangle_{\eq}$ not only removes the ambiguity in the definition of the Hamiltonian of mean force and the reduced partition function, but it also leads to all thermodynamic variables at equilibrium depending solely on the system's degrees of freedom, through the properties of its reduced equilibrium state. Because of this, we shall refer to $H^\sharp_{\rm S}$ as the ``intrinsic'' Hamiltonian of mean force. In addition, the usual equivalence between informational and thermodynamic entropy is recovered via the von Neumann expression, and being its form universal, instead of writing $S^\sharp$ we denote it just by $S\equiv S^\sharp$.

\subsubsection{Consistency with weak-coupling case}
We can check that the previously defined thermodynamic variables approach the standard ones in the weak-coupling limit. In this limit, the steady state becomes the usual Gibbs state
\begin{equation}\label{Gibbs_sys}
    \rho_{\rm S,\eq}\xrightarrow{V_{\rm SR}\rightarrow 0}\rho_{\rm S,\beta}= \frac{e^{-\beta H_{\rm S}}}{Z_{\rm S}},
\end{equation}
and hence the von Neumann entropy becomes $S_{\rm vN,S}(\beta)=\beta\langle H_{\rm S}\rangle_\beta + \log Z_{\rm S}(\beta)$, with $\langle H_{\rm S}\rangle_\beta:=\Tr(\rho_{\rm S,\beta}H_{\rm S})$. Therefore
\begin{align}
   \int_{\beta}^\infty  \frac{S_{\rm vN,S}(\alpha)}{\alpha^2}d\alpha&=\int_{\beta}^\infty \left[\frac{\Tr(H_{\rm S} e^{-\alpha H_{\rm S}})}{\alpha Z_{\rm S}(\alpha)}+\frac{\log Z_{\rm S}(\alpha)}{\alpha^2}\right]d\alpha\nonumber \\
    &=\int_{\beta}^\infty \left[-\frac{\partial_\alpha Z_{\rm S}(\alpha)}{\alpha Z_{\rm S}(\alpha)}+\frac{\log Z_{\rm S}(\alpha)}{\alpha^2}\right]d\alpha\nonumber \\
    &=-\int_{\beta}^\infty \partial_\alpha [\alpha^{-1}\log Z_{\rm S}(\alpha)]d\alpha\nonumber \\
    &=\beta^{-1}\log Z_{\rm S}(\beta)=-F_{\rm S}(\beta).
\end{align}
This result in \eqref{FHashS} leads to $F_{\rm S}^{\sharp}(\beta)=F_{\rm S}^{\sharp}(\infty)+F_{\rm S}(\beta)$. Finally, using this equation and $-\beta^{-1}\log(\rho_{\rm S,\beta})=H_{\rm S}+\beta^{-1}\log Z_{\rm S}$ in \eqref{HashS} we straightforwardly obtain $H_{\rm S}^{\sharp}=H_{\rm S}+F^{\sharp}(\infty)$. Consequently, as expected, in the weak-coupling limit, the Hamiltonian of mean force and its associated free energy equal the system Hamiltonian and standard free energy, respectively, up to an arbitrary and irrelevant additive constant that we shall take $F_{\rm S}^\sharp(\infty)=0$ from now on.

\subsubsection{Connection between global and system changes}\label{sec:isothermal}
As noted previously in Sec.~\ref{Sec:IIA}, the total free energy change between two system-reservoir equilibrium states for a process where only the system Hamiltonian is modified $H_{\rm S}(0)\to H_{\rm S}(1)$ equals $\Delta F^*_{\rm S}$ \eqref{DeltaF}. Since $Z_{\rm S}^*\neq Z_{\rm S}^{\sharp}$, this free energy change is no longer associated exclusively to a system free energy change in the intrinsic Hamiltonian of mean force approach. The reason is that, although only the system Hamiltonian is modified during the process, the strong system-reservoir coupling distributes the free energy change across the entire system-reservoir composite. 

More specifically, if $S_{\rm SR}(\beta)=k_B S_{\mathrm{vN}, \rm SR}(\beta)$ denotes the entropy of the joint system-reservoir equilibrium state, the total free energy change can be written in terms of the internal energy and entropy changes as
\begin{equation}
    \Delta F_{\rm SR}=\Delta E_{\rm SR}-T\Delta S_{\rm SR},
\end{equation}
with $\Delta E_{\rm SR}=\langle H(1)\rangle_\beta-\langle H(0)\rangle_\beta$. Writing the total entropy using the quantum conditional entropy $S_{\rm R|S}:=S_{\rm SR}-S_{\rm S}$ of the reservoir conditioned on the system, namely the remaining uncertainty about the reservoir state after accounting for the uncertainty in the system, we obtain
\begin{equation}\label{F_parts}
    \Delta F_{\rm SR}=\Delta F_{\rm S}^{\sharp} + \Delta F_{\rm R|S}^{\sharp},
\end{equation}
where
\begin{align}\label{DFR|S}
    \Delta F_{\rm R|S}^{\sharp}:= \Delta E_{\rm R|S}^\sharp -T\Delta S_{\rm R|S}.
\end{align}
Here, $\Delta E_{\rm R|S}^\sharp := \Delta E_{\rm SR} - \Delta E_{\rm S}^{\sharp}$ is the change in the \emph{external energy} $E_{\rm R|S}^\sharp$; i.e., it represents the portion of the total energy change that is not accounted for by the change in the system's internal energy. Likewise, $\Delta S_{\rm R|S}$ accounts for the contribution to the total entropy variation \eqref{F_parts} that is not associated with changes in the system's entropy.

Thus, $\Delta F_{\rm SR} \neq \Delta F^{\sharp}_{\rm S}$ because, although the external driving modifies only the system Hamiltonian, the strong interaction can simultaneously alter the external energy, the reservoir entropy, and the system-reservoir correlations, resulting in $\Delta F^{\sharp}_{\rm R|S} \neq 0$. We recover $\Delta F_{\rm SR} = \Delta F^{\sharp}_{\rm S}$ only when the change in the reservoir's conditional entropy is exactly compensated by the change in the external energy, or when both trivially vanish, as in the weak-coupling limit.

It is interesting to further decompose the conditional free energy $\Delta F_{\rm R|S}^{\sharp}$ by introducing the reservoir internal energy $E_{\rm R}^{\sharp}$ [formally defined as the analog of Eq.~\eqref{EUeq} under the substitution ${\rm S}\to{\rm R}$], as well as the ``correlation'' (or ``interaction'') internal energy
\begin{equation}\label{Ecorr}
    E_{\rm corr}^{\sharp}:=E_{\rm SR}-E_{\rm S}^{\sharp}-E_{\rm R}^{\sharp}.
\end{equation}
Using the identity $S_{\rm SR}=S_{\rm S}+S_{\rm R}-I_{\rm SR}$, where $I_{\rm SR}$ is the quantum mutual information, which quantifies the amount of correlations shared between the system and the reservoir, we obtain
\begin{equation}\label{FEdecomp}
    \Delta F_{\rm SR}=\Delta F_{\rm S}^{\sharp} + \Delta F_{\rm R}^{\sharp} +\Delta F^\sharp_{\rm corr},
\end{equation}
where
\begin{align}
    \Delta F_{\rm R}^{\sharp}&:=\Delta E_{\rm R}^{\sharp} - T\Delta S_{\rm R},\label{DFR}\\
    \Delta F_{\rm corr}^{\sharp}&:=\Delta E_{\rm corr}^{\sharp} +T\Delta I_{\rm SR}.\label{DFC}
\end{align}
Here, $\Delta F_{\rm R}^{\sharp}$ is the change in the reservoir free energy, whereas $\Delta F_{\rm corr}^{\sharp}$ represents an additional contribution to the total free energy change arising from variations in the system-reservoir correlations. Notice the positive sign of the last term in $\Delta F_{\rm corr}^{\sharp}$, in contrast to the negative entropic contributions appearing in $\Delta F_{\rm S}^{\sharp}$ and $\Delta F_{\rm R}^{\sharp}$. The correlation free energy therefore increases with the mutual information, indicating that correlations constitute a source, rather than a sink, of free energy, unlike entropy or disorder. This provides a clear information-theoretic interpretation of the decomposition of the equilibrium free energy \eqref{FEdecomp}.

Solving $\beta^2\partial_\beta F_{\rm SR}(\beta)=S_{\rm SR}(\beta)$ and using the entropic identities above, the conditional, reservoir, and correlation free energies also admit the integral representations:
\begin{align}
    \Delta F_{\rm R|S}^{\sharp}&:=-\int_{\beta}^\infty  \frac{\Delta S_{\rm vN, R|S}(\alpha)}{\alpha^2}d\alpha,\label{DFR|S2}  \\
    \Delta F_{\rm R}^{\sharp}&:=-\int_{\beta}^\infty  \frac{\Delta S_{\rm vN,R}(\alpha)}{\alpha^2}d\alpha,\label{DFR2}\\
    \Delta F_{\rm corr}^{\sharp}&:=\int_{\beta}^\infty  \frac{\Delta I_{\rm SR}(\alpha)}{\alpha^2}d\alpha.\label{DFC2}
\end{align}
Needless to say, all expressions involving the reservoir energies or entropy should be understood in a formal sense. Although their variations may remain finite, these quantities are, strictly speaking, divergent because the reservoir is an exceedingly large system, potentially with a continuous spectrum of degrees of freedom.

\subsubsection{Heat capacity}
The fact that the thermodynamic entropy equals the von Neumann expression using the intrinsic Hamiltonian of mean force allows us to recover a familiar result in standard thermodynamics. The heat capacity, i.e. the rate of change of internal energy (heat) per temperature increment can be written as
\begin{equation}\label{Heat-cap}
    C_{\rm S}^\sharp(\beta):=\frac{\partial E^\sharp_{\rm S}}{\partial T}=-k_B \beta^2 \frac{\partial E^\sharp_{\rm S}}{\partial \beta}=-\beta \frac{\partial S_{\rm S}}{\partial \beta},
\end{equation}
where the last equality, satisfied also in standard thermodynamics, follows here after \eqref{EUeq}. 
Note that since the von Neumann entropy takes some finite value at zero temperature, and $S_{\rm S}(\beta)=S_{\rm S}(\infty)+\int_{\beta}^\infty \frac{1}{\alpha}C_{\rm S}^\sharp(\alpha) d\alpha$, the heat capacity must vanish at zero temperature, in analogy with standard thermodynamics. These results express a strong-coupling version of the Einstein statement of the third law \cite{Klimenko2012}.

In addition, using the Snider-Wilcox formula \cite{Snider64,Wilcox1967} for the parametric derivative of an exponential, it is not difficult to obtain the equivalent equation $C_{\rm S}^\sharp(\beta)=
(\Delta H^\sharp_{\rm S})^2_{\eq} +\beta\frac12 \big\langle \partial_{\beta}\big(H^{\sharp2}_{\rm S}\big)\big\rangle_{\eq}$. 

\subsection{Density of states at strong coupling}
In the weak-coupling equilibrium situation, the entropy provides the average level of uncertainty (or information) about the internal energy state of the system, which follows the Boltzmann distribution at equilibrium. In the general strong coupling case, one can ask which ``internal energy states'' of the system the entropy actually quantifies the uncertainty of. The answer is provided by the density of states distribution $\varrho^\sharp(\epsilon)$, introduced by the Laplace transformation formula:
\begin{equation}\label{DOS}
    Z_{\rm S}^{\sharp}(\beta)=:\int_0^\infty \varrho^\sharp(\varepsilon) e^{-\beta \varepsilon} d\varepsilon.
\end{equation}
In the standard, weak-coupling case, $Z_{\rm S}^{\sharp}(\beta)=Z_{\rm S}(\beta)=\Tr(e^{-\beta H_{\rm S}})$ and so $\varrho^\sharp(\varepsilon)\equiv\varrho(\varepsilon)=\sum_{n}{\rm deg}(\varepsilon_n)\delta(\varepsilon-\varepsilon_n)$, with $\varepsilon_n$ the eigenvalues of the system Hamiltonian $H_{\rm S}$ and ${\rm deg}(\varepsilon_n)$ their degeneracies. In the strong-coupling case, the density of states is generally obtained from the inverse Laplace transform of Eq.~\eqref{DOS}. An explicit example of this calculation is given in Sec.~\ref{sec:example_DOS}.

One of the most appealing features of the density of states is that it allows us to formulate statistical expressions for thermodynamic quantities that remain valid regardless of the strength of the coupling. Specifically, the internal energy \eqref{EUeq} can be written as
\begin{align}
E_{\rm S}^{\sharp}(\beta)&=\frac{1}{Z_{\rm S}^\sharp(\beta)}\int_0^\infty \varepsilon  e^{-\beta \varepsilon} \varrho^\sharp(\varepsilon) d\varepsilon\nonumber\\
&=\int_0^\infty \varepsilon  P_B(\epsilon) \varrho^\sharp(\varepsilon) d\varepsilon,
\end{align}
which corresponds to a mean value of the energy variable $\varepsilon$ in the Boltzmann distribution $P_B(\varepsilon):=e^{-\beta \varepsilon}/ Z_{\rm S}^\sharp(\beta)$ weighted by the density of states $\varrho^\sharp(\varepsilon)$. Similarly, the entropy \eqref{Shash} can be expressed as
\begin{equation}
    S_{\rm S}(\beta)=-k_B\int_0^\infty P_B(\varepsilon)\log [P_B(\varepsilon)] \varrho^\sharp(\varepsilon) d\varepsilon.
\end{equation}
Thus, the formal structure of these thermodynamic expressions is unchanged; all strong-coupling effects are encoded in the modified density of states $\varrho^\sharp(\varepsilon)$.

\section{Strong-coupling nonequilibrium thermodynamics}\label{Sec:IV}
In this section, we extend the above equilibrium strong-coupling thermodynamic formulation to the nonequilibrium case. To this end, we first consider the system and thermal reservoir to be initially in the equilibrium state \eqref{TotalGibbs}, and for $t>0$ the total Hamiltonian becomes time-dependent, $H(t)$. This corresponds to a physical situation in which certain parameters $\lambda$, on which the Hamiltonian depends, $H(\lambda)$, are externally driven in time, i.e., $\lambda \equiv \lambda_t$, so that we can write $H(t) \equiv H(\lambda_t)$.

The simplest situation arises when only the system Hamiltonian depends on the external driving, in which case the total Hamiltonian reads
\begin{equation}\label{Hamiltonian-t}
    H(t)=H_{\rm S}(t)+H_{\rm R}+V_{\rm SR}.
\end{equation}
The interaction term $V_{\rm SR}$ may also, in general, depend on time, for instance to describe situations such as the switching on or off of the system-reservoir interaction. The following discussion remains valid in this more general case, with the only difference being the experimental procedure required to determine work changes, which we comment on below.

\subsection{Work definition and fluctuation theorem}\label{sec:work-definition}
In accordance with the standard convention, we regard work as the energy cost of modifying the system's structural or dynamical constraints—effectively deforming the potential landscape—as opposed to the energy changes associated with the dynamical response, which are identified with heat and dissipation. 

Since the total system-reservoir composite is a closed system undergoing unitary evolution, its von Neumann entropy remains constant, and any change in its total energy is entirely attributed to work. More specifically, the total work applied to the closed system-reservoir composite, initially at equilibrium, satisfies the Jarzynski fluctuation equality \cite{StrasbergBook,EspositoReview}
\begin{equation}\label{Jarzynski_total}
    \braket{e^{-\beta w}}_{\{t,0\}}=e^{-\beta \Delta F_{\rm SR}(t)},
\end{equation}
where $w=\epsilon_{t}-\epsilon_0$ is a stochastic variable corresponding to the difference between the outcomes $\epsilon_t$ and $\epsilon_0$ of projective filtering measurements of the total Hamiltonian at times $t$ and $0$, respectively. The symbol $\braket{\cdot }_{\{t,0\}}$ denotes the statistical average over such two-time measurement results, and $\Delta F_{\rm SR}(t)=F_{\rm SR}(t)-F_{\rm SR}(0)$ is the difference between the equilibrium free energies associated with $H(t)$ and $H(0)$. 

The fact that only equilibrium free energy differences appear on the right-hand side of \eqref{Jarzynski_total} allows us to introduce the decomposition \eqref{F_parts}. Noticing that $\Delta F_{\rm R|S}^{\sharp}(t)$ is a macroscopic, deterministic quantity (i.e., it does not fluctuate with the outcomes of the quantum measurements), we can multiply both sides by $e^{\beta \Delta F_{\rm R|S}^{\sharp}(t)}$ and bring this factor inside the expectation value, leading to
\begin{equation}\label{Jarzynski_system}
    \braket{e^{-\beta w^\sharp}}_{\{t,0\}}=e^{-\beta \Delta F_{\rm S}^\sharp(t)},
\end{equation}
where $w^\sharp :=w -\Delta F_{\rm R|S}^{\sharp}(t)$.  This fluctuation relation strongly suggests that the work performed on (or by) the system is given by the total work minus the conditional free energy change $\Delta F_{\rm R|S}^{\sharp}(t)$.

Therefore, if the total instantaneous power is
\begin{equation}\label{power}
    \dot{W}(t)=\frac{d \langle H(t)\rangle }{dt}=\Tr[\rho_{\rm SR}(t)\dot{H}(t)],
\end{equation}
total work is
\begin{equation}\label{work}
    W(t)=\int_0^t \Tr[\rho_{\rm SR}(t')\dot{H}(t')]dt',
\end{equation}
and the system work is given by
\begin{equation}\label{W-sharp}
    W_{\rm S}^\sharp(t)=W(t)-\Delta F_{\rm R|S}^{\sharp}(t).
\end{equation}
Using Jensen’s inequality in \eqref{Jarzynski_system}, together with the identity $\braket{w}_{\{t,0\}}=W(t)$ \cite{StrasbergBook,EspositoReview}, we obtain the second law in the form
\begin{equation}\label{II-law0}
    W_{\rm S}^\sharp(t)\geq \Delta F_{\rm S}^\sharp(t).
\end{equation}
Crucially, both sides of this inequality can be determined by controlling and measuring the system alone, without requiring microscopic control of the reservoir. As discussed, $\Delta F_{\rm S}^\sharp$ depends only on the system state [see, e.g., \eqref{FHashS}], and the total work $W$ also depends only on the system state if the driving acts exclusively on the system Hamiltonian \eqref{Hamiltonian-t}, 
\begin{equation}\label{work_2}
    W(t)=\int_0^t \Tr[\rho_{\rm SR}(t')\dot{H}(t')]dt'=\int_0^t \Tr[\rho_{\rm S}(t')\dot{H}_{\rm S}(t')]dt'.
\end{equation}
To determine the correction $\Delta F_{\rm R|S}^{\sharp}$—which depends only on the initial and final Hamiltonians and not on the specific driving protocol $\lambda_t$—we may consider driving the system very slowly, such that the timescale of variation of $\lambda_t$ is much larger than the relaxation time. In this quasistatic limit, the system follows a trajectory of instantaneous equilibrium states, $\rho_{\rm S}(t)=\rho_{\rm S,eq}(t)$, and the total work \eqref{work} becomes
\begin{align} \label{W_rev}
    W_{\rm rev}(t)&=\int_0^t \Tr[\rho_{\rm S,eq}(t')\dot{H}_{\rm S}(t')]dt'\nonumber\\
    &=\int_0^t \Tr[\rho_{\rm SR,\beta}(t')\dot{H}(t')]dt'\nonumber\\
    &=-\beta^{-1}\int_0^t\frac{\dot{Z}_{\rm SR}(t')}{Z_{\rm SR}(t')} dt'=\Delta F_{\rm SR}(t)
\end{align}
where the penultimate equality follows, for example, from applying the Snider–Wilcox formula \cite{Snider64,Wilcox1967} for the parametric derivative of an exponential. Thus, since $\Delta F_{\rm SR}$ and $\Delta F_{\rm S}^\sharp$ depend solely on quantities defined on the system's Hilbert space, the correction $\Delta F_{\rm R|S}^{\sharp}=\Delta F_{\rm SR}-\Delta F_{\rm S}^\sharp=\int_0^t \Tr\{\rho_{\rm S,eq}(r)[\dot{H}_{\rm S}(r)-\dot{H}^\sharp_{\rm S}(r)]\}dr$ can be determined by controlling the system degrees of freedom alone.

Moreover, for quasistatic driving, inserting \eqref{W_rev} into \eqref{W-sharp} yields
\begin{equation}
    W^\sharp_{\rm S,rev}(t)=\Delta F_{\rm S}^{\sharp}(t),
\end{equation}
thereby saturating the inequality \eqref{II-law0}, as expected for a reversible quasistatic transformation. Now, if we write the total work as 
\begin{equation}\label{W-Wdiss}
    W(t)=\Delta F_{\rm SR}(t)+W_{\rm diss}(t),
\end{equation}
it follows from \eqref{W-sharp} that the system work takes the analogous form
\begin{equation}\label{Ws-Wdiss}
    W_{\rm S}^\sharp(t)=\Delta F_{\rm S}^\sharp(t)+W_{\rm diss}(t).
\end{equation}
In other words, the system work consists of the energy required to traverse the intrinsic equilibrium landscape ($\Delta F_{\rm S}^\sharp$) together with the thermodynamic friction generated in the entire system-bath complex ($W_{\rm diss}$). 

Complementarily, using \eqref{FEdecomp}--\eqref{DFC}, Eqs.~\eqref{W-sharp} and \eqref{Ws-Wdiss} can also be rewritten as
\begin{multline}\label{W-referee}
    W_{\rm S}^\sharp(t)=W(t) - [\Delta E_{\rm R}^\sharp(t) + \Delta E_{\rm corr}^\sharp(t)] \\
     + T \Delta S_{\rm R}(t) - T \Delta I_{\rm SR}(t).
\end{multline}
This expression makes transparent the relation between the total work and its system contribution. The changes in the reservoir and correlation equilibrium internal energies---which together constitute the change in the equilibrium external energy $\Delta E_{\rm R|S}(t)$---are subtracted from the total work because they are associated with degrees of freedom other than the system. Likewise, the energetic cost associated with reducing the equilibrium reservoir entropy and increasing the equilibrium system-reservoir correlations is not attributed to the system itself. Consequently, the terms $-T\Delta S_{\rm R}(t)$ and $T\Delta I_{\rm SR}(t)$ are also removed from the total work in order to isolate the contribution associated exclusively with the system. In the weak-coupling limit, because the instantaneous global equilibrium state factorizes and the driving acts only on the system, the equilibrium state of the reservoir remains constant in time. Consequently, $\Delta E_{\rm R}^\sharp(t)\to 0$, $\Delta E_{\rm corr}^\sharp(t)\to 0$, $\Delta S_{\rm R}(t)\to 0$, and $\Delta I_{\rm SR}(t)\to 0$, naturally recovering $W_{\rm S}^\sharp(t)\to W(t)$.

Similar expressions to \eqref{Ws-Wdiss} and \eqref{W-referee} are straightforwardly obtained for the reservoir work $W_{\rm R}^\sharp$ (exchanging ${\rm S}\leftrightarrow{\rm R}$) and the correlation work $W_{\rm corr}^\sharp$, satisfying $W(t)=W_{\rm S}^\sharp(t)+W_{\rm R}^\sharp(t)+W_{\rm corr}^\sharp(t)$.

Finally, if the interaction term becomes also explicitly driving dependent, $V_{\rm SR}\to V_{\rm SR}(t)$, the above reasoning still applies [provided that $V_{\rm SR}(t)$ remains quasilocal to ensure that the dynamics stays within the thermodynamic regime in which the composite system is expected to thermalize in the absence of driving, Eq. \eqref{TotalGibbs}]. However, in that case the total work $W$, and therefore the system work $W_{\rm S}^\sharp$, are no longer functions solely of the reduced system state. Their determination becomes more demanding, but it remains possible provided that the observer can measure the energy variation of the external work system (e.g., a battery, a light field, etc.) supplying or consuming the energy exchanged with the thermodynamic setup.

\subsection{Nonequilibrium first and second laws}\label{Sec:IVB}
In order to formulate fully general nonequilibrium thermodynamic laws, consider the relation
\begin{equation}
\tilde{F}(t)=\tilde{E}(t)-TS(t)
\end{equation}
usually employed in standard weak-coupling thermodynamics to define the nonequilibrium free energy $\tilde{F}(t)$ associated with a general state $\rho(t)$, where $\tilde{E}(t)$ and $S(t)$ are its nonequilibrium internal energy and entropy, respectively. This equation can be equivalently rewritten as
\begin{equation}\label{Ftilde-F}
\tilde{F}(t)-F(t)= \beta^{-1}D[\rho(t)\Vert\rho_{\rm eq}(t)].
\end{equation}
Thus, the difference between the nonequilibrium and equilibrium free energies at time $t$ corresponds to the ``difference''---as measured by the relative entropy---between the corresponding nonequilibrium and equilibrium states.

Relation \eqref{Ftilde-F} is particularly appealing because it admits a straightforward generalization to the strong-coupling case. Indeed, the actual state $\rho(t)$ and the corresponding equilibrium state $\rho_{\rm eq}(t)$ are universally well-defined objects, the equilibrium free energy is already provided by $F_{\rm S}^\sharp(t)$, and the quantum relative entropy functional $D[\cdot\Vert\cdot]$ retains its standard form provided we retain the von Neumann expression for the entropy in the nonequilibrium regime,
\begin{equation}
    S_{\rm S}(t)=-k_{B}\Tr[\rho_{\rm S}(t)\log\rho_{\rm S}(t)].
\end{equation}
Therefore, the nonequilibrium free energy is naturally extended to arbitrary coupling strengths as
\begin{equation}\label{tildeF}
    \tilde{F}_{\rm S}^\sharp(t):=F_{\rm S}^\sharp(t)+\beta^{-1}D[\rho_{\rm S}(t)\Vert\rho_{\rm S,eq}(t)].
\end{equation}
Assuming the usual thermodynamic relation
\begin{equation}\label{FTS}
    \tilde{F}_{\rm S}^\sharp(t)=\tilde{E}^\sharp_{\rm S}(t) -T S_{\rm S}(t),
\end{equation}
Equation~\eqref{tildeF} leads directly to the following definition for the nonequilibrium internal energy:
\begin{equation}\label{tildeE}
    \tilde{E}^\sharp_{\rm S}(t):= \Tr[\rho_{\rm S}(t)H^{\sharp}_{\rm S}(t)].
\end{equation}
The first law is then written as
\begin{align}\label{first_law}
    \dot{\tilde{E}}_{\rm S}^\sharp(t)=\dot{Q}_{\rm S}^\sharp(t)+\dot{W}_{\rm S}^\sharp(t),
\end{align}
where the heat is given by
\begin{align}\label{Heat-t}
    Q^{\sharp}_{\rm S}(t):= \Tr[\rho_{\rm S}(t)H^{\sharp}_{\rm S}(t)]-\Tr[\rho_{\rm S}(0)H^{\sharp}_{\rm S}(0)]- W_{\rm S}^\sharp(t).
\end{align}

To formulate the second law in terms of nonequilibrium thermodynamic potentials, we first note that 
\begin{multline}\label{IILaw_aux0}
   D[\rho_{\rm SR}(t)\Vert\rho_{\rm SR,\beta}(t)]-D[\rho_{\rm SR}(0)\Vert\rho_{\rm SR,\beta}(0)] \\
   =\beta[W(t)-\Delta F_{\rm SR}(t)]. 
\end{multline}
Here, we have used the fact that the system–reservoir composite is a closed system, consequently, its von Neumann entropy remains constant, and any change in its energy equals the work, $W(t)=\Tr[\rho_{\rm SR}(t) H(t)]-\Tr[\rho_{\rm SR}(0) H(0)]$. Equation \eqref{IILaw_aux0}, together with \eqref{W-sharp}, allows us to write the system work as
\begin{equation}
    W_{\rm S}^\sharp(t)= \Delta F_{\rm S}^\sharp(t) + \beta^{-1}\Delta D_{\rm SR}(t),
\end{equation}
where $\Delta D_{\rm SR}(t)=D_{\rm SR}(t)-D_{\rm SR}(0)$ and we have adopted the shorthand notation $D_{\rm SR}(t)\equiv D[\rho_{\rm SR}(t)\Vert\rho_{\rm SR,\beta}(t)]$. 

In addition, from \eqref{tildeF}, we  write $\Delta \tilde{F}_{\rm S}^\sharp = \Delta F_{\rm S}^\sharp + \beta^{-1}\Delta D_{\rm S}$, where the analogous shorthand $D_{\rm S}(t)\equiv D[\rho_{\rm S}(t)\Vert\rho_{\rm S,eq}(t)]$ has been used. Therefore, we obtain
\begin{multline}\label{II-law_aux1}
        \beta \left[W_{\rm S}^\sharp(t)-\Delta \tilde{F}_{\rm S}^\sharp(t)\right] = \Delta D_{\rm SR}(t)-\Delta D_{\rm S}(t)\\
        =[D_{\rm SR}(t)-D_{\rm S}(t)]-[D_{\rm SR}(0)-D_{\rm S}(0)].
\end{multline}
From this expression, the second law in the form
\begin{equation}\label{II-law}
    W_{\rm S}^\sharp(t)\geq\Delta \tilde{F}_{\rm S}^\sharp(t)
\end{equation}
follows provided that $\Delta D_{\rm SR}(t)-\Delta D_{\rm S}(t)\geq 0$. This condition holds under general thermodynamic initial conditions, as we shall discuss shortly.

Alternatively, using \eqref{FTS} together with the first law \eqref{first_law}, Eq.~\eqref{II-law} can be rewritten as the positive entropy production inequality
\begin{equation}\label{II-lawEntropy}
    \Sigma_{\rm S}(t):=\Delta S_{\rm S}(t)-\frac{Q_{\rm S}^\sharp(t)}{T}\geq0.
\end{equation}

\subsubsection{Equilibrium initial condition under Hamiltonian driving} 
Consider first the same initial condition used in Sec. \ref{sec:work-definition}, where the system and reservoir start in the equilibrium state:
\begin{equation}\label{equi-ic}
    \rho_{\rm SR,\beta}=\frac{e^{-\beta H(0)}}{Z_{\rm SR}(0)},
\end{equation}
and the system Hamiltonian (or, more generally, the interaction term as well) is subsequently varied for $t>0$. Under this initial condition, we have $D_{\rm SR}(0)=D_{\rm S}(0)=0$, so that the second law \eqref{II-law} follows from \eqref{II-law_aux1} due to the monotonicity of the quantum relative entropy under partial trace \cite{Lindblad1975,Uhlmann1977}, namely $D_{\rm SR}(t)\ge D_{\rm S}(t)$. Note that \eqref{II-law} is a tighter inequality than \eqref{II-law0} due to the positivity of the relative entropy in \eqref{tildeF}. 

\subsubsection{Product initial condition}
Consider now the case where the system and reservoir are brought into contact at time $t=0$, so that the initial condition is a product state $\rho_{\rm SR}(0)=\rho_{\rm S}(0)\otimes\rho_{\rm R,\beta}$, as in \eqref{rhoSRt}. In the strong-coupling regime, this initial condition is thermodynamically ambiguous because the fact that the interaction term is nonzero during the evolution does not clarify what its value is at the very initial time $t=0$. This is a critical point in the strong-coupling regime, as the contribution of the interaction cannot be neglected in the energy balance. 

To make sense of the product initial condition in strong-coupling thermodynamics, we must assume that $V(0)=0$ and account for the non-negligible thermodynamic cost of switching on the interaction. Under this situation, $\rho_{\rm SR,\beta}(0)=\rho_{\rm S,\beta}\otimes\rho_{\rm R,\beta}$ and therefore
\begin{align}\label{D_prod}
    D_{\rm SR}(0)&=D[\rho_{\rm S}(0)\otimes \rho_{\rm R,\beta}(0)\Vert\rho_{\rm S,\beta}(0)\otimes \rho_{\rm R,\beta}(0)]\nonumber \\
    &=D[\rho_{\rm S}(0)\|\rho_{\rm R,\beta}(0)]=D_{\rm S}(0)
\end{align}
where we have used the identity $D(\rho_1 \otimes \rho_2 \| \sigma_1 \otimes \sigma_2) = D(\rho_1 \| \sigma_1) + D(\rho_2 \| \sigma_2)$. Hence, Eq.~\eqref{D_prod} jointly with the monotonicity of the relative entropy under partial trace, $D_{\rm SR}(t)\ge D_{\rm S}(t)$, implies the validity of the second law \eqref{II-law}, with the work given by the general expression \eqref{W-sharp}. 

A particularly interesting case arises when the interaction is suddenly switched on at the initial time via a quench $V(t)=V\theta(t)$. In this case, the system work just after the quench becomes
\begin{align}\label{Wquench}
    W_{\rm S}^\sharp(0^+)=\Tr[\rho_{\rm S}(0)\otimes \rho_{\rm R,\beta}V]-\Delta F_{\rm R|S}^{\sharp}(0^+).
\end{align}
If, as commonly assumed \cite{RivasHuelga}, the first moment of the interaction vanishes for the reservoir Gibbs state, $\Tr_{\rm R}[\rho_{\rm R,\beta}V]=0$, the system work reduces to $W_{\rm S}^\sharp(0^+)=-\Delta F_{\rm R|S}^{\sharp}(0^+)$. Thus, even though no external energy is injected during the sudden quench (on average), the system effectively behaves as if work had been extracted from it, because a portion of the system internal energy must be spent to establish the quantum correlations with the reservoir required by the post-quench Hamiltonian $H$. 

In the weak-coupling limit, $W_{\rm S}^\sharp(0^+)=0$ in any case, and the standard thermodynamics of the weak coupling follows. The entropy production \eqref{II-lawEntropy} becomes
\begin{equation}
      \Delta S_{\rm S}(t)-\frac{Q_{\rm S}(t)}{T}\geq0,
\end{equation}
with $Q_{\rm S}(t)=\int_0^t\Tr[\dot{\rho}_{\rm S}(t')H_{\rm S}(t')]dt'$. Note that this does not imply the positivity of the entropy production rate  $\dot{S}_{\rm S}(t)-\dot{Q}_{\rm S}(t)/T\geq0$ even in this limit. Ensuring a positive rate typically requires slow driving in comparison to the reservoir correlation time \cite{Alicki79,StrasbergBook}.

\subsubsection{General thermodynamic initial conditions}
One cannot assume that arbitrary initial conditions for the system and reservoir will produce dynamics satisfying the second law \eqref{II-law}. Classically, this can be illustrated \cite{StrasbergBook} by a scenario where a rare fluctuation causes all the velocities of an equilibrium gas to suddenly align in the same direction, resulting immediately afterward in a state of lower entropy. The second law is expected to be satisfied only under specific initial conditions for the system and reservoir, requiring the latter to behave as an invariant source of thermal entropy that drives the system to equilibrate with it at the same temperature. This property defines what can be called ``thermodynamic initial conditions'' and corresponds quantitatively to those fulfilling the equality
\begin{equation}\label{Petzequality}
    D[\rho_{\rm SR}(0)\|\rho_{\rm SR,\beta}(0)] = D[\rho_{\rm S}(0)\|\rho_{\rm S,eq}(0)].
\end{equation}
In other words, under thermodynamic initial conditions, the entropic difference, as measured by the relative entropy, between the system-reservoir state $\rho_{\rm SR}(0)$ and the global equilibrium state $\rho_{\rm SR,\beta}(0)$ reduces entirely to the system's contribution, $D[\rho_{\rm S}(0)\Vert\rho_{\rm S,eq}(0)]$. The reservoir part adds nothing, making its net effect identical to the thermal equilibrium case, leaving the system as the sole independent source of nonequilibrium in the composite.

Now, the equality \eqref{Petzequality} is achieved if and only if (see \cite{Petz1986,Hayden2004}) the total state $\rho_{\rm SR}$ is connected to its reduced part $\rho_{\rm S}$ by the Petz adjoint map $\mathcal{P}$ associated with the partial trace:
\begin{align}\label{Petz_map}
    \rho_{\rm SR}(0) &= \mathcal{P}[\rho_{\rm S}(0)] \nonumber\\
    &= \rho_{\rm SR,\beta}^{1/2}(0) \Big[\rho_{\rm S,eq}^{-1/2}(0) \rho_{\rm S}(0) \rho_{\rm S,eq}^{-1/2}(0) \otimes \mathds{1}_{\rm R} \Big] \rho_{\rm SR,\beta}^{1/2}(0).
\end{align}
The product and the equilibrium initial conditions are particular cases for which this relation is satisfied: the first corresponds to an arbitrary $\rho_{\rm S}(0)$ with $V(0)=0$, and the second to $\rho_{\rm S}(0)=\rho_{\rm S,\eq}(0)$.

To better visualize the physical meaning of \eqref{Petz_map}, it is useful to consider the classical case in which all density matrices are diagonal, with eigenvalues given by the probabilities of obtaining a particular pure state (or microstate) labeled by $j$ for the system and $k$ for the reservoir: $p_{\rm SR}(j,k)$, $p_{\rm SR,\beta}(j,k)$, $p_{\rm S}(j)$, and $p_{\rm S,eq}(j)$, respectively. Since diagonal matrices commute, \eqref{Petz_map} simplifies in this case to
\begin{equation}\label{Petz_classical}
    p_{\rm SR}(j,k) = \frac{p_{\rm SR,\beta}(j,k)}{p_{\rm S,eq}(j)}p_{\rm S}(j) = p_{\rm S}(j)p_{\rm R,\beta}(k|j),
\end{equation}
i.e. the probability of finding the reservoir in a specific pure state $k$, conditional on the system being in pure state $j$, is governed exactly by the equilibrium conditional probability $p_{\rm R,\beta}(k|j)$. To prepare one of these states, we simply construct a statistical mixture after measuring the system part of a joint equilibrium state, $p_{\rm SR,\beta}(j,k) = p_{\rm S,eq}(j)p_{\rm R,\beta}(k|j)$. Namely, starting from a large ensemble of these systems all initially in the joint equilibrium state, we measure the system part of each and group them by the outcome. We then purposely filter these groups—discarding or keeping specific proportions—such that the relative frequency of each group perfectly matches the desired target system probability $p_{\rm S}(j)$.

The condition \eqref{Petzequality} is the noncommutative quantum counterpart of this classical framework, defining the thermodynamic initial states when quantum coherences are present. In that case, if one considers decompositions of the system Hilbert space $\mathcal{H}_{\rm S}=\bigoplus_{j=1}^J\mathcal{H}_{{\rm S}^{\rm A}_j}\otimes \mathcal{H}_{{\rm S}^{\rm B}_j}$, and writes the Gibbs state as
\begin{equation}\label{rhoSRbeta-decomposition}
    \rho_{\rm SR,\beta}(0)=\bigoplus_{j=1}^J p_{\eq,j} \sigma_{{\rm S}^{\rm A}_j}\otimes \sigma_{{\rm S}^{\rm B}_j {\rm R}}
\end{equation}
with $p_{\eq,j}$ a probability distribution, $\sigma_{{\rm S}^{\rm A}_j}$ a density matrix in the Hilbert space $\mathcal{H}_{{\rm S}^{\rm A}_j}$ and $\sigma_{{\rm S}^{\rm B}_j{\rm R}}$ a density matrix in the Hilbert space $\mathcal{H}_{{\rm S}^{\rm B}_j}\otimes \mathcal{H}_{\rm R}$, the condition \eqref{Petzequality} is equivalent to requiring an initial state that can be written in the form:
\begin{equation}\label{rhoSR0-decomposition}
    \rho_{\rm SR}(0)=\bigoplus_{j=1}^Jp_j \rho_{{\rm S}^{\rm A}_j}\otimes \sigma_{{\rm S}^{\rm B}_j {\rm R}}
\end{equation}
with $\rho_{{\rm S}^{\rm A}_j}$ being an arbitrary density matrix in $\mathcal{H}_{{\rm S}^{\rm A}_j}$. This is a consequence of the results in \cite{Petz1986,Hayden2004} but, for the sake of completeness, we write down the explicit proof in Appendix \ref{app:A0}.

Thus, the variety of allowed thermodynamic initial conditions $\rho_{\rm SR}(0)$ in the quantum case depends on how many ways the state $\rho_{\rm SR,\beta}(0)$ can be decomposed in the form \eqref{rhoSRbeta-decomposition}. For $V(0)=0$—or in the weak-coupling limit $V\simeq 0$—we have $\rho_{\rm SR,\beta}(0)=\rho_{\rm S,\beta}\otimes \rho_{\rm R,\beta}$ [which corresponds to $J=1$, $\mathcal{H}_{{\rm S}^{\rm A}_1}=\mathcal{H}_{\rm S}$, and $\dim(\mathcal{H}_{{\rm S}^{\rm B}_1})=1$], and there are an infinite number of thermodynamic initial conditions of the form $\rho_{\rm SR}(0)=\rho_{\rm S}(0)\otimes\rho_{\rm R,\beta}$. In the opposite extreme, if the only way to decompose the Gibbs state is the trivial one $\big[J=1$, $\dim(\mathcal{H}_{{\rm S}^{\rm A}_1})=1$, $\mathcal{H}_{{\rm S}^{\rm B}_1}=\mathcal{H}_{\rm S}$, yielding $p_1=1$, $\sigma_{{\rm S}^{\rm A}_1}=1$, and $\sigma_{{\rm S}^{\rm B}_1 {\rm R}}=\rho_{\rm SR,\beta}(0)\big]$, then $\rho_{\rm SR}(0)=\rho_{\rm SR,\beta}(0)$ is the only possible thermodynamic initial condition.

The classical result \eqref{Petz_classical} naturally emerges from this structure when $J=\dim(\mathcal{H}_{\rm S})$ and $\dim(\mathcal{H}_{{\rm S}^{\rm A}_j})=\dim(\mathcal{H}_{{\rm S}^{\rm B}_j})=1$. In this scenario, the quantum freedom inside the blocks vanishes, leaving only the classical variables $p_j \equiv p_{\rm S}(j)$ and the diagonal reservoir states $\sigma_{{\rm S}^{\rm B}_j\rm R}$ with eigenvalues $p_{\rm R,\beta}(k|j)$.

In analogy with the classical case, any valid quantum thermodynamic initial condition \eqref{rhoSR0-decomposition} can be prepared starting from the joint equilibrium state $\rho_{\rm SR,\beta}(0)$ by utilizing only local operations on the system $\rm S$. First, one performs a non-destructive projective measurement on the system to determine the classical block label $j$. Because the joint equilibrium state is block-diagonal, this measurement does not disturb the correlations with the reservoir. Once the system is projected into block $j$, the ${\rm S}^{\rm A}_j$ part is tensor-factored from the reservoir. Therefore, one can apply arbitrary local quantum operations (e.g., measurements, unitaries, or re-preparations) strictly within $\mathcal{H}_{{\rm S}^{\rm A}_j}$ to transform the equilibrium state $\sigma_{{\rm S}^{\rm A}_j}$ into the desired target state $\rho_{{\rm S}^{\rm A}_j}$. Finally,  by statistically filtering these conditional states such that their relative frequencies match the target $p_j$, one perfectly constructs $\rho_{\rm SR}(0)$.

Ultimately, this reveals a profound feature of quantum thermodynamics: it is the necessity to preserve genuine quantum correlations in the equilibrium state that acts as the fundamental source of restriction on the initial states. Unlike classical correlations, which can be freely conditioned and filtered, arbitrary local operations on a quantum system will generally destroy non-classical correlations such as entanglement and discord shared with the reservoir. Therefore, valid thermodynamic initial conditions are strictly confined to manipulating the entirely ``disentangled'' subspaces $\mathcal{H}_{{\rm S}^{\rm A}_j}$, while avoiding the creation of quantum coherences between different classical blocks $j$.

\subsubsection{Initial system operations on general thermodynamic states}
In the cases considered so far, the entropy of the system-reservoir composite remains invariant. One may instead consider situations in which some general thermodynamic state \eqref{Petz_map} is prepared and, before any driving protocol, the system is subject to an arbitrary quantum operation [not necessarily preserving the block structure \eqref{rhoSR0-decomposition}] represented by a completely positive and trace-preserving (CPTP) map $\mathcal{E}$ and assumed instantaneous for the sake of simplicity (e.g., a non-selective measurement), changing the state according to
\begin{equation}\label{postmeas}
\rho_{\rm SR}(0^+)=\mathcal{E}\otimes\mathcal{I}[\rho_{\rm SR}(0)].
\end{equation}
Here, $\mathcal{I}$ denotes the identity map on the reservoir part. Since any quantum map is the result of unitary evolution on an enlarged space that includes auxiliary degrees of freedom (e.g., a measurement apparatus), this scenario can be formally embedded into our previous paradigm by simply redefining the thermodynamic system to include these extra variables (see \cite{Strasberg2019} for an example of using this idea in quantum thermodynamics). However, it is highly desirable to determine the conditions under which the second law remains valid for the system alone, without requiring an explicit entropic accounting of these auxiliary degrees of freedom.

To this end, note that the equilibrium free energy does not change during the instantaneous initial operation; so from \eqref{tildeF}, we find
\begin{equation}\label{meas1}
\Delta\tilde{F}_{\rm S}^{\sharp}(0^+)=\beta^{-1}\Delta D_{\rm S}(0^+).
\end{equation}
On the other hand, since $\Delta F_{\rm R|S}(0^+)=0$, a straightforward calculation yields
\begin{align}\label{meas2}
    \beta^{-1}\Delta D_{\rm SR}(0^+)&=-T\Delta S_{\rm SR}(0^+)+W(0^{+})\nonumber\\
    &=-T\Delta S_{\rm SR}(0^+)+W_{\rm S}^{\sharp}(0^{+}),
\end{align}
where 
\begin{equation}
    \Delta S_{\rm SR}(0^+)=k_B S_{\rm vN}[\rho_{\rm SR}(0^+)]- k_B S_{\rm vN}[\rho_{\rm SR}(0)],
\end{equation}
is the total entropy change during the operation $\mathcal{E}$, and $W(0^+)=\Tr\{[\rho_{\rm SR}(0^+)-\rho_{\rm SR}(0)] H(0)\}$ is the work performed. Because we assume that the local action of $\mathcal{E}$ on the system is instantaneous compared to the system-reservoir dynamics, no heat is exchanged during this process. Hence, subtracting \eqref{meas1} from \eqref{meas2}, we find that during the operation $\mathcal{E}$:
\begin{multline}\label{meas3}
    \Delta D_{\rm SR}(0^+)-\Delta D_{\rm S}(0^+)\\=\beta\left[W_{\rm S}^{\sharp}(0^{+})-\Delta \tilde{F}_{\rm S}^{\sharp}(0^+)-T\Delta S_{\rm SR}(0^+)\right].
\end{multline}
Now, taking the state \eqref{postmeas} at $t=0^+$ as the initial condition for the subsequent unitary evolution, the same steps used to derive \eqref{II-law_aux1} now yield
\begin{equation}\label{meas4}
        \Delta D_{\rm SR}(t,0^+)-\Delta D_{\rm S}(t,0^+)=\beta \left[W_{\rm S}^\sharp(t,0^{+})-\Delta\tilde{F}_{\rm S}^\sharp(t,0^+)\right],
\end{equation}
where
\begin{equation}
    W_{\rm S}^\sharp(t,0^+)=\int_{0^+}^t \Tr[\rho_{\rm SR}(t')\dot{H}(t')]dt'-\Delta F_{\rm R|S}^{\sharp}(t,0^+),
\end{equation}
and we have introduced the shorthand notation $\Delta X(t,0^+):=X(t)-X(0^+)$. Therefore, since $\Delta X(t,0^+)+\Delta X(0^+)=\Delta X(t)=X(t)-X(0)$, adding \eqref{meas3} and \eqref{meas4} we arrive at
\begin{equation}\label{meas5}
        \Delta D_{\rm SR}(t)-\Delta D_{\rm S}(t)=\beta \left[W_{\rm S}^\sharp(t)-\Delta \tilde{F}_{\rm S}^\sharp(t)-T\Delta S_{\rm SR}(0^+)\right],
\end{equation}
where $W_{\rm S}^\sharp(t)=W_{\rm S}^\sharp(t,0^+)+W_{\rm S}^{\sharp}(0^{+})$ is the total accumulated system work, including the work performed by $\mathcal{E}$. Finally, as we have assumed that the state before the action of $\mathcal{E}$ is a thermodynamic state \eqref{Petz_map}, we conclude that
\begin{equation}\label{IILaw_aux3}
    W_{\rm S}^\sharp(t)\geq\Delta \tilde{F}_{\rm S}^\sharp(t)+T\Delta S_{\rm SR}(0^+).
\end{equation}
To recover the second law in the form \eqref{II-law}, certain conditions on the operation $\mathcal{E}$ are required. Consider, for instance, a non-selective projective measurement, for which $\mathcal{E}(\rho)=\sum_{x}\Pi_x\rho\Pi_x$. It is well known that this action does not decrease the von Neumann entropy, so $\Delta S_{\rm SR}(0^+)\geq0$, and thus \eqref{II-law} follows directly from \eqref{IILaw_aux3}.

More generally, the only maps for which $\Delta S_{\rm SR}(0^+)\geq0$ for all states are the unital ones, satisfying $\mathcal{E}(\mathds{1})=\mathds{1}$ (this is a straightforward consequence of the monotonicity of the relative entropy under CPTP maps \cite{Lindblad1975,Uhlmann1977}). In terms of the Kraus operators $M_x$ of $\mathcal{E}$, this condition requires $\sum_{x}M_xM_x^\dagger=\sum_x M_x^\dagger M_x=\mathds{1}$. Consider now a general POVM with effects $E_x=M_x^\dagger M_x$. The polar decomposition allows us to write $M_x=U_x\sqrt{E_x}$, where $U_x$ is a unitary operator on the system (or, in the general infinite-dimensional case, a partial isometry). If the operators $U_x$ are independent of $x$, then the map is unital and $\Delta S_{\rm SR}(0^+)\geq0$, leading again to the second law. However, if this is not the case, the operations $U_x$ represent a form of post-measurement physical operation on the system that is explicitly conditioned on the outcome $x$. In other words, this feedback formally acts as a Maxwell's demon, preventing the derivation of the usual second law $W_{\rm S}^\sharp(t)\geq\Delta \tilde{F}_{\rm S}^\sharp(t)$ in general. In such situations, the only consistent strategy is to include the measurement apparatus as part of the thermodynamic description to properly account for its entropic changes.

It should be emphasized that this derivation ensures the validity of the second law over the entire interval $[0, t]$. However, this does not imply that the second law must hold if one considers $t=0^+$ as the starting point. In fact, unless the measurement preserves the block structure \eqref{rhoSR0-decomposition}, the post-measurement state $\rho_{\rm SR}(0^+)$ will generally not be a thermodynamic initial state. Consequently, the dynamics in the interval $[0^+, t]$ could exhibit apparent violations of the standard second law, which are only compensated for by the entropic and energetic cost of the measurement.

\section{Worked example: oscillator coupled to a composite bosonic reservoir}\label{sec:example}
\begin{figure}[t]
	\includegraphics[width=0.9\columnwidth]{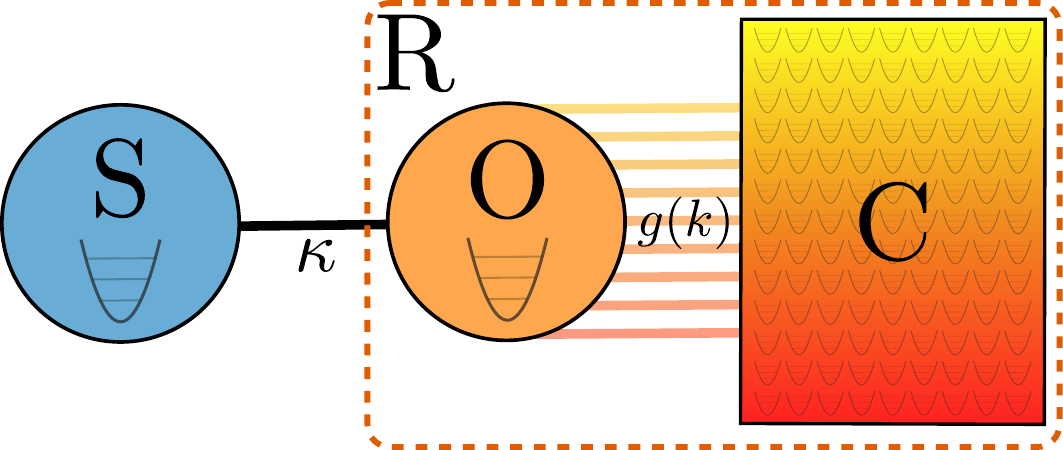}
	\caption{Schematics of the model. The system oscillator $\mathrm{S}$ is potentially strongly coupled, via the constant $\kappa$, to one of the oscillators $\mathrm{O}$ in the reservoir $\mathrm{R}$, which in turn is weakly coupled, via $g(k)$, to the rest of the continuum reservoir oscillators $\mathrm{C}$. See the main text for a detailed description.}
	\label{Fig:1}
\end{figure}

To illustrate the utility of the previous approach, we apply it to describe the thermodynamic properties of an oscillator system strongly coupled to a structured reservoir. More specifically, we consider a quantum harmonic oscillator $\mathrm{S}$ with Hamiltonian $H_{\rm S}=\hbar\omega_0 a^\dagger a$ coupled to a composite thermal reservoir consisting of a discrete oscillator $\mathrm{O}$ with Hamiltonian $H_{\rm O}=\hbar\omega_0 d^\dagger d$, and a continuum $\mathrm{C}$ of one-dimensional oscillator modes with Hamiltonian $H_{\rm C}=\hbar\int dk  \omega(k) b^\dagger(k)b(k)$. Here $a$, $d$ and $b(k)$ stand for the annihilation operators for the system, the discrete oscillator and a $k$-wavevector mode of the continuum, respectively. From now on we shall take units of $\hbar=k_B=1$. The discrete oscillator is potentially strongly coupled to $\mathrm{S}$ via the interaction 
\begin{equation}
    V_{\rm SO}=\kappa (a^\dagger d+a d^\dagger),
\end{equation}
and weakly coupled to the continuum by
\begin{equation}
V_{\rm OC}=\int dk g(k) (d+d^\dagger )[b(k)+b^\dagger(k)],
\end{equation}
with $g(k)$ some coupling function. Thus, the system $\mathrm{S}$ and the continuum part $\mathrm{C}$ only interact through the discrete oscillator $\mathrm{O}$ (see Fig. \ref{Fig:1}). This type of composite environment arises naturally in cavity QED and optomechanical setups (see, e.g., \cite{Aspelmeyer2014}), and, in its fermionic counterpart, in the description of impurities in solid-state environments (e.g., \cite{Ludovico2014}). Furthermore, a large class of other physical environments can be mapped to these models using the pseudomode, or reaction coordinate, method \cite{Garraway1997,Tamascelli2018,Pleasance2020}. 

Since the coupling between $\mathrm{O}$ and $\mathrm{C}$ is assumed to be weak, we can apply a Markovian treatment with the initial condition
\begin{equation}
    \rho_{\rm SR}(0)\simeq \rho_{\rm SO}(0)\otimes \rho_{\rm C,\beta},
\end{equation}
and derive a master equation for the dynamics of the $\mathrm{SO}$ composite. This is done following the usual Davies procedure \cite{Davies1,BrPe02,alicki-book,RivasHuelga}. First, we diagonalize the Hamiltonian 
\begin{align}\label{SO-Hamil}
    H_{\rm SO}&=H_{\rm S}+H_{\rm O}+V_{\rm SO}=\begin{bmatrix}
        a^{\dagger} & d^{\dagger}
    \end{bmatrix}
    \begin{bmatrix}
        \omega_0 & \kappa \\
        \kappa & \omega_0
    \end{bmatrix}
    \begin{bmatrix}
        a \\
        d 
    \end{bmatrix} 
    \nonumber \\
    & = \begin{bmatrix}
        c_1^{\dagger} & c_2^{\dagger}
    \end{bmatrix}
    \begin{bmatrix}
        \omega_1 & 0 \\
        0 & \omega_2
    \end{bmatrix}
    \begin{bmatrix}
        c_1 \\
        c_2 
    \end{bmatrix}=\omega_1 c^\dagger_1 c_1 +\omega_2 c^\dagger_2 c_2,
\end{align}
where $\omega_1=\omega_0+\kappa$, $\omega_2=\omega_0-\kappa$, and
\begin{equation}\label{adc1c2}
    \begin{bmatrix}
        a \\
        d
    \end{bmatrix} = \frac{1}{\sqrt{2}}
    \begin{bmatrix}
        1 & 1 \\
        1 & -1
    \end{bmatrix}
    \begin{bmatrix}
        c_1 \\
         c_2
    \end{bmatrix}.
\end{equation}
In terms of $c_1$ and $c_2$, the interaction between the discrete oscillator and the continuum of modes is written as
\begin{equation}
    V_{\rm OC}=\int dk \frac{g(k)}{\sqrt2} (c_1+c_1^\dagger-c_2-c_2^\dagger)[b(k)+b^\dagger(k)].
\end{equation}
Since $c_1$ and $c_2$ are the eigenoperators of $H_{\rm SO}$, with Bohr frequencies $-\omega_1$ and $-\omega_2$, respectively, $[H_{\rm SR},c_j]=-\omega_j c_j$, and $\omega_1-\omega_2=2\kappa\neq 0$, the master equation reads
\begin{align}\label{master_eq}
    \frac{d\rho_{\rm SO}}{dt}&=-\ii[H_{\rm SO},\rho_{\rm SO}]+\sum_{j=1}^2\Big\{\tfrac{\gamma}{2} [\bar{n}_j+1]\big(c_j\rho_{\rm SO}c_j^\dagger\nonumber \\
    &-\tfrac{1}{2}\{c_j^\dagger c_j,\rho_{\rm SO}\}\big)+\tfrac{\gamma}{2} \bar{n}_j \big(c_j^\dagger \rho_{\rm SO}c_j-\tfrac{1}{2}\{c_j c_j^\dagger,\rho_{\rm SO}\}\big)\Big\}.
\end{align}
Here, $\gamma=\gamma_j=2\pi J(\omega_j)$, and $\bar{n}_j=[\exp(\beta \omega_j)-1]^{-1}$ is the mean number of quanta with frequency $\omega_j$ of the continuum. The spectral density of the continuum is given by $J(\omega):=g^2[k(\omega)]\frac{dk(\omega)}{d\omega}$, with $k(\omega)$ the inverted function of the ``dispersion'' relation $\omega(k)$. For the sake of simplification, we take a flat spectral density in the frequency range that includes $\omega_1$ and $\omega_2$, so $\gamma_j=\gamma$. Furthermore, we neglect the Lamb shift term (or assume that $\omega_{1,2}$ have already been renormalized).

\subsection{Equilibrium steady state}\label{Sec:VA}
The master equation \eqref{master_eq} describes the relaxation process $\rho_{\rm SO}(0)\to \rho_{\rm SO,\beta}$, with $\rho_{\rm SO,\beta}=\exp(-\beta H_{\rm SO})/Z_{\rm SO}$ the equilibrium steady state. From the total system-reservoir composite perspective, the convergence \eqref{TotalGibbs} can be expressed, for this model, as
\begin{equation}\label{convExample}
\rho_{\rm SO} \otimes \rho_{\rm C,\beta}\xrightarrow{t\to\infty}\rho_{\rm SO,\beta}\otimes \rho_{\rm C,\beta},
\end{equation}
because the continuum $\mathrm{C}$ is weakly coupled to the oscillator $\mathrm{O}$. In addition, when the coupling between $\mathrm{S}$ and the reservoir $\mathrm{R}=\mathrm{OC}$ is weak [i.e. for $\kappa\to 0$ keeping $\kappa/\gamma$ fixed to not jeopardize the secular approximation in \eqref{master_eq}], the steady state factorizes to the product of usual Gibbs states $\rho_{\rm SO,\beta}\otimes \rho_{\rm C,\beta}\simeq \rho_{\rm S,\beta} \otimes \rho_{\rm O,\beta} \otimes \rho_{\rm C,\beta}\simeq \rho_{\rm S,\beta}\otimes \rho_{\rm OC,\beta}$.

In the general case, \eqref{convExample} implies that the equilibrium state of $\mathrm{S}$ is obtained by tracing out the discrete oscillator $\mathrm{O}$ from $\rho_{\rm SO,\beta} = e^{-\beta H_{\rm SO}} / Z_{\rm SO}$. Since the state is Gaussian, it is enough to determine the values of its covariance matrix
\begin{equation}\label{covariance_matrix}
    \sigma = 
    \begin{bmatrix}
        2(\braket{q^2} - \braket{q}^2) & \braket{qp+pq}-2\braket{q}\braket{p}  \\
        \braket{qp+pq}-2\braket{q}\braket{p} & 2(\braket{p^2} - \braket{p}^2)
    \end{bmatrix},
\end{equation}
where the position $q$ and momentum $p$ operators can be written as
\begin{align}
    q &= \frac{1}{\sqrt{2}}(a+a^{\dagger})=\frac{1}{2}(c_1+c_1^{\dagger}+c_2+c_2^{\dagger})\label{SHO_q}, \\
    p &= \frac{1}{\ii\sqrt{2}}(a-a^{\dagger}) =\frac{1}{2\ii}(c_1-c_1^{\dagger}+c_2-c_2^{\dagger}),\label{SHO_p}
\end{align}
in terms of the eigenoperators of $H_{\rm SO}$. A mechanical calculation with the state $\rho_{\rm SO,\beta}$ leads to $\braket{q}=\braket{p}=\braket{qp+pq}=0$ and
\begin{align}\label{CovM-eq}
    \sigma&=2\begin{bmatrix}
        \braket{q^2} & 0 \\
        0 & \braket{p^2}
    \end{bmatrix}\nonumber\\
    &= \Bigg(\frac{1}{e^{\beta(\omega_0+\kappa)}-1}+\frac{1}{e^{\beta(\omega_0-\kappa)}-1}+1\Bigg) \begin{bmatrix}
        1 & 0 \\
        0 & 1 
    \end{bmatrix}.
\end{align}
The von Neumann entropy of a single mode Gaussian state $\rho$ can be calculated as \cite{adesso-review,Weedbrook12,Serafini23},
\begin{align}\label{vn_entropy_gauss}
    S_{\rm vN}(\rho)=\frac{(\nu+1)}{2}\log \left(\frac{\nu + 1}{2}\right)-\frac{(\nu - 1)}{2}\log \left(\frac{\nu - 1}{2}\right)
\end{align}
where $\nu$ is the symplectic eigenvalue of $\sigma$ given by 
\begin{align}\label{nu_simplectic}
    \nu&=\sqrt{\det\sigma}=2\sqrt{\langle q^2\rangle \langle p^2\rangle}\nonumber\\
    &=\Bigg(\frac{1}{e^{\beta(\omega_0+\kappa)}-1}+\frac{1}{e^{\beta(\omega_0-\kappa)}-1}+1\Bigg).
\end{align} 

\begin{figure}[t]
	\includegraphics[width=\columnwidth]{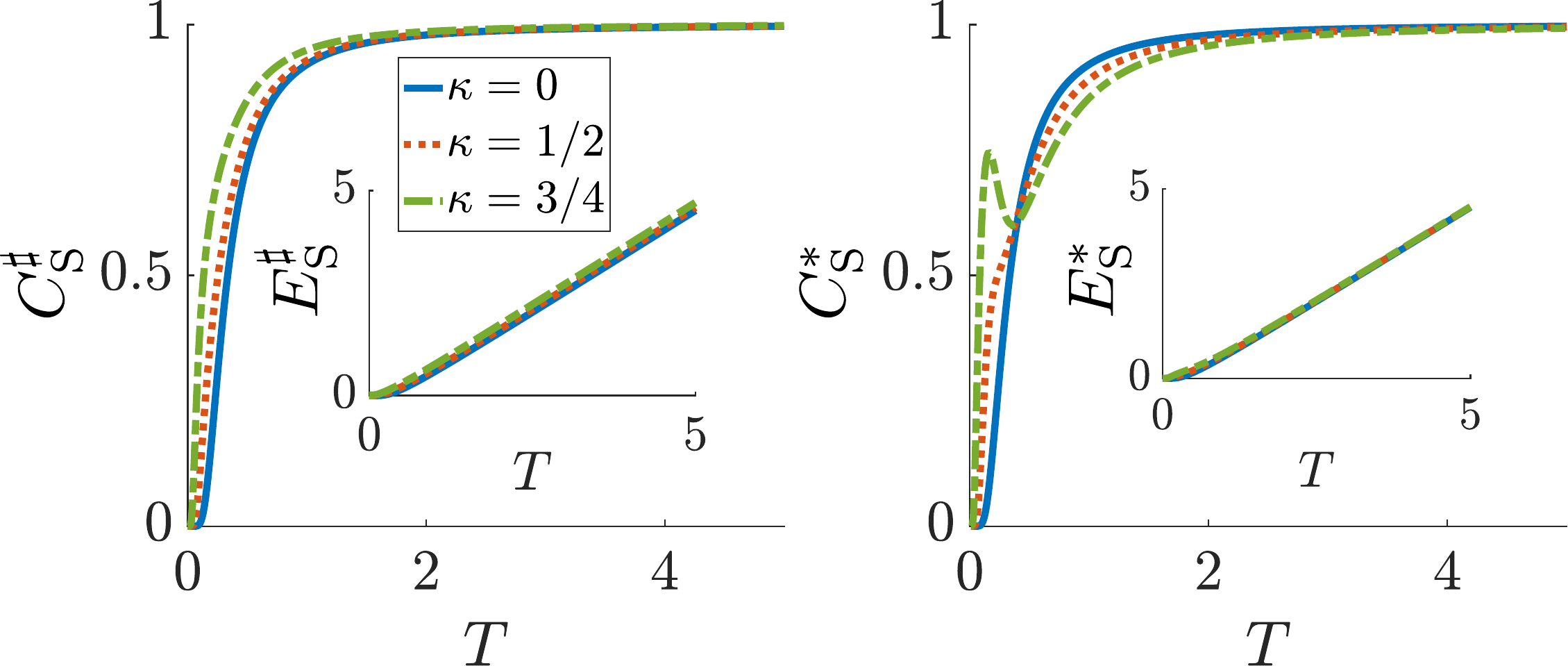}
	\caption{Equilibrium energy (inset) and heat capacity as a function of temperature for the intrinsic (left) and usual (right) Hamiltonian of mean force. Plots for different values of the coupling $\kappa$ are shown. All quantities are in units of $\omega_0=1$.}
	\label{Fig:2}
\end{figure}

\subsection{Equilibrium internal energy and heat capacity}
Using \eqref{EUeq} we obtain the internal energy from the von Neumann entropy, and by differentiation, the heat capacity \eqref{Heat-cap}. Figure \ref{Fig:2} shows a comparison of the value of the heat capacity and the internal energy (inset plot) as a function of the coupling constant and temperature. The behavior with $T$ is similar to the standard case corresponding to $\kappa=0$, both quantities are monotonically increasing functions of the temperature. On the other hand, increasing the coupling $\kappa$ leads to higher values of both $C_{\rm S}^\sharp$ and $E_{\rm S}^\sharp$. 

For the sake of comparison, the result for the usual Hamiltonian of mean force partition function $Z^*_{\rm S}$ is also plotted on the right side of Fig. \ref{Fig:2}. This is calculated from \eqref{E*_Z} with
\begin{equation}
    Z_{\rm S}^*=\frac{Z_{\rm SR}}{Z_{\rm R}}\simeq \frac{Z_{\rm SO}}{Z_{\rm O}}=\frac{1-e^{-\beta \omega_0}}{(1-e^{-\beta \omega_1})(1-e^{-\beta \omega_2})}, 
\end{equation}
where we have used again that the coupling between $\mathrm{O}$ and $\mathrm{C}$ is negligible. In this case, the heat capacity shows an oscillation for small values of $T$, and larger couplings lead to smaller $C^*$ for large temperatures, i.e. the opposite behavior to $C^\sharp$.

\begin{figure}[t]
	\includegraphics[width=\columnwidth]{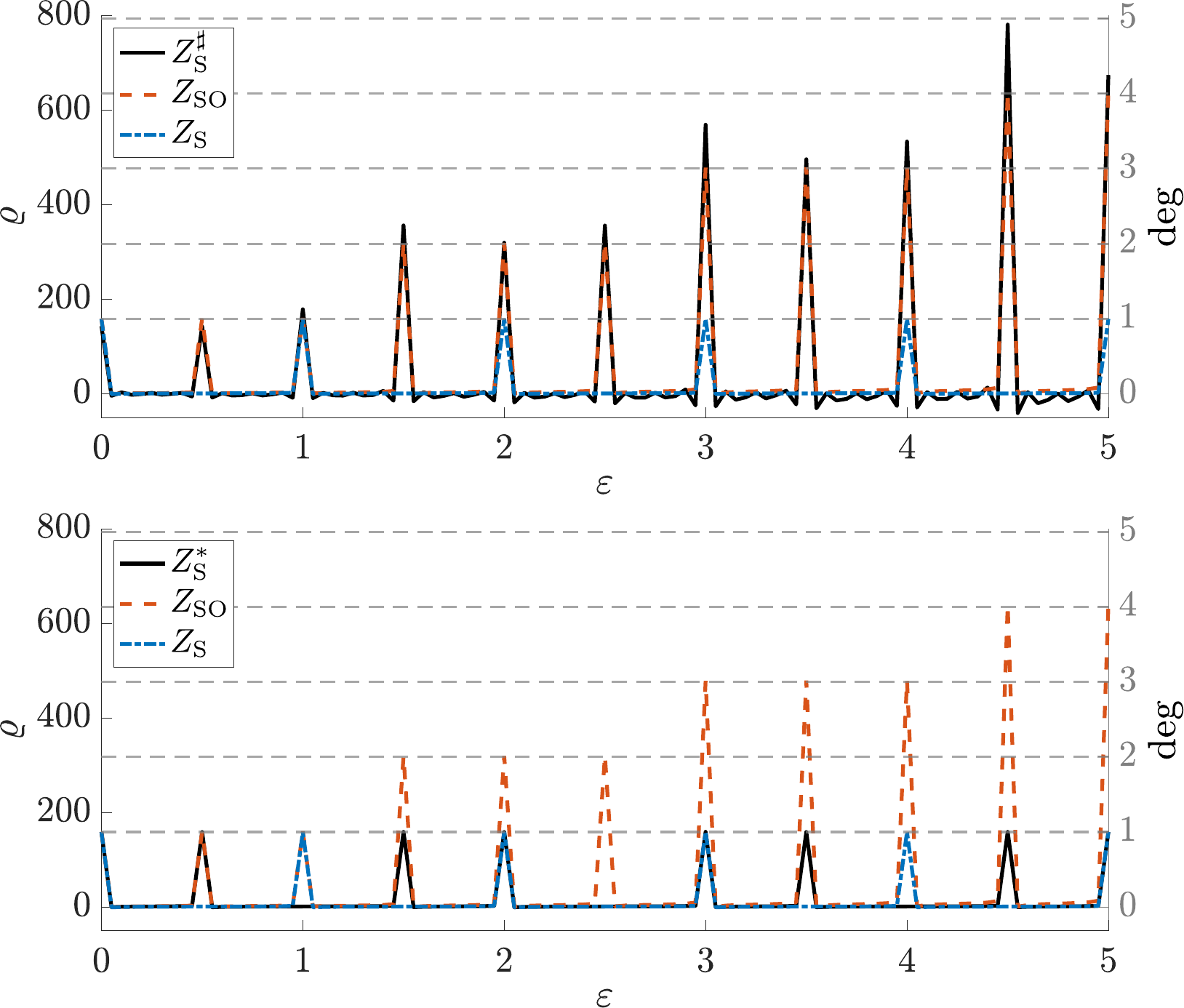}
	\caption{Comparison of the density of states for $Z_{\rm S}^{\sharp}$ (top) and $Z_{\rm S}^{*}$ (bottom) with the ones obtained with $Z_{\rm S}$ and $Z_{\rm SO}$. In these plots $\kappa=1/2$. The right axis indicates the degeneracy of each energy level corresponding to the height of the approximated delta functions. All quantities are in units of $\omega_0=1$.}
	\label{Fig:3}
\end{figure}

\subsection{Density of states}\label{sec:example_DOS}
To calculate the density of states $\varrho^\sharp(\varepsilon)$ we must invert the Laplace transform in \eqref{DOS}. This is difficult because, on the one hand, there is no analytical closed form for $Z^\sharp_{\rm S}$ for this model, so the integral in \eqref{Zhash} must be calculated numerically. On the other hand, the inverse Laplace transform of $Z_{\rm S}^\sharp(\beta)$ can be highly singular, with a proper meaning only in the distributional sense. Thus, in order to numerically calculate the inverse Laplace transform, $\mathfrak{L}^{-1}$, we introduce a regularizing factor and calculate
\begin{equation}
    \varrho^\sharp_{\epsilon}(\varepsilon)=\mathfrak{L}^{-1}\left[\frac{(1 - e^{-\epsilon\beta})}{\epsilon\beta}Z_{\rm S}^{\sharp}(\beta)\right]
\end{equation} 
for some small $\epsilon$. In the limit of vanishing $\epsilon$, $\varrho^\sharp_{\epsilon}(\varepsilon)$ approaches $\varrho^\sharp(\varepsilon)$ in the distributional sense, in similar fashion as 
\begin{multline}
    \lim_{\epsilon\downarrow0}\mathfrak{L}^{-1}\left[\frac{(1 - e^{-\epsilon\beta})}{\epsilon\beta}e^{-\varepsilon_0\beta}\right]\\    =\lim_{\epsilon\downarrow0}\frac{\chi_{[\varepsilon_0,\varepsilon_0-\epsilon]}(\varepsilon)}{\epsilon}=\delta(\varepsilon-\varepsilon_0).
\end{multline}

\begin{figure}[t]
	\includegraphics[width=\columnwidth]{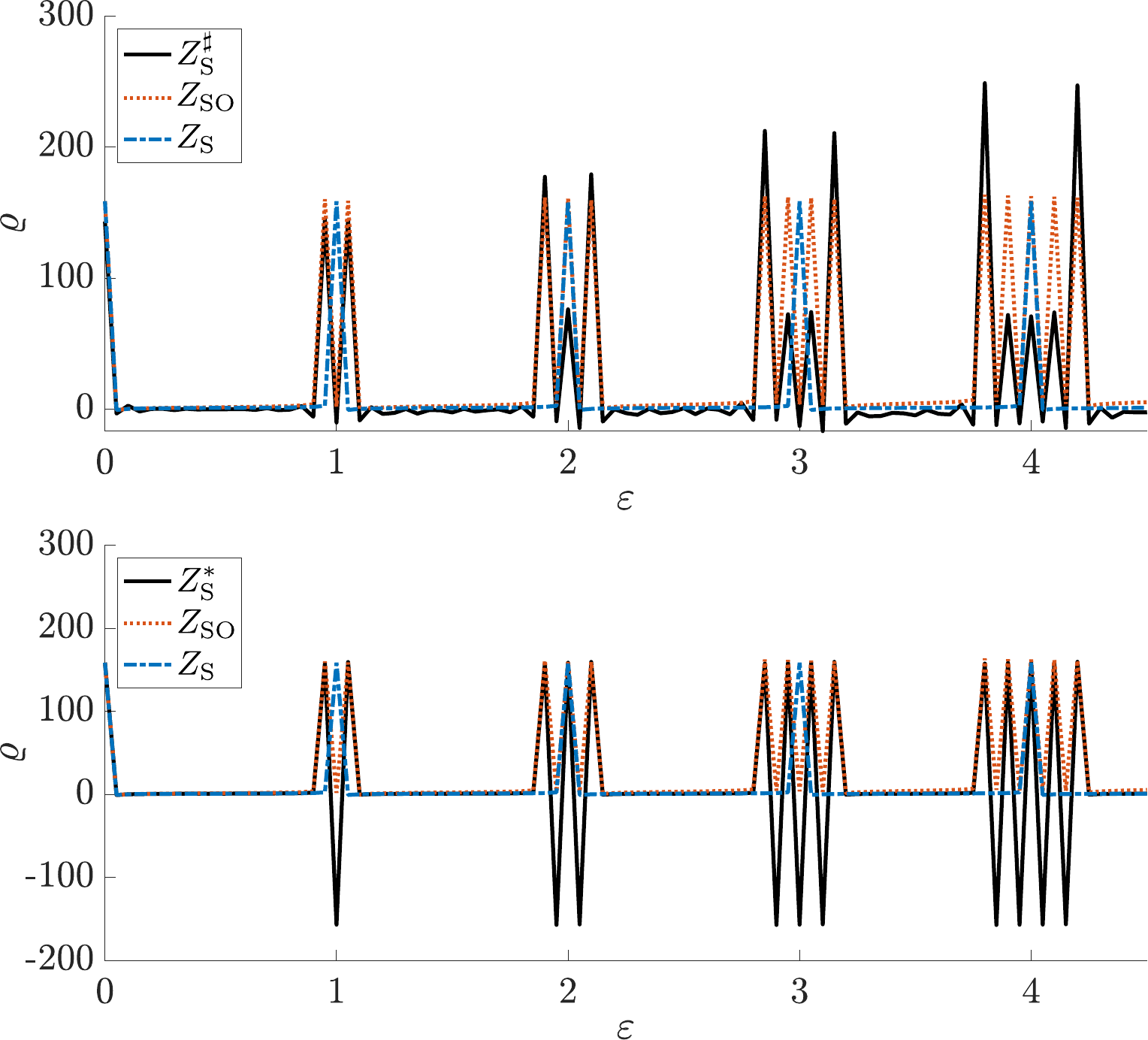}
	\caption{Comparison of the density of states for $Z_{\rm S}^{\sharp}$ (top) and $Z_{\rm S}^{*}$ (bottom) with the ones obtained with $Z_{\rm S}$ and $Z_{\rm SO}$. In these plots $\kappa=1/20$. All quantities are in units of $\omega_0=1$.}
	\label{Fig:4}
\end{figure}

In Figs. \ref{Fig:3} (for $\kappa=\omega_0/2$) and \ref{Fig:4} (for $\kappa=\omega_0/20$) we show the result of this approximate numerical inverse Laplace transform of $Z_{\rm S}^{\sharp}(\beta)$, for $\epsilon=10^{-3}$, obtained by the Piessens algorithm \cite{Piessens}. For the sake of comparison, the same approximate numerical inverse Laplace transform applied to $Z_{\rm SO}(\beta)$, $Z_{\rm S}(\beta)$ (i.e. $\kappa=0$) and $Z_{\rm S}^{*}(\beta)$ is included. However, in these last three cases, the result can be obtained analytically:
\begin{align}
 &Z_{\rm SO}(\beta)=\frac{1}{(1-e^{-\beta \omega_1})(1-e^{-\beta \omega_2})}=\sum_{n,m}e^{-\beta(n\omega_1+m\omega_2)} \nonumber \\
 &\Rightarrow \varrho_{\rm SO}(\varepsilon)=\mathfrak{L}^{-1}[Z_{\rm SO}(\beta)]=\sum_{n,m=0}^{\infty}\delta[\varepsilon-(n\omega_1+m\omega_2)],\label{DSO}
\end{align}
similarly, 
\begin{align}\label{D*}
    Z_{\rm S}^*(\beta)
    &= \frac{(1-e^{-\beta \omega_0})}
    {(1-e^{-\beta \omega_1})(1-e^{-\beta \omega_2})}
    \Rightarrow \varrho^*(\varepsilon)
    = \mathfrak{L}^{-1}[Z_{\rm S}^*(\beta)]\nonumber \\
    &= \begin{aligned}[t]
        &\sum_{n,m=0}^{\infty}
        \delta[\varepsilon-(n\omega_1+m\omega_2)]\\
        &-\sum_{n,m=0}^{\infty}
        \delta[\varepsilon-(n\omega_1+m\omega_2+\omega_0)]
    \end{aligned}
\end{align}
and ${\varrho}_{\rm S}(\varepsilon)=\mathfrak{L}^{-1}[Z_{\rm S}(\beta)]=\sum_{n=0}^{\infty}\delta(\varepsilon-n\omega_0)$, as it is well-known for a single harmonic oscillator. 

We may see in both figures that the density of states for $Z_{\rm S}^{\sharp}(\beta)$, $\varrho^\sharp$, is sensitive to the dressed Hamiltonian levels of $\mathrm{S}$ due to the strong coupling with $\mathrm{O}$, showing the same peaks with some variations in height as the density of states of $\mathrm{SO}$, $\varrho_{\rm SO}$. In fact, in Fig. \ref{Fig:3} the degeneracy of every peak of $\varrho_{\rm SO}$ and $\varrho^\sharp$ has been indicated on the right axis for comparison. The result for $Z_{\rm S}^{*}(\beta)$ does not have the same peak structure, with missing peaks with respect to $\varrho_{\rm SO}$ in Fig. \ref{Fig:3} and becoming rather exotic with negative deltas in Fig. \ref{Fig:4}, as can also be seen directly from the exact form \eqref{D*}. As a point of interest, the quantity $\varrho^*$ was previously calculated for a damped harmonic oscillator and no indication of negative values was found \cite{Hanke1995}. 

As a check, in Appendix \ref{app:A}, we calculate analytically $\varrho^\sharp(\varepsilon)$ to the first nontrivial order in $\kappa$ showing it has the correct weak-coupling limit.

\subsection{Nonequilibrium thermodynamics for product initial conditions}
We now focus on the nonequilibrium thermodynamics. We will choose the system to be initially prepared in a vacuum or squeezed vacuum state $\rho_{\rm S}=\ketbra{\xi}{\xi}$, with squeezing parameter $\xi=re^{\ii\theta}$, so that the initial state of $\mathrm{SO}$ is given by
\begin{equation}\label{xirhobeta}
\rho_{\rm SO}(0)=\ketbra{\xi}{\xi}\otimes \rho_{\rm O,\beta}.
\end{equation}
As discussed in Sec. \ref{Sec:IVB}, in order to make this initial condition coherent with the thermodynamic formalism, we shall consider the interaction $V_{\rm SO}$ to be suddenly turned on at $t=0$. Since the initial state does not have enough time to change due to this sudden quench, the only thermodynamical effect during this process will be a finite system work after the quench given by \eqref{Wquench}, which in this case reads
\begin{multline}
    W_{\rm S}^\sharp(0^+)=-\Delta F_{\rm R|S}^{\sharp}=-\Delta F_{\rm SR}+\Delta F_{\rm S}^{\sharp}\\
    =\beta^{-1}\log\dfrac{Z_{\rm SR}(0^{+})}{Z_{\rm SR}(0)}-\beta^{-1}\log\dfrac{Z_{\rm S}^{\sharp}(0^+)}{Z_{\rm S}^{\sharp}(0)}\\
    \simeq\beta^{-1}\log\dfrac{Z_{\rm SO}(0^{+})}{Z_{\rm SO}(0)}-\beta^{-1}\log\dfrac{Z_{\rm S}^{\sharp}(0^+)}{Z_{\rm S}^{\sharp}(0)}\\
    =\beta^{-1}\log\dfrac{Z_{\rm SO}(0^{+})}{Z_{\rm S}(0)Z_{\rm O}(0)}-\beta^{-1}\log\dfrac{Z_{\rm S}^{\sharp}(0^+)}{Z_{\rm S}(0)}\\
    =\beta^{-1}\log\dfrac{1-e^{-\beta\omega_0}}{(1-e^{-\beta\omega_1})(1-e^{-\beta\omega_2})}-\int_{\beta}^{\infty}\dfrac{S_{\rm vN}(\alpha)}{\alpha^2}d\alpha,
\end{multline}
where we have used that the interaction between O and C is weak, and \eqref{Zhash}, with $S_{\rm vN}(\alpha)$ given by \eqref{vn_entropy_gauss} and \eqref{nu_simplectic}.

To solve the dynamics of the system from $t=0^{+}$ onward, we apply the master equation \eqref{master_eq} with this initial condition and trace out the oscillator's degrees of freedom. Again, since the initial state $\rho_{\rm SO}(0)$ is Gaussian, and Gaussianity is preserved by the dynamics of \eqref{master_eq}, we only need to study the time dependence of the first moments and the covariance matrix $\sigma(t)$, Eq. \eqref{covariance_matrix}. This is done in detail in Appendix \ref{app:B}, obtaining
\begin{equation}\label{firstmoments_vanish}
\braket{q}_t=\braket{p}_t=0,    
\end{equation}
and
\begin{align}\label{sigmat}
    \sigma(t)= \begin{bmatrix}
        \sigma_{11}(t) & \sigma_{12}(t) \\
        \sigma_{21}(t) & \sigma_{22}(t)
    \end{bmatrix}
\end{align}
with 
\begin{multline}\label{sigma1122}
    \sigma_{jj}(t)=e^{-\frac{\gamma}{2}t}\bigg\{2\cos^2(\kappa t)\sinh^2 r+2\frac{\sin^2(\kappa t)}{e^{\beta \omega_0}-1}\\
    +(-1)^j \cos ^2(\kappa t) \cos(2\omega_0 t-\theta) \sinh 2r\bigg\}\\
    +(1-e^{-\frac{\gamma}{2}t})(\bar{n}_1+\bar{n}_2)+1,
\end{multline}
and
\begin{equation}\label{sigma1221}
    \sigma_{12}(t)=\sigma_{21}(t)=e^{-\frac{\gamma}{2}t} \cos ^2(\kappa t)\sin(2\omega_0 t-\theta) \sinh 2r.
\end{equation}
The entropy is then calculated by \eqref{vn_entropy_gauss} with the symplectic eigenvalue
\begin{equation}\label{nu_general}
    \nu(t)=\sqrt{\det \sigma(t)}=\sqrt{\sigma_{11}(t)\sigma_{22}(t)-\sigma_{12}^2(t)}.
\end{equation}

\begin{figure}[t]
 	\includegraphics[width=\columnwidth]{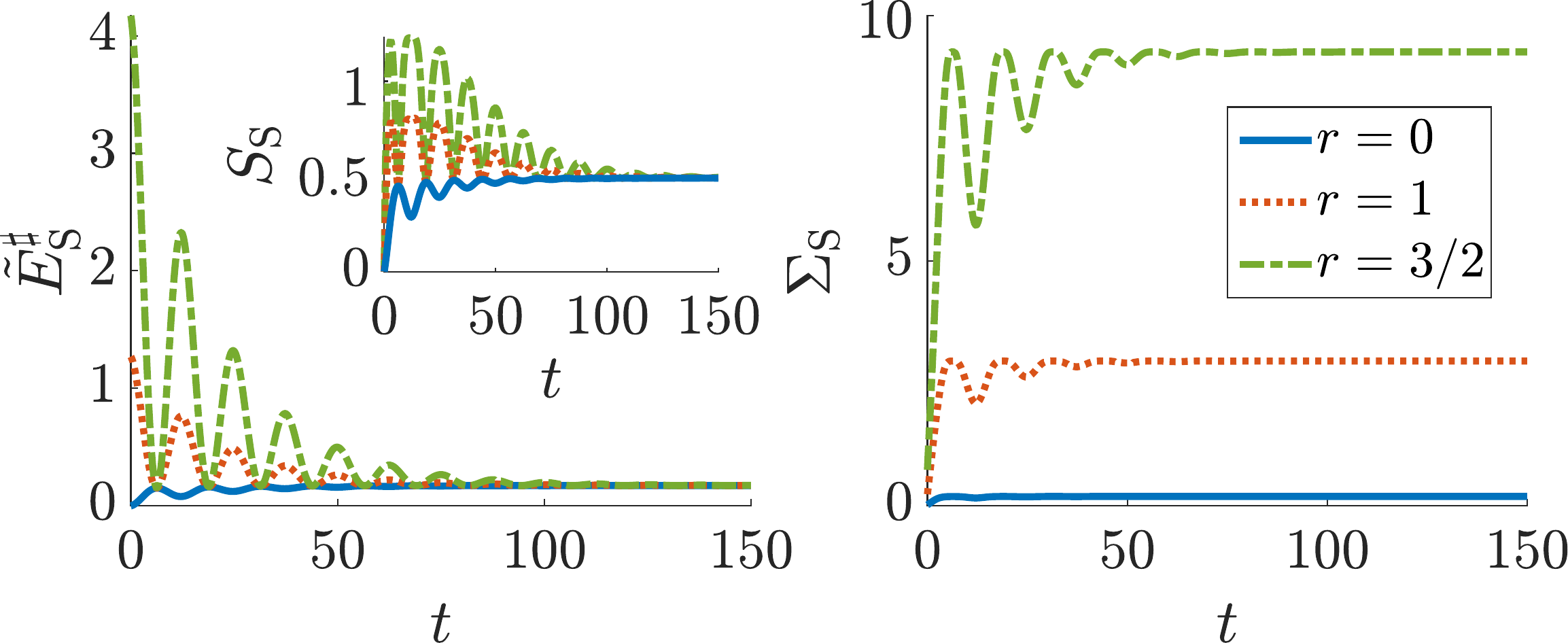}
 	\caption{Internal energy (left), von Neumann entropy (inset) and entropy production (right) for initially squeezed vacuums with squeezing parameter $r$ (in all cases $\theta=0$). The values $\kappa=1/4$, $\gamma=1/10$ and $T=1/2$ were chosen. All quantities are in units of $\omega_0=1$.}
 	\label{Fig:5}
 \end{figure}

In order to calculate the internal energy, we note that since $\rho_{\rm S,eq}$ has the single symplectic eigenvalue \eqref{nu_simplectic}, it can be seen as a thermal state $\rho_{\rm S,eq}=\rho_\phi=e^{-\beta H_\phi}/Z_\phi$, where ${H_{\phi}=\phi a^{\dagger}a}$, ${Z_\phi=\Tr\big[e^{-\beta H_\phi}\big]=\big\{1-\exp[-\beta\,\phi]\big\}^{-1}}$, and $\phi$ is defined by
\begin{equation}
    \frac{2}{e^{\beta\phi}-1}:=\frac{1}{e^{\beta\omega_1}-1}+\frac{1}{e^{\beta\omega_2}-1}.
\end{equation}
Hence, from \eqref{Hash} we find,
\begin{align}\label{HashHzeta}
    H_{\rm S}^{\sharp}=H_\phi-\beta^{-1}\log \frac{Z_{\rm S}^{\sharp}(0^+)}{Z_\phi}.
\end{align}
Thus, inserting this result in \eqref{tildeE}, we obtain
\begin{align}  
\tilde{E}^\sharp_{\rm S}(t)&=\Tr\big[\rho_{\rm S}(t)H_{\rm S}^{\sharp}\big]\nonumber \\
&=\frac{\phi}{2}(\braket{q^2}_t+\braket{p^2}_t-1)-\beta^{-1}\log \frac{Z_{\rm S}^{\sharp}(0^+)}{Z_\phi}\nonumber \\
&=\frac{\phi}{2}\left\{\frac{\Tr[\sigma(t)]}{2}-1\right\}-\beta^{-1}\log \frac{Z_{\rm S}^{\sharp}(0^+)}{Z_\phi}\nonumber \\
&=\phi\bigg\{e^{-\frac{\gamma}{2}t}\bigg[\cos^2(\kappa t)\sinh^2 r+\frac{\sin^2(\kappa t)}{e^{\beta \omega_0}-1}\bigg]\nonumber \\
    &\quad +\tfrac12(1-e^{-\frac{\gamma}{2}t})(\bar{n}_1+\bar{n}_2)\bigg\}-\beta^{-1}\log \frac{Z_{\rm S}^{\sharp}(0^+)}{Z_\phi},
\end{align}
where we have used $a^{\dagger}a=\frac12(q^2+p^2-1)$, \eqref{firstmoments_vanish} and \eqref{sigma1122}.

With the entropy and the internal energy, we can obtain the entropy production $\Sigma_{\rm S}(t)$, Eq.\eqref{II-lawEntropy}, where the heat $Q^{\sharp}_{\rm S}(t)$ is given by \eqref{Heat-t}.

The values of entropy, internal energy, and entropy production are depicted in Fig. \ref{Fig:5} for different values of the squeezing parameter $r$. The internal energy increases with squeezing, which is expected since squeezing has an energetic cost. For the vacuum initial state ($r=0$), the internal energy and entropy increase non-monotonically with $t$, reaching their maximum values at long times, once the system equilibrates. The oscillations become more pronounced in the presence of squeezing, where an initially higher internal energy relaxes to a lower equilibrium value. The entropy rises rapidly from zero at $t=0$ in a highly oscillatory pattern before decaying to its equilibrium value. The entropy production also increases and exhibits larger oscillations as the amount of squeezing grows, reaching its maximum value once the system relaxes, in accordance with the second law.

\begin{figure}[t]
	\includegraphics[width=\columnwidth]{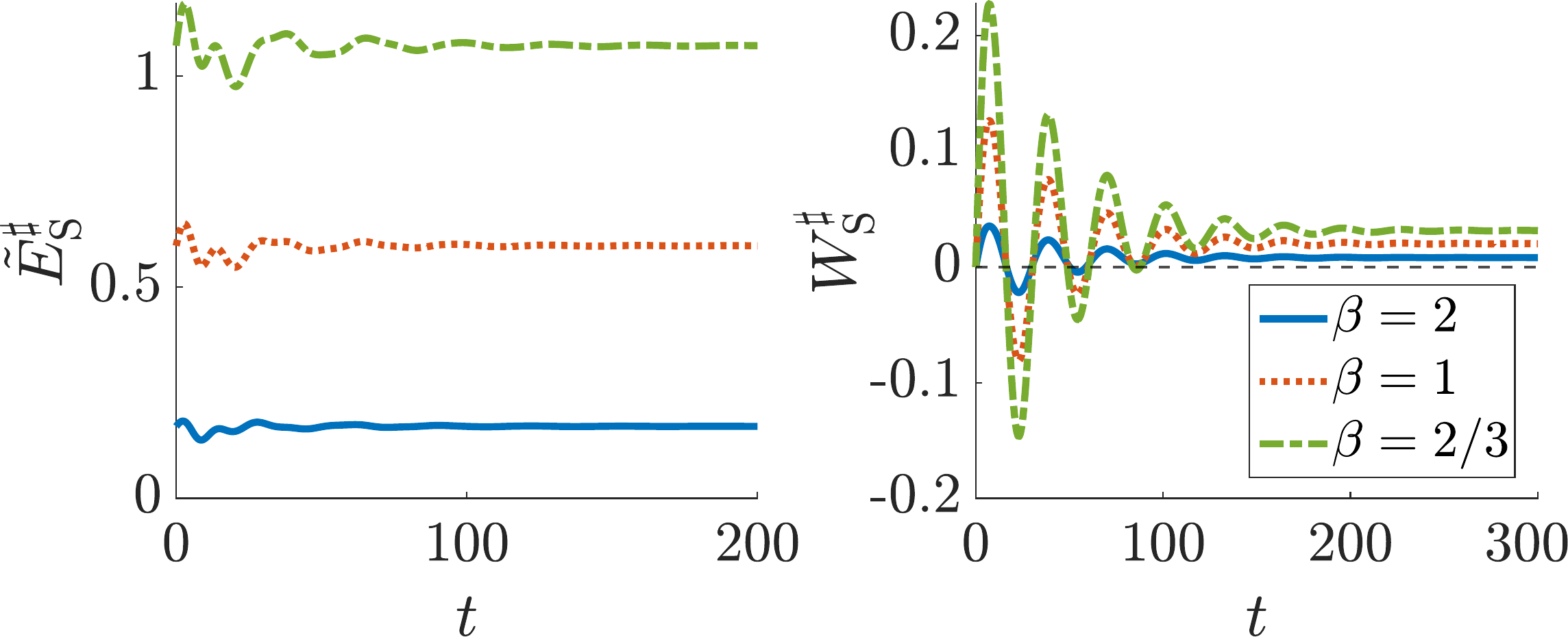}
	\caption{Internal energy (left) and system work (right) for initial equilibrium states at inverse temperature $\beta$, with a damped driving \eqref{driving} ($\eta=1/40$). The values $\kappa=1/4$, $\gamma=1/10$, $\lambda=1/4$ and $\Omega=1/5$ were chosen. All quantities are in units of $\omega_0=1$.}
	\label{Fig:6}
\end{figure}

\subsection{Nonequilibrium thermodynamics for equilibrium initial conditions}\label{sec:example_time_dep}
Let us modify the previous situation by considering an initial equilibrium state $\rho_{\rm SO}(0)=\rho_{\rm SO,\beta}$, and by allowing the system Hamiltonian to be externally driven. In particular, we consider a modulation of the system frequency, $H_{\rm S}(t) = \omega(t) a^{\dagger}a$. As a result, the system-oscillator Hamiltonian is given by
\begin{align}\label{HSO(t)}
    H_{\rm SO}(t)&=\omega(t)a^{\dagger}a+\omega_0 d^{\dagger}d+\kappa(a^{\dagger}d+d^{\dagger}a) \nonumber\\
    &=\begin{bmatrix}
    a^{\dagger} & d^{\dagger}    
    \end{bmatrix}
    \begin{bmatrix}
        \omega(t) & \kappa \\
        \kappa & \omega_0
    \end{bmatrix}
    \begin{bmatrix}
        a \\
        d
    \end{bmatrix}\nonumber \\
    &=\omega_1(t)c^\dagger_1(t) c_1(t)+\omega_2(t)c^\dagger_2(t) c_2(t),
\end{align}
where
\begin{equation}
    \omega_{1,2}(t)=\frac{\omega(t)+\omega_0}{2}\pm\sqrt{\left[\frac{\omega(t)-\omega_0}{2}\right]^2+\kappa^2}
\end{equation}
and 
\begin{figure}[t]
	\includegraphics[width=\columnwidth]{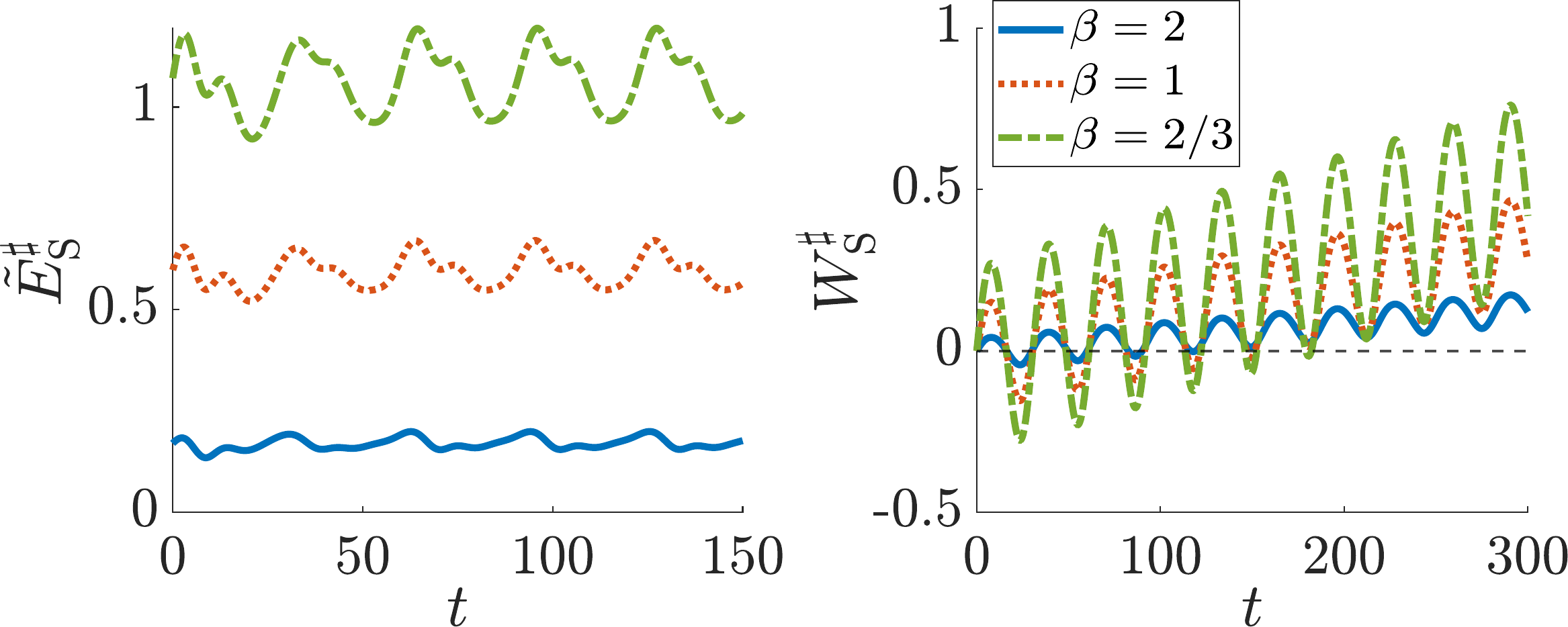}
	\caption{Internal energy (left) and system work (right) for initial equilibrium states at inverse temperature $\beta$, with a periodic driving \eqref{driving} ($\eta=0$). The values $\kappa=1/4$, $\gamma=1/10$, $\lambda=1/4$ and $\Omega=1/5$ were chosen. All quantities are in units of $\omega_0=1$.}
	\label{Fig:7}
\end{figure}
\begin{equation}\label{c(t)M(t)a}
    \begin{bmatrix}
        c_1(t) \\
        c_2(t)
    \end{bmatrix} = \bm{Y}(t)
    \begin{bmatrix}
        a \\
         d
    \end{bmatrix}
\end{equation}
are the normal mode frequencies and annihilation operators, respectively. These are obtained after diagonalization of the matrix of \eqref{HSO(t)} by the matrix $\bm{Y}(t)$, which is orthogonal, symmetric, i.e. $\bm{Y}^2(t)=\mathds{1}$, and traceless (see details in Appendix \ref{app:C}).  
We assume that the rate of change of $\omega(t)$ is slow compared with the relaxation time of the continuum so that the master equation \eqref{master_eq} is now modified \cite{DaviesSpohn,Alicki79} by replacing $\omega_{j}$ and $c_j$ by $\omega_{j}(t)$ and $c_j(t)$, respectively: 
\begin{align}\label{master_eq_driving}
    &\frac{d\rho_{\rm SO}}{dt}=\mathcal{L}_t[\rho_{\rm SO}]=-\ii[H_{\rm SO}(t),\rho_{\rm SO}]\nonumber\\
    &+\sum_{j=1}^2\tfrac{\gamma}{2} [\bar{n}_j(t)+1]\big[c_j(t)\rho_{\rm SO}c_j^\dagger(t)-\tfrac{1}{2}\{c_j^\dagger(t) c_j(t),\rho_{\rm SO}\}\big]\nonumber \\
    &\ \ +\sum_{j=1}^2\tfrac{\gamma}{2} \bar{n}_j(t) \big[c_j^\dagger(t) \rho_{\rm SO}c_j(t)  -\tfrac{1}{2}\{c_j(t) c_j^\dagger(t),\rho_{\rm SO}\}\big].
\end{align}
For the driving function, we shall consider the following specific form
\begin{equation}\label{driving}
    \omega(t)=\omega_0+\lambda e^{-\eta t}\sin\left(\Omega t\right),
\end{equation}
with $\eta,\Omega,\lambda\geq0$. Thus, $\omega(0)=\omega_0$ and, if the damping factor is finite $\eta>0$, $\lim_{t\to\infty}\omega(t)=\omega_0$. For $\eta=0$ we obtain a periodic modulation.

\begin{figure}[t]
	\includegraphics[width=\columnwidth]{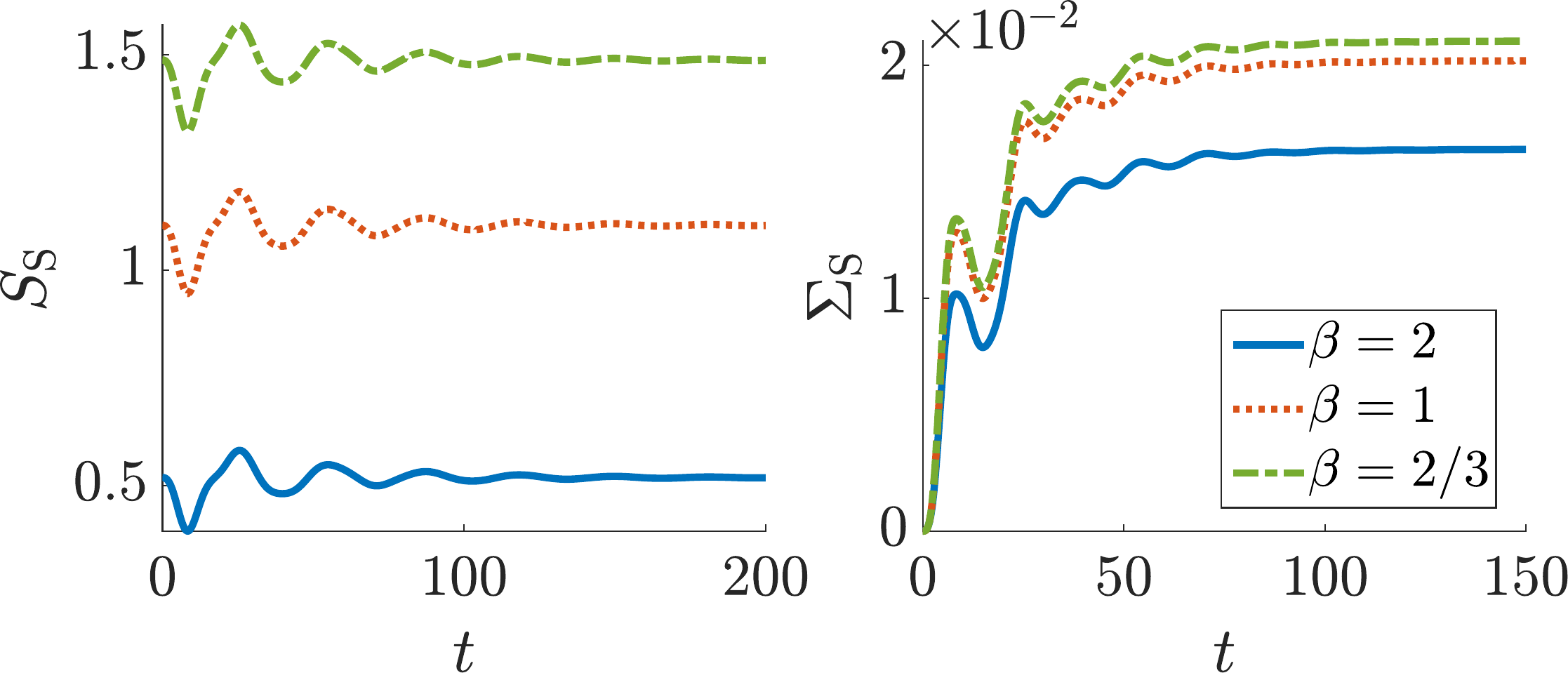}
	\caption{Entropy (left) and entropy production (right) for initial equilibrium states at inverse temperature $\beta$, with a damped driving \eqref{driving}. The values $\kappa=1/4$, $\gamma=1/10$, $\lambda=1/4$, $\eta=1/40$ and $\Omega=1/5$ were chosen. All quantities are in units of $\omega_0=1$.}
	\label{Fig:8}
\end{figure}
The details of the solution of \eqref{master_eq_driving}, including the calculation of covariance matrices, internal energy, and work are devoted to Appendix \ref{app:C}.

In Figs. \ref{Fig:6} and \ref{Fig:7} we show the results for internal energy and work for a system initially prepared in an equilibrium state $\rho_{\rm SO,\beta}$, under damped $(\eta>0)$ and periodic ($\eta=0$) driving, respectively, as a function of time for different values of the inverse temperature $\beta$. As we can see, the higher the temperature the more energetic the system is, as expected. However, after the transient period, the long-time behavior is markedly different. Under damped driving, the internal energy relaxes to its initial value, 
consistent with $\lim_{t\rightarrow\infty}\omega(t)=\omega(0)$. Under periodic driving the internal energy settles to an oscillatory pattern around the initial value. This arises from the work applied to or performed by the system during each driving cycle. As we can see, the greater the initial temperature, the more work is required to shift the system frequency $\omega(t)$ away from its initial value. In the damped case (Fig. \ref{Fig:6}), the power vanishes at long times and the system eventually relaxes to the initial equilibrium mean force Gibbs state. For the periodic case $\eta = 0$ (Fig.~\ref{Fig:7}), the system work oscillates after relaxation with a similar frequency as the internal energy, with a superimposed net increase on this oscillatory behavior, thereby preventing its extraction in a cyclical manner.

\begin{figure}[t!]
	\includegraphics[width=\columnwidth]{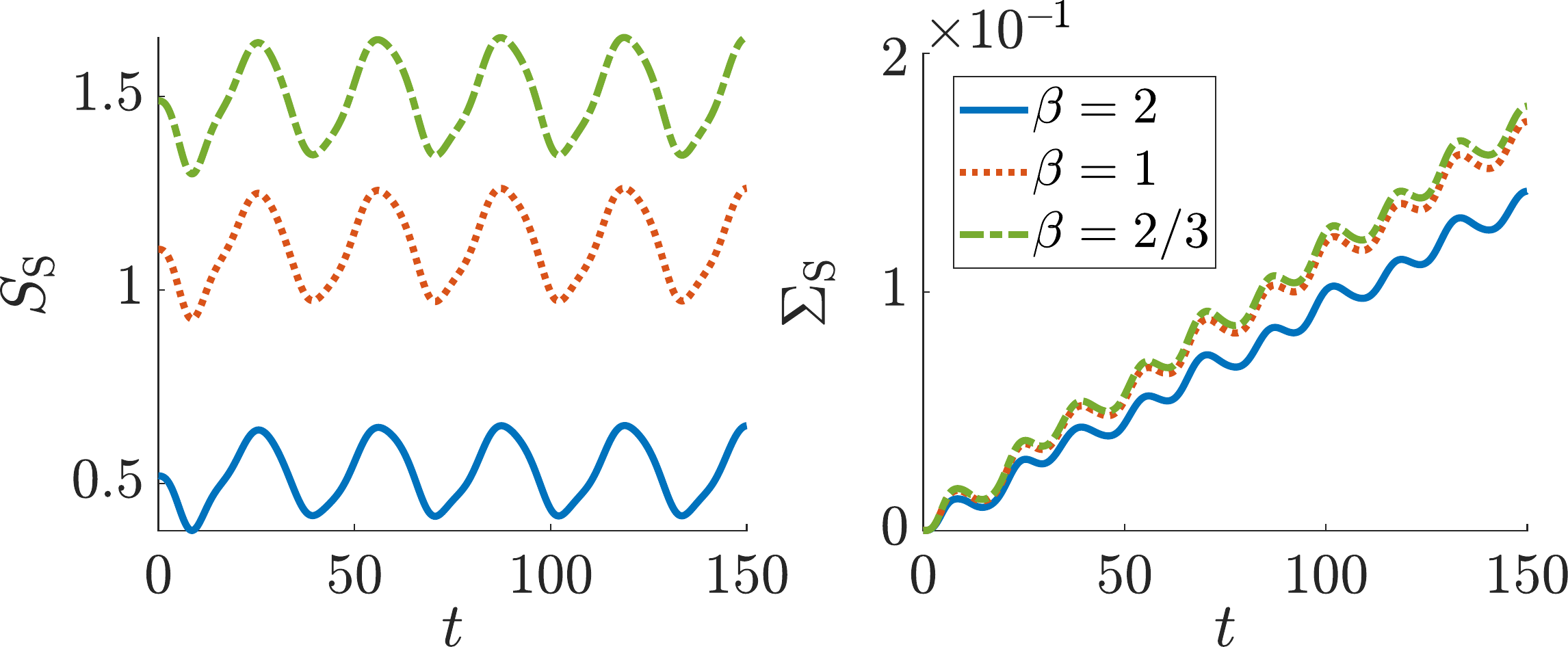}
	\caption{Entropy (left) and entropy production (right) for initial equilibrium states at inverse temperature $\beta$, with a periodic driving \eqref{driving} ($\eta=0$). The values $\kappa=1/4$, $\gamma=1/10$, $\lambda=1/4$ and $\Omega=1/5$ were chosen. All quantities are in units of $\omega_0=1$.}
	\label{Fig:9}
\end{figure}

The values of entropy $S_{\rm S}(t)$ and entropy production $\Sigma_{\rm S}(t)$, Eq.~\eqref{II-lawEntropy}, with the heat given by $Q^\sharp_{\rm S}(t)=E_{\rm S}^\sharp(t,\beta)-E_{\rm S}^\sharp(0,\beta)-W_{\rm S}^{\sharp}(t)$, are shown in Figs. \ref{Fig:8} and \ref{Fig:9} for damped and periodic drivings, respectively. In the damped driving case we observe that, after initial oscillations, the system entropy relaxes to the initial value, again due to the fact that $\lim_{t\rightarrow\infty}\omega(t)=\omega(0)$. However, the entropy production, despite having similar initial oscillations, converges to some positive value, which is larger for higher temperatures. This shows that, even though the system returns to its initial state, this process is irreversible. In the case of periodic driving, the system entropy oscillates around the initial value for long times, while the entropy production increases linearly at the same time as it oscillates. This shows that in each driving cycle the system produces a positive amount of entropy, and hence the driving is not reversible. The slope of this linear behavior of the entropy increases with temperature.

\subsection{Weak-coupling limit}
We may check that the appropriate result is recovered in the weak-coupling limit $\kappa\rightarrow0$. To see this, we evaluate the quantity
\begin{equation}\label{WKL_measure}
    \max_{t}\lvert \tilde{E}_{\rm S}^{\sharp}(t) - \braket{H_{\rm S}(t)}\rvert
\end{equation}
which indicates uniform convergence if approaching zero as the coupling is decreased. In order to accomplish this without jeopardizing the validity of the Davies weak-coupling treatment, we adopt the parametrization ${\kappa=\omega_0/(4\zeta)}$ and $\gamma=\omega_0/(10\zeta)$, so that the weak-coupling condition is approached by increasing $\zeta$ while keeping the ratio $\kappa/\gamma$ constant. This preserves the validity of the secular approximation in the Davies generator. The results of this analysis are presented in Fig.~\ref{Fig:11}, for an initial product state \eqref{xirhobeta} without external driving (left), and for initial equilibrium states with a periodic driving (right). The value of \eqref{WKL_measure} clearly decreases with $\zeta$.

\subsection{Comparison with other strong-coupling nonequilibrium thermodynamic approaches}
Finally, it is interesting to compare the entropy production we obtained with the one calculated according to other strong-coupling thermodynamic approaches. In particular, this model is a prototypical case where the reaction coordinate (RC) thermodynamic framework can be applied \cite{Binder2018, Strasberg2016}. However, it is important to note that in the RC framework, the thermodynamic description must be extended to include the RC (the oscillator O in our case), which is physically a degree of freedom of the environment. By contrast, in our approach, the thermodynamic system is strictly taken to be the oscillator S alone, while O is treated entirely as part of the thermal reservoir (Fig.~\ref{Fig:1}). Consequently, RC thermodynamics and the intrinsic Hamiltonian of mean force describe the thermodynamic properties of different systems—SO in the former and S in the latter—and thus address different operational paradigms. Yet, they remain fundamentally compatible. By this, we mean that the entropy production in the intrinsic Hamiltonian of mean force approach is bounded from below by the one obtained via the RC approach, namely:
\begin{figure}[t]
	\includegraphics[width=\columnwidth]{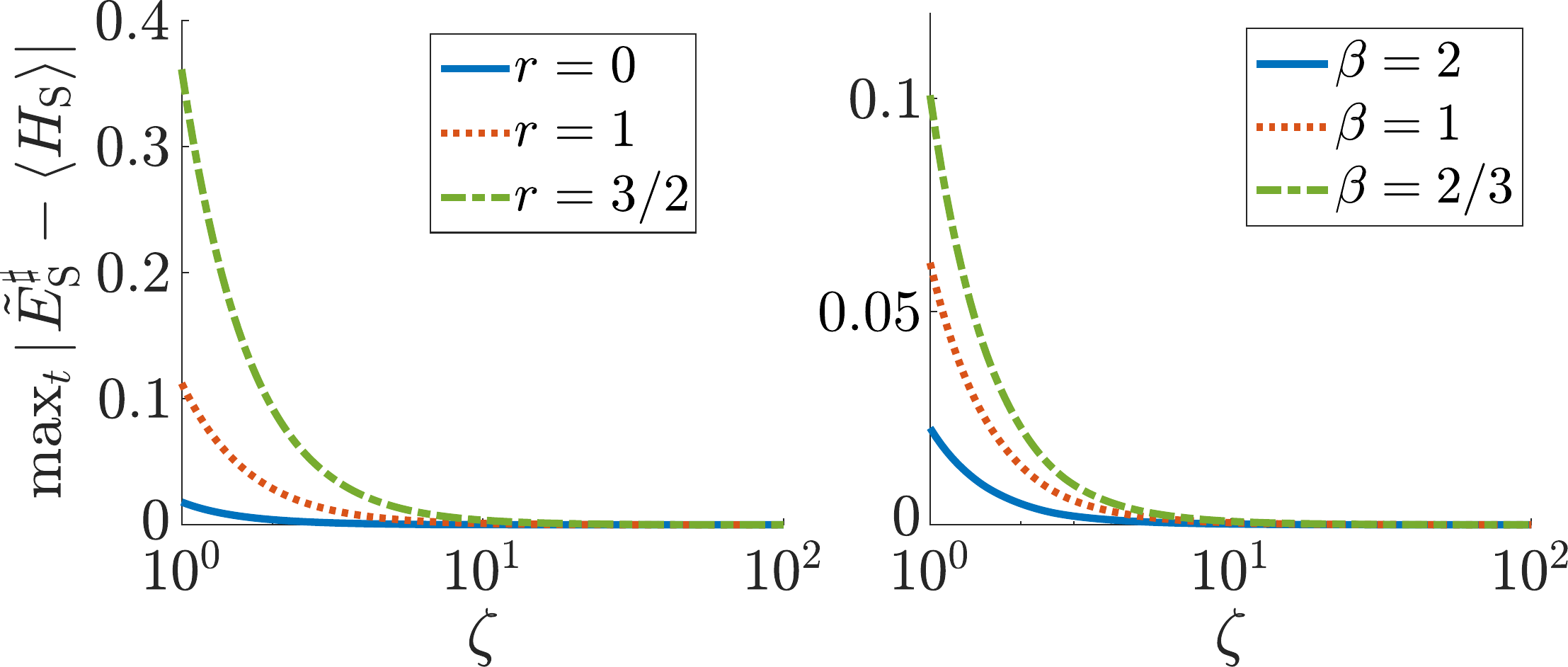}
	\caption{Measure \eqref{WKL_measure} for initially squeezed vacuums with squeezing parameter $r$ (in all cases $\theta=0$) and no driving (left), and for initial equilibrium states at inverse temperature $\beta$ with periodic driving (right), as a function of the parameter $\zeta$ which accounts for the weakness of the coupling (see main text). In both cases the values $\kappa=1/(4\zeta)$ and $\gamma=1/(10\zeta)$ were chosen. For the no driving case, $T=1/2$. For the periodic case (right), $\lambda=1/4$ and $\Omega=1/5$.  All quantities are in units of $\omega_0=1$.}
	\label{Fig:11}
\end{figure}
\begin{equation}\label{S>SO}
    \Sigma_{\rm S}\geq \Sigma_{\rm SO}.
\end{equation}
Indeed, since S is a subsystem of SO, the entropy production cannot be smaller for S than for the composite SO. Otherwise, an observer restricted to manipulating and measuring only S would somehow be able to execute a process with greater thermodynamic reversibility than an observer with full control over both S and O. From an information-theoretic standpoint, tracing out the oscillator O discards the information encoded in both the S-O correlations and the internal state of O. Because information is a thermodynamic resource, this unavoidable loss of information strictly manifests as an additional source of irreversibility—and thus higher entropy production—for the local observer. The specific proof of \eqref{S>SO} is given in Appendix \ref{app:E}.

On the other hand, this model has recently been examined in \cite{Albarelli2025} using the thermodynamic approach of Refs.~\cite{Esposito2010,Strasberg2017,Tanimura2016}. As discussed in Sec.~\ref{Sec:I}, in this framework the interaction Hamiltonian is counted as part of the system’s internal energy—defined by the expectation value $\langle H_{\rm S}+V_{\rm SO}\rangle$—while the system entropy is still evaluated using the local von Neumann expression. Under this partition, the positivity of the entropy production can be proven for initial product states.

\begin{figure}[t]
	\includegraphics[width=\columnwidth]{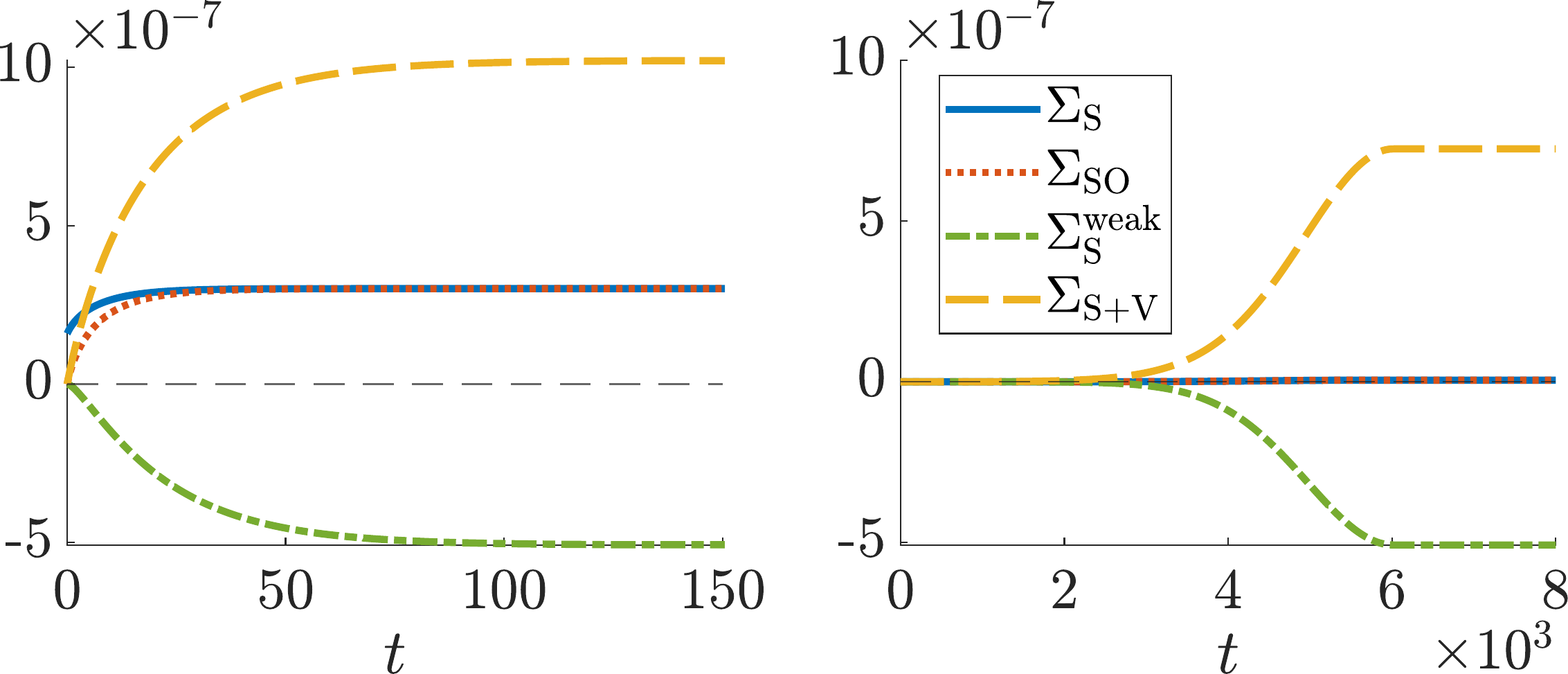}
	\caption{Entropy production according to the intrinsic HMF approach (blue), RC thermodynamics (orange), the naive weak-coupling expression (green) and the approach including the bare interaction energy (yellow), for an initial state $\rho_{\rm SO}(0)=\rho_{\rm S,\beta}\otimes\rho_{\rm O,\beta}$ and driving \eqref{driving2}. The values for the driving time $\tau_{\rm d}\rightarrow 0$ (left) and $\tau_{\rm d}= 6\cdot10^3$(right) are chosen, corresponding to the quench and quasistatic limits respectively. For both plots $T=1/20$, $\kappa_0=1/4$ and $\gamma=1/10$. All quantities are in units of $\omega_0=1$.}
	\label{Fig:12}
\end{figure}

In Fig.~\ref{Fig:12}, we plot the entropy production evaluated across four distinct frameworks: our intrinsic HMF approach ($\Sigma_{\rm S}$), RC thermodynamics ($\Sigma_{\rm SO}$), the approach including the interaction Hamiltonian in the system energy ($\Sigma_{\rm S+V}$), and the bare weak-coupling approach ($\Sigma_{\rm S}^{\rm weak}$), which takes the internal energy to be $\langle H_{\rm S}\rangle$. These quantities are calculated for an initial product of Gibbs states, $\rho_{\rm S,\beta}\otimes\rho_{\rm O,\beta}$---a state that is stationary only in the vanishingly weak-coupling limit. The coupling strength is parametrized as the sigmoid function
\begin{equation}\label{driving2}
\kappa(t)=\begin{cases}
0 & t < 0, \\
\kappa_{0}\sin^2\!\left(\frac{\pi t}{2\tau_{\rm d}}\right) & t \in [0,\tau_{\rm d}], \\
\kappa_0 & t > \tau_{\rm d},
\end{cases}
\end{equation}
such that, in the quench limit ($\tau_{\rm d} \to 0$), the interaction is switched on instantaneously at $t=0$ and remains constant at $\kappa_0$ for all $t > 0$, whereas in the limit $\tau_{\rm d} \to \infty$, the driving becomes quasistatic and maintains instantaneous equilibrium $\rho_{\rm SO}(t) = \rho_{\rm SO,\beta}(t)$.

In the left panel of Fig.~\ref{Fig:12} (quench limit, $\tau_{\rm d} \to 0$), we verify that the bound in Eq.~\eqref{S>SO} is strictly satisfied. Notably, $\Sigma_{\rm S+V}$ exhibits the largest values among all frameworks. This is physically counterintuitive from an operational standpoint: evaluating $\Sigma_{\rm S+V}$ requires tracking changes in $\langle V_{\rm SO}\rangle$, which implies that an observer must have simultaneous measurement access to both S and O—a level of control comparable to that required in RC thermodynamics. Yet, while the smaller value of $\Sigma_{\rm SO}$ reflects the fact that this extended access allows irreversibility to be reduced, the large values of $\Sigma_{\rm S+V}$ suggest that this framework significantly overestimates it.

This behavior is further confirmed in the right panel, corresponding to the quasistatic regime ($\tau_{\rm d}$ large). Here, while $\Sigma_{\rm S}(t) = \Sigma_{\rm SO}(t) \simeq 0$, as expected for reversible dynamics, $\Sigma_{\rm S+V}(t) > 0$. Indeed, one can verify analytically that, in the quasistatic limit $\tau_{\rm d} \to \infty$, the relation $\Sigma_{\rm S+V}(t) = D[\rho_{\rm O}(t) \Vert \rho_{\rm O,\beta}(0)] + I_{\rm SO}(t) > 0$ holds.

Furthermore, we observe that the weak-coupling expression $\Sigma_{\rm S}^{\rm weak}$ yields unphysical negative values in both driving regimes, underscoring the fundamental breakdown of weak-coupling thermodynamics in this context. For the initial state considered here, this follows directly from the general analytical result $\Sigma_{\rm S}^{\rm weak} = - D[\rho_{\rm S}(t) \Vert \rho_{\rm S}(0)] < 0$.

\section{Conclusion}
In this work, we establish a universal thermodynamic framework for quantum systems that may be strongly coupled to thermal sources. In contrast to other proposals, our framework focuses on thermodynamic variables that depend strictly on the system state and are therefore accessible through microscopic control of the system alone. This bridges a critical gap between theoretical generality and experimental feasibility, enabling one to answer concrete operational questions—such as determining the minimum work required to drive a strongly coupled system from an initial reduced state $\rho_{\rm S}(0)$ to a target state $\rho_{\rm S}(t)$ within a given time $t$. Importantly, it provides a consistent description of both equilibrium (including thermostatics) and nonequilibrium properties, preserving the same gauge freedoms as in weak-coupling thermodynamics, and retaining the von Neumann entropy as the thermodynamic entropy. Thus, it ensures compatibility with the principles of statistical mechanics and quantum information theory. Moreover, we have clarified the exact requirements for initial conditions to produce thermodynamic behavior, characterizing this set as those states where the system acts as the sole independent source of initial nonequilibrium.

We have benchmarked the proposed approach by applying it to a paradigmatic case of a strongly coupled environmental model consisting of a structured bosonic reservoir. This not only allows us to illustrate the internal consistency of the formalism—including a direct comparison with alternative strong-coupling thermodynamic approaches—but also its ability to yield meaningful predictions for thermodynamic equilibrium quantities, e.g., heat capacity and density of states, and nonequilibrium dynamical behavior, e.g., entropy production.

One should note that, from a theoretical perspective, the nonequilibrium thermodynamics proposed here does not introduce substantial difficulties beyond solving the dynamics of an open quantum system strongly coupled to a thermal reservoir. Once the dynamics is known, determining the thermodynamic variables reduces to calculating the relevant expectation values. From an experimental perspective, however, nonlinear functions of the system state—such as the von Neumann entropy or the logarithm involved in defining the intrinsic Hamiltonian of mean force—typically require tomographic methods. Nevertheless, several recent works have developed and experimentally implemented efficient techniques to determine the entanglement Hamiltonian \cite{Dalmonte2018,Kokail2021,Kokail2021b,Mueller2023,Joshi2023}, which is essentially the same type of logarithmic state function required in our approach.

In light of these results, several directions for future research can be considered. First, one could examine a possible extension to multiple thermal reservoirs. Additionally, it would be valuable to investigate mathematical questions, such as the conditions under which the positivity of the density of states—derived from the intrinsic Hamiltonian of mean force—can be rigorously proven, or, less ambitiously, the positivity of the heat capacity. Furthermore, exploring the implications of this framework in the context of quantum measurement and feedback control could provide new insights into the thermodynamic cost of information processing. Finally, applying the proposed framework on current experimental platforms—such as trapped ions, superconducting circuits, or optomechanical systems—appears feasible and would offer a critical test of its practical applicability, potentially guiding the development of next-generation quantum technologies.

\begin{acknowledgments}
The authors acknowledge support from Spanish MICIN grants PID2021-122547NB-I00 and EUR2024-153545, from the TEC-2024/COM-84 QUITEMAD-CM project funded by Comunidad de Madrid, and from the ``MADQuantum-CM'' project funded by Comunidad de Madrid and by the Recovery, Transformation and Resilience Plan – Funded by the European Union - NextGenerationEU. I.G. acknowledges support from the MICIN contract PRE2022-101824 (MICIN/AEI/FSE,UE). S.C. was supported by the Margarita Salas fellowship funded by the European Union (NextGenerationEU). 
\end{acknowledgments}

\appendix

\section{Proof of the general form of thermodynamic initial conditions} \label{app:A0}
To check that \eqref{Petzequality} is satisfied for \eqref{rhoSRbeta-decomposition} and \eqref{rhoSR0-decomposition} is straightforward, so we shall prove the converse, i.e. that \eqref{Petzequality} requires \eqref{rhoSRbeta-decomposition} and \eqref{rhoSR0-decomposition}.

First, let us introduce an ancillary qubit system A, and consider the joint state
\begin{equation}\label{rhoASR}
\rho'_{\rm ASR} = \tfrac{1}{2}|0\rangle_{\rm A}\langle0| \otimes \rho_{\rm SR}(0) + \tfrac{1}{2}|1\rangle_{\rm A}\langle1| \otimes \rho_{\rm SR,\beta}(0).
\end{equation}
Defining the marginals $\rho'_{\rm A} := \Tr_{\rm SR}(\rho'_{\rm ASR})$, $\rho'_{\rm SR} := \Tr_{\rm A}(\rho'_{\rm ASR})$, $\rho'_{\rm AS} := \Tr_{\rm R}(\rho'_{\rm ASR})$, and $\rho'_{\rm S} := \Tr_{\rm AR}(\rho'_{\rm ASR})$, the monotonicity of the relative entropy with respect to the partial trace implies:
\begin{equation}\label{Dmon_1}
D(\rho'_{\rm ASR}\| \rho'_{\rm A}\otimes \rho'_{\rm SR}) \geq D(\rho'_{\rm AS}\| \rho'_{\rm A}\otimes \rho'_{\rm S}).
\end{equation}
If \eqref{Petzequality} is fulfilled, the CPTP map given in \eqref{Petz_map} satisfies \cite{Petz1986,Hayden2004}:
\begin{align}
&\rho_{\rm SR}(0) = \mathcal{P}[\rho_{\rm S}(0)],\\
&\rho_{\rm SR,\beta}(0) = \mathcal{P}[\rho_{\rm S,eq}(0)].
\end{align}
Since
\begin{align}
&\rho'_{\rm AS} = \tfrac{1}{2}|0\rangle_{\rm A}\langle0| \otimes \rho_{\rm S}(0) + \tfrac{1}{2}|1\rangle_{\rm A}\langle1| \otimes \rho_{\rm S,\eq}(0),\\
&\rho'_{\rm SR} = \tfrac{1}{2} \rho_{\rm SR}(0) + \tfrac{1}{2} \rho_{\rm SR,\beta}(0),\\
&\rho'_{\rm S} = \tfrac{1}{2} \rho_{\rm S}(0) + \tfrac{1}{2} \rho_{\rm S,eq}(0),
\end{align}
the linearity of $\mathcal{P}$ ensures that
\begin{align}
&\rho'_{\rm SR} = \mathcal{P}(\rho'_{\rm S}),\\
&\rho'_{\rm ASR}=\mathcal{I}_{\rm A}\otimes \mathcal{P}(\rho'_{\rm AS}).
\end{align}
Therefore, the monotonicity of the relative entropy under the CPTP map $\mathcal{I}_{\rm A}\otimes \mathcal{P}$ leads to:
\begin{equation}
D(\rho'_{\rm ASR}\| \rho'_{\rm A}\otimes \rho'_{\rm SR}) \leq D(\rho'_{\rm AS}\| \rho'_{\rm A}\otimes \rho'_{\rm S}).
\end{equation}
This result, together with \eqref{Dmon_1}, implies strict equality:
\begin{equation}
D(\rho'_{\rm ASR}\| \rho'_{\rm A}\otimes \rho'_{\rm SR}) = D(\rho'_{\rm AS}\| \rho'_{\rm A}\otimes \rho'_{\rm S}).
\end{equation}
Theorem 6 in \cite{Hayden2004} asserts that this equality is satisfied if and only if there exists a partition of the system Hilbert space $\mathcal{H}_{\rm S} = \bigoplus_{j=1}^J \mathcal{H}_{{\rm S}^{\rm A}_j} \otimes \mathcal{H}_{{\rm S}^{\rm B}_j}$ such that
\begin{equation}\label{sigma-block1}
\rho'_{\rm ASR}= \bigoplus_{j=1}^J q_j \sigma_{{\rm AS}^{\rm A}_j}\otimes \sigma_{{\rm S}^{\rm B}_j {\rm R}}  
\end{equation}
with states $\sigma_{{\rm AS}^{\rm A}_j}$ on $\mathcal{H}_{\rm A}\otimes \mathcal{H}_{{\rm S}^{\rm A}_j}$ and $\sigma_{{\rm S}^{\rm B}_j {\rm R}}$ on $\mathcal{H}_{{\rm S}^{\rm B}_j}\otimes \mathcal{H}_{\rm R}$ and $q_j$ a probability distribution.

From \eqref{rhoASR}, we observe that there are no coherences in the qubit basis for A. Therefore, the most general form for every individual block is
\begin{align}
\sigma_{{\rm AS}^{\rm A}_j} = c_{0|j} \Big( |0\rangle_{\rm A}\langle0| \otimes \rho_{{\rm S}^{\rm A}_j} \Big) + c_{1|j} \Big( |1\rangle_{\rm A}\langle1| \otimes \sigma_{{\rm S}^{\rm A}j} \Big)
\end{align}
where $c_{i|j} \geq 0$, $\sum_{i} c_{i|j} = 1$, and $\rho_{{\rm S}^{\rm A}_j}$ and $\sigma_{{\rm S}^{\rm A}_j}$ are valid states on $\mathcal{H}_{{\rm S}^{\rm A}_j}$. Substituting this result into \eqref{sigma-block1}, we obtain:
\begin{multline}
\rho'_{\rm ASR} = |0\rangle_{\rm A}\langle0| \otimes \bigoplus_{j=1}^J q_j c_{0|j} \rho_{{\rm S}^{\rm A}_j} \otimes \sigma_{{\rm S}^{\rm B}_j {\rm R}} \\
+ |1\rangle_{\rm A}\langle1| \otimes \bigoplus_{j=1}^J q_j c_{1|j} \sigma_{{\rm S}^{\rm A}j} \otimes \sigma_{{\rm S}^{\rm B}_j {\rm R}}
\end{multline}
Comparison with \eqref{rhoASR} straightforwardly leads to \eqref{rhoSRbeta-decomposition} and \eqref{rhoSR0-decomposition} and with $p_j = 2q_j c_{0|j}$ and $p_{{\rm eq},j} = 2q_j c_{1|j}$.\\

\section{Power expansion of the density of states} \label{app:A}
Expanding Eq. \eqref{vn_entropy_gauss}, with $\nu$ given by \eqref{nu_simplectic}, to second order in $\kappa$,
\begin{multline}
    S_{\rm vN}(\beta)\simeq\frac{\beta \omega_0 e^{\beta \omega_0}}{e^{\beta\omega_0}-1}-\log \left(e^{\beta \omega_0}-1\right)\\
    +\frac{\beta ^3 \omega_0  e^{\beta  \omega_0 } \left(e^{\beta  \omega_0 }+1\right)}{2 \left(e^{\beta  \omega_0 }-1\right)^3}\kappa^2+\ldots
\end{multline}
Under this approximation
\begin{multline}
    \int_{\beta}^{\infty}\frac{S_{\rm vN}(\alpha)}{\alpha^2}d\alpha \simeq \omega_0-\frac{\log \left(e^{\beta  \omega_0}-1\right)}{\beta }\\
    +\frac{e^{\beta  \omega_0} (\beta \omega_0+1)-1}{2 \omega_0 \left(e^{\beta  \omega_0}-1\right)^2}\kappa^2+\ldots
\end{multline}
and so, according to \eqref{Zhash},
\begin{multline}\label{ZsharpO2}
    Z^{\sharp}_{\rm S}(\beta) \simeq \frac{1}{1-e^{-\beta \omega_0}}\\
    +\frac{\beta  \left(e^{-\beta  \omega_0 } (\beta  \omega_0 +1)-e^{-2 \beta  \omega_0 }\right)}{2 \omega_0  \left(1-e^{-\beta  \omega_0 }\right)^3}\kappa^2+\ldots
\end{multline}
By expanding the geometric series, it is easy to prove that
\begin{align}
    \frac{1}{\left(1-e^{-\beta  \omega_0 }\right)^3}&=\left(\sum_{n=0}^\infty e^{-\beta n \omega_0 }\right)^3\nonumber \\
    &=\sum_{n=0}^\infty \frac{(n+2)(n+1)}{2} e^{-\beta n \omega_0 }.
\end{align}

\begin{widetext}
\noindent This expression in \eqref{ZsharpO2} yields 
\begin{align}
   Z^{\sharp}_{\rm S}(\beta) &\simeq \sum_{n=0}^\infty e^{-\beta n \omega_0}
    +\sum_{n=0}^\infty\frac{(n+2)(n+1)}{4\omega_0} \beta  \left[e^{-\beta  \omega_0 } (\beta  \omega_0 +1)-e^{-2 \beta  \omega }\right]  e^{-\beta n \omega } \kappa^2+\ldots\nonumber\\
    &\simeq \sum_{n=0}^\infty e^{-\beta n \omega_0}
    +\sum_{n=0}^\infty\frac{(n+2)(n+1)}{4\omega_0}   \left[\beta  \left(e^{-\beta (n+1)  \omega_0 }-e^{- \beta (n+2)  \omega_0 }\right)+\beta ^2 \omega_0  e^{-\beta (n+1) \omega_0 }\right] \kappa^2+\ldots\nonumber\\
    &\simeq \sum_{n=0}^\infty e^{-\beta n \omega_0}
    +\left[\frac{1}{4} \sum_{n=1}^\infty n(n+1) \beta^2 e^{-\beta n \omega_0}+  \frac{1}{2\omega_0} \sum_{n=2}^\infty n \beta e^{-\beta n\omega_0}\right] \kappa^2+\ldots
\end{align}
where we have simplified the summation
\begin{align}
    \sum_{n=0}^{\infty}(n+2)(n+1)e^{-\beta (n+1)\omega_0}-\sum_{n=0}^{\infty}(n+2)(n+1)e^{-\beta (n+2)\omega_0}=\sum_{n=2}^\infty 2 n e^{-\beta n\omega_0} .
\end{align}
Using the result of the Laplace transform for the delta derivatives
\[
\int_0^\infty \delta^{(n)}(\varepsilon-\varepsilon_0)e^{-\beta \varepsilon}d\varepsilon=\beta^n e^{-\beta \varepsilon_0},
\]
we finally obtain
\begin{align}
\varrho^{\sharp}(\varepsilon)=\sum_{n=0}^{\infty}\delta(\varepsilon-n\omega_0)+\left[\sum_{n=1}^\infty \frac{n(n+2)}{4}\delta''(\varepsilon-n\omega_0)+\sum_{n=2}^\infty \frac{n}{2\omega_0}\delta'(\varepsilon-n\omega_0)\right]\kappa^2+\ldots
\end{align}
On the other hand, expanding \eqref{DSO} formally to second order in $\kappa$:
\begin{align}
    \varrho_{\rm SO}(\varepsilon)&=\sum_{n,m=0}^{\infty}\left\{\delta[\varepsilon - (m+ n)\omega_0]+(m-n) \delta'[\varepsilon-(m+n)\omega_0]\kappa  +\frac{(m-n)^2}{2} \delta''[\varepsilon-(m+n)\omega_0]\kappa ^2+\ldots\right\}
\end{align}
The second term is clearly odd in the double sum, so it vanishes. Changing the labels $m+n=k$, $m-n=l$ the first term becomes
\begin{align}
    \sum_{k=0}^{\infty}\sum_{\substack{l=-k \\ \text{step } 2}}^{k}\delta(\varepsilon - k\omega_0)=\sum_{k=0}^{\infty}(k+1)\delta(\varepsilon - k\omega_0),
\end{align}
and for the third term
\begin{align}
    \sum_{k=0}^{\infty}\sum_{\substack{l=-k \\ \text{step } 2}}^{k}\frac{l^2}{2} \delta''(\varepsilon-k\omega_0)=\sum_{k=1}^{\infty}\frac{k(k+1)(k+2)}{6}\delta''(\varepsilon-k\omega_0),
\end{align}
leading to
\begin{align}\label{DSO_2nd}
      \varrho_{\rm SO}(\varepsilon)=\sum_{n=0}^\infty (n+1)\delta(\varepsilon-n\omega_0)+\sum_{n=1}^\infty\frac{n(n+1)(n+2)}{6}\delta''(\varepsilon-n\omega_0)\kappa^2+\ldots
\end{align}
In addition, using this result for $\varrho_{\rm SO}(\varepsilon)$, it is straightforward to obtain 
from \eqref{D*}:
\begin{align}
    \varrho^*(\varepsilon)&=\sum_{n=0}^\infty (n+1)\delta(\varepsilon-n\omega_0)-\sum_{n=0}^\infty (n+1)\delta[\varepsilon-(n+1)\omega_0]\nonumber \\
    &+\left\{\sum_{n=1}^\infty\frac{n(n+1)(n-1)}{6}\delta''(\varepsilon-n\omega_0)-\sum_{n=1}^\infty\frac{n(n+1)(n-1)}{6}\delta''[\varepsilon-(n+1)\omega_0]\right\}\kappa^2+\ldots\nonumber \\
    &=\sum_{n=0}^\infty\delta(\varepsilon-n\omega_0)+\sum_{n=1}^\infty\frac{n(n+1)}{2}\delta''(\varepsilon-n\omega_0)\kappa^2+\ldots
\end{align}
As expected, in the weak-coupling limit $\varrho^\sharp(\varepsilon)$ and $\varrho^*(\varepsilon)$ approach the correct density of states of an isolated quantum harmonic oscillator . On the other hand, as mentioned in the main text, the peaks of $\varrho^\sharp(\varepsilon)$ coincide, up to numerical precision, with those of $\varrho_{\rm SO}(\varepsilon)$, which approaches the density of states of two identical uncoupled oscillators for vanishing $\kappa$, leading to the degeneracy factor $(n+1)$ in the first term of \eqref{DSO_2nd}. In all cases, the first nontrivial order contribution is the second-order term in $\kappa$. However, due to the presence of exotic delta derivatives in the expansion, it is difficult to infer a clear behavior from these equations.
\end{widetext}

\section{Dynamics of the covariance matrix in the absence of external driving}\label{app:B}
The covariance matrix $\sigma(t)$ of $\mathrm{S}$ is given by the expectation values
\begin{equation}\label{covariance_time_dependent}
    \sigma(t) = 
    \begin{bmatrix}
        2(\braket{q^2}_t - \braket{q}_t^2) & \braket{qp+pq}_t-2\braket{q}_t\braket{p}_t  \\
        \braket{qp+pq}_t-2\braket{q}_t\braket{p}_t & 2(\braket{p^2}_t - \braket{p}_t^2)
    \end{bmatrix}.
\end{equation}
It is convenient to solve the dynamics in the Heisenberg picture, i.e., for any $\mathrm{SO}$ operator $X$, 
\begin{align} 
\braket{X}_t=\Tr\big[\rho_{\rm SO}(t)X \big] =\text{Tr}\big[\rho_{\rm SO}\,X_{\rm H}(t)\big]
\end{align}
where $X_{\rm H}(t)$ represents $X$ in the Heisenberg picture. The master equation (\ref{master_eq}) in this picture reads
\begin{multline} \label{master_eq_Heisenberg}   
\frac{d X_{\rm H}(t)}{dt}=\mathcal{L}^\star[X_{\rm H}(t)]=\ii[H_{\rm SO},X_{\rm H}(t)]\\
+\sum_{j=1}^2\bigg\{\tfrac{\gamma}{2} [\bar{n}_j+1]\bigg(c_j^\dagger X_{\rm H}(t) c_j \ -\tfrac{1}{2}\{c_j^\dagger c_j,X_{\rm H}(t)\}\bigg) \\
     +\tfrac{\gamma}{2} \bar{n}_j \bigg(c_j X_{\rm H}(t) c_j^\dagger-\tfrac{1}{2}\{c_j c_j^\dagger,X_{\rm H}(t)\}\bigg)\bigg\}\\
     =\Big(\mathcal{L}^\star[X]\Big)_{\rm H} (t).
\end{multline}
We easily obtain
\begin{align}
    \mathcal{L}^\star[c_j]&=-\bigg(\ii\omega_j+\frac{\gamma}{4}\bigg)c_j, \nonumber \\
    \mathcal{L}^\star[c_jc_k]&=-\bigg[\ii(\omega_j+\omega_k)+\frac{\gamma}{2}\bigg]c_jc_k, \\
    \mathcal{L}^\star[c_j^{\dagger}c_k]&= -\bigg[\ii(\omega_k-\omega_j)+\frac{\gamma}{2}\bigg]\Big(c_j^{\dagger}c_k-\bar{n}_j\delta_{jk}\Big),\nonumber
\end{align}
for $j=1,2$. Therefore, as a result of \eqref{master_eq_Heisenberg}, we have
\begin{align}
    (c_j)_{\rm H}(t)&=e^{-\big(\ii\omega_j+\frac{\gamma}{4}\big)t}c_j, \label{cjt}\\
    (c_jc_k)_{\rm H}(t)&=e^{-\big[\ii(\omega_j+\omega_k)+\frac{\gamma}{2}\big]t}c_jc_k, \label{cjckt}\\
    (c_j^{\dagger}c_k)_{\rm H}(t)&=e^{-\big[\ii(\omega_k-\omega_j)+\frac{\gamma}{2}\big]t}c_j^{\dagger}c_k+\bar{n}_j\big(1-e^{-\frac{\gamma}{2}t}\big)\delta_{jk}. \label{cdjckt}
\end{align}
For the initial state \eqref{xirhobeta}, the system squeezed vacuum satisfies
\begin{equation}\label{<a's>xi}
    \begin{cases}
        \braket{a}=0,\\
        \braket{a^2}=-e^{\ii\theta}\sinh r\cosh r,\\
        \braket{a^{\dagger}a}=\sinh^2 r.
    \end{cases}
\end{equation}
and the Gibbs state for the oscillator,
\begin{equation}\label{<d's>beta}
    \begin{cases}
        \braket{d}=0,\\
        \braket{d^2}=0,\\
        \braket{d^{\dagger}d}=[e^{\beta \omega_0}-1]^{-1}.
    \end{cases}
\end{equation}
which implies [see \eqref{adc1c2}],
\begin{equation}\label{cs(0)}
    \begin{cases}
        \braket{c_j}=\frac{1}{\sqrt{2}}\big(\braket{a}-(-1)^{j}\braket{d}\big)=0,\\
        \braket{c_jc_k}=\frac12 \braket{a^2},\\
        \braket{c_j^{\dagger}c_k}=\frac12 \big(\braket{a^\dagger a}+(2\delta_{jk}-1)\braket{d^\dagger d}\big).
    \end{cases}
\end{equation}
Using this result after taking mean values in \eqref{cjt}
\begin{align}
    \braket{c_j}_t=e^{-\big(\ii\omega_j+\frac{\gamma}{4}\big)t}\braket{c_j}=0.
\end{align}
Similarly for \eqref{cjckt},
\begin{align}
    \braket{c_jc_k}_t=-\frac12 e^{-\big[\ii(\omega_j+\omega_k)+\frac{\gamma}{2}\big]t}e^{\ii\theta}\sinh r\cosh r,
\end{align}
and \eqref{cdjckt},
\begin{multline}
    \braket{c^\dagger_jc_k}_t=\frac12 e^{-\big[\ii(\omega_k-\omega_j)+\frac{\gamma}{2}\big]t}\bigg(\sinh^2{r}+\frac{2\delta_{jk}-1}{e^{\beta \omega_0}-1}\bigg)\\
    +\bar{n}_j\big(1-e^{-\frac{\gamma}{2}t}\big)\delta_{jk}.
\end{multline}
Therefore, for the mean value of position and momentum,
\begin{align}
    \braket{q}_t&=\sqrt{2}\,\mathrm{Re}\braket{a}_t=\mathrm{Re}\big(\braket{c_1}_t+\braket{c_2}_t\big)=0, \label{qt}\\
    \braket{p}_t&=\sqrt{2}\,\mathrm{Im}\braket{a}_t=\mathrm{Im}\big(\braket{c_1}_t+\braket{c_2}_t\big)=0.\label{pt}
\end{align}
For the second order moments:
\begin{align}
    &\braket{qp+pq}_t=2\,\mathrm{Im}\braket{a^2}_t=\mathrm{Im}\left(\braket{c_1^2}_t+2\braket{c_1c_2}_t+\braket{c_2^2}_t\right)\nonumber \\
    &\ =-\mathrm{Im}\Bigg\{\Bigg[e^{-\ii\omega_1 t}+e^{- \ii \omega_2 t}\Bigg]^2\frac{e^{\ii\theta}}{2}\Bigg\}e^{-\frac{\gamma}{2}t}\sinh r\cosh r\nonumber \\
    &\ =\cos^2(\kappa t)\sin(2\omega_0t-\theta)e^{-\frac{\gamma}{2}t}\sinh 2r,
\end{align}
\begin{align}
    &\braket{q^2}_t =\braket{a^{\dagger}a}_t+\mathrm{Re}\braket{a^2}_t+\frac{1}{2}\nonumber \\
    &\ =\frac{1}{2}\bigg(\braket{c_1^{\dagger}c_1}_t+\braket{c_2^{\dagger}c_2}_t+\braket{c_1^{\dagger}c_2}_t+\braket{c_2^{\dagger}c_1}_t+ 1 \bigg) \nonumber \\
    &\ \ -\frac{1}{2}\mathrm{Re}\Bigg\{\Bigg[e^{-\ii\omega_1 t}+e^{- \ii \omega_2 t}\Bigg]^2\frac{e^{\ii\theta}}{2}\Bigg\}e^{-\frac{\gamma}{2}t}\sinh r\cosh r\nonumber\\
    &= \frac{1}{2} \Big[e^{-\frac{\gamma}{2}t}\bigg(\sinh^2 r+\frac{1}{e^{\beta \omega_0}-1}\bigg)\nonumber \\
 &\ \ +e^{-\frac{\gamma}{2}t}\cos \big(2\kappa t\big)\bigg(\sinh^2 r-\frac{1}{e^{\beta \omega_0}-1}\bigg)\nonumber \\
    &\ \ +(\bar{n}_1+\bar{n}_2)(1-e^{-\frac{\gamma}{2}t})+1\Big]\nonumber \\
    &\ \ -\frac{1}{2}\cos^2(\kappa t)\cos[2\omega_0t-\theta]e^{-\frac{\gamma}{2}t}\sinh 2r,
\end{align}
\begin{align}
    &\braket{p^2}_t =\braket{a^{\dagger}a}_t-\mathrm{Re}\braket{a^2}_t+\frac{1}{2}\nonumber\\
    &= \frac{1}{2} \Big[e^{-\frac{\gamma}{2}t}\bigg(\sinh^2 r+\frac{1}{e^{\beta \omega_0}-1}\bigg)\nonumber \\
 &\ \ +e^{-\frac{\gamma}{2}t}\cos \big(2\kappa t\big)\bigg(\sinh^2 r-\frac{1}{e^{\beta \omega_0}-1}\bigg)\nonumber \\
    &\ \ +(\bar{n}_1+\bar{n}_2)(1-e^{-\frac{\gamma}{2}t})+1\Big]\nonumber \\
    &\ \ +\frac{1}{2}\cos^2(\kappa t)\cos[2\omega_0t-\theta]e^{-\frac{\gamma}{2}t}\sinh 2r,
\end{align}
which lead to the covariance matrix elements \eqref{sigma1122} and \eqref{sigma1221}.

\section{Dynamics of the covariance matrix in the presence of external driving}\label{app:C}
The unitary matrix that diagonalizes the Hamiltonian \eqref{HSO(t)} is given by $\bm{Y}(t)=\left[\frac{\bm{v}_{1}(t)}{\norm{\bm{v}_{1}(t)}},\frac{\bm{v}_{2}(t)}{\norm{\bm{v}_{2}(t)}}\right]$ with unnormalized orthogonal eigenvectors
\begin{equation}
    \bm{v}_{1,2}(t)=\begin{pmatrix}
        \sqrt{\left(\frac{\omega(t)-\omega_0}{2\kappa}\right)^2+1}\pm\frac{\omega(t)-\omega_0}{2\kappa}\\
        \pm1
    \end{pmatrix}.
\end{equation}
With this choice of the phases of the eigenvectors, $Y_{12}(t)=Y_{21}(t)$, and $\bm{Y}(t)$ is a symmetric, orthogonal, and traceless $2\times2$ matrix, i.e. $\bm{Y}^2(t)=\mathds{1}$ and $Y_{11}(t)=-Y_{22}(t)$, as it can be easily checked. Using these properties, after taking the time derivative in \eqref{c(t)M(t)a}, we find
\begin{equation}\label{cdot}
    \begin{bmatrix}
        \dot{c}_1(t)\\
        \dot{c}_2(t)
    \end{bmatrix}=\mu(t)\begin{bmatrix}
        - c_2(t)\\ c_1(t)
    \end{bmatrix},
\end{equation}
where
\begin{equation}
    \mu(t):=Y_{11}(t)\dot{Y}_{12}(t)-\dot{Y}_{11}(t)Y_{12}(t).
\end{equation}
Now, if $\Lambda_t$ is the dynamical map generated by $\mathcal{L}_t$ in \eqref{master_eq_driving}, and $X(t)$ is a time-dependent operator in the Schr\"odinger picture, its evolution in the Heisenberg picture $X_{\rm H}(t)=\Lambda_t^{\star}\big[X(t)\big]$ satisfies the equation
\begin{widetext}
\begin{multline}    
\frac{d X_{\rm H}(t)}{dt}=\Lambda_t^{\star}\big\{\mathcal{L}_t^\star\big[X(t)\big]\big\}+\Lambda_t^{\star}\big[\dot{X}(t)\big]=\ii[H_{\rm SO}(t),X(t)]_{\rm H}+\sum_{j=1}^2\bigg\{\tfrac{\gamma}{2} [\bar{n}_j(t)+1]\big[c_j^\dagger(t) X(t) c_j(t) -\tfrac{1}{2}\{c_j^\dagger(t) c_j(t),X(t)\}\big]_{\rm H} \\
+\tfrac{\gamma}{2} \bar{n}_j(t) \big[c_j(t) X(t) c_j^\dagger(t)-\tfrac{1}{2}\{c_j(t) c_j^\dagger(t),X(t)\}\big]_{\rm H}\bigg\}+ \big[\dot{X}(t)\big]_{\rm H}.
\end{multline}
\end{widetext}
From this and \eqref{cdot} we may obtain the following system of coupled differential equations,
\begin{align}\label{cdotH}
    \dfrac{d}{dt}\begin{bmatrix}
        c_{1\rm H}(t) \\
        c_{2\rm H}(t)
    \end{bmatrix}=\bm{L_1}(t)\begin{bmatrix}
        c_{1\rm H}(t) \\
        c_{2\rm H}(t)
    \end{bmatrix},
\end{align}
with
\begin{equation}
    \bm{L_1}(t)=\begin{bmatrix}
        -\frac{\gamma}{4}-i\omega_1(t) & -\mu(t)\\
        \mu(t) & -\frac{\gamma}{4}-i\omega_2(t)
    \end{bmatrix}.
\end{equation}
The solution of this equation reads 
\begin{equation}\label{cTc}
    \begin{bmatrix}
        c_{1\rm H}(t) \\
        c_{2\rm H}(t)
    \end{bmatrix}=\tilde{\bm{A}}(t)\begin{bmatrix}
        c_1(0) \\
        c_2(0)
    \end{bmatrix},
\end{equation}
with $\tilde{\bm{A}}(t)=\mathcal{T}\exp\big[\int_0^t\bm{L_1}(s)ds\big]$. By applying $\Lambda_t^\star$ on \eqref{c(t)M(t)a} and using \eqref{cTc} we obtain
\begin{align}\label{atHa}
\begin{bmatrix}
        a_{\rm H}(t) \\
        d_{\rm H}(t)
    \end{bmatrix}=\bm{Y}(t)
    \begin{bmatrix}
        c_{1\rm H}(t) \\
        c_{2\rm H}(t)
    \end{bmatrix}=\bm{A}(t)
    \begin{pmatrix}
        a \\
        d
    \end{pmatrix},
\end{align}
with $\bm{A}(t)=\bm{Y}(t)\tilde{\bm{A}}(t)\bm{Y}(0)$. 

Similarly, for the product of two annihilation operators we find
\begin{align}\label{ccdotH}
    \dfrac{d}{dt}\begin{bmatrix}
        (c_1^2)_{\rm H}(t) \\
        (c_1c_2)_{\rm H}(t) \\
        (c_2^2)_{\rm H}(t)
    \end{bmatrix}=\bm{L_2}(t)\begin{bmatrix}
        (c_1^2)_{\rm H}(t) \\
        (c_1c_2)_{\rm H}(t) \\
        (c_2^2)_{\rm H}(t)
    \end{bmatrix},
\end{align}
where we have written $(c_ic_j)_{\rm H}(t)\equiv [c_i(t)c_j(t)]_{\rm H}(t)$ to avoid a cumbersome notation, and 
\begin{equation}
  \bm{L_2}(t)=\left\{\begin{smallmatrix}
        -\left[\frac{\gamma}{2}+\ii 2\omega_1(t)\right] & -2\mu(t) & 0\\
        \mu(t) & -\frac{\gamma}{2}-\ii[\omega_1(t)+\omega_2(t)] & -\mu(t)\\
        0 & 2\mu(t) & -\left[\frac{\gamma}{2}+\ii 2\omega_2(t)\right]
    \end{smallmatrix}\right\}.
\end{equation}
The solution of this equation is
\begin{equation}\label{cctH}
    \begin{bmatrix}
        (c_1^2)_{\rm H}(t) \\
        (c_1c_2)_{\rm H}(t) \\
        (c_2^2)_{\rm H}(t)
    \end{bmatrix}=\tilde{\bm{B}}(t)\begin{bmatrix}
        c_1^2(0) \\
        c_1(0)c_2(0) \\
        c_2^2(0)
        \end{bmatrix},
\end{equation}
with $\tilde{\bm{B}}(t)=\mathcal{T}\exp\big[\int_0^t\bm{L_2}(s)ds\big]$. From \eqref{c(t)M(t)a},
\begin{equation}\label{c^2a^2}
    \begin{bmatrix}
        c_1^2(t) \\
        c_1(t)c_2(t) \\
        c_2^2(t)
        \end{bmatrix}=\bm{R}(t)\begin{pmatrix}
        a^2 \\
        ad \\
        d^2
        \end{pmatrix},
\end{equation}
where
\begin{equation}
    \bm{R}(t)=\left\{\begin{smallmatrix}
        [Y_{11}(t)]^2 & 2Y_{11}(t) Y_{12}(t) & [Y_{12}(t)]^2\\
        Y_{11}(t) Y_{21}(t) & Y_{11}(t) Y_{22}(t)+Y_{12}(t) Y_{21}(t) & Y_{12}(t) Y_{22}(t) \\
        [Y_{21}(t)]^2 & 2Y_{21}(t) Y_{22}(t) & [Y_{22}(t)]^2
    \end{smallmatrix}\right\}.
\end{equation}
Using that $\bm{Y}(t)$ is symmetric and orthogonal, one easily concludes that the same is true for $\bm{R}(t)$, $\bm{R}^2(t)=\mathds{1}$. The application of $\Lambda_t^\star$ on \eqref{c^2a^2} gives
\begin{equation}
    \begin{bmatrix}
        (c_1^2)_{\rm H}(t) \\
        (c_1c_2)_{\rm H}(t) \\
        (c_2^2)_{\rm H}(t)
    \end{bmatrix}=\bm{R}(t)\begin{bmatrix}
        a^2_{\rm H}(t) \\
        (ad)_{\rm H}(t) \\
        d^2_{\rm H}(t)
        \end{bmatrix}.
\end{equation}
Introducing this equation in \eqref{cctH},
\begin{equation}\begin{bmatrix}\label{aatHa}
        a^2_{\rm H}(t) \\
        (ad)_{\rm H}(t) \\
        d^2_{\rm H}(t)
        \end{bmatrix}=\bm{B}(t)\begin{pmatrix}
        a^2 \\
        ad \\
        d^2
        \end{pmatrix},
\end{equation}
where $\bm{B}(t)=\bm{R}(t)\tilde{\bm{B}}(t)\bm{R}(0)$. By taking mean values we conclude that $\braket{a^2}_t=\braket{ad}_t=\braket{d^2}_t=0$ for the states where $\braket{a^2}=\braket{ad}=\braket{d^2}=0$. Among the considered ones in Sec. \ref{sec:example_time_dep}, only the squeezed state does not satisfy this property, \eqref{<a's>xi}. To solve that case, using \eqref{cctH} in \eqref{ccdotH}, and taking mean values:
\begin{equation}\label{auxL2_1}
    \frac{d\tilde{\bm{B}}(t)}{dt}\begin{pmatrix}
        \braket{c_1^2} \\
        \braket{c_1c_2} \\
        \braket{c_2^2}
        \end{pmatrix}=\bm{L_2}(t)\tilde{\bm{B}}(t)\begin{pmatrix}
        \braket{c_1^2} \\
        \braket{c_1c_2} \\
        \braket{c_2^2}
        \end{pmatrix}.
\end{equation}
This equality must be satisfied for any initial state, so the mean values are arbitrary and we conclude
\begin{equation}\label{auxL2_2}
    \frac{d\tilde{\bm{B}}(t)}{dt}=\bm{L_2}(t)\tilde{\bm{B}}(t).
\end{equation}
This equation, with the initial condition $\tilde{\bm{B}}(0)=\mathds{1}$ is easily numerically solved, for instance by using Runge-Kutta methods. 

Lastly, we find
\begin{align}
    \dfrac{d}{dt}\begin{bmatrix}
        (c_1^{\dagger}c_1)_{\rm H}(t)\\
        (c_1^{\dagger}c_2)_{\rm H}(t)\\
        (c_2^{\dagger}c_1)_{\rm H}(t)\\
        (c_2^{\dagger}c_2)_{\rm H}(t)
    \end{bmatrix}=\bm{L_3}(t)\begin{bmatrix}
        (c_1^{\dagger}c_1)_{\rm H}(t)\\
        (c_1^{\dagger}c_2)_{\rm H}(t)\\
        (c_2^{\dagger}c_1)_{\rm H}(t)\\
        (c_2^{\dagger}c_2)_{\rm H}(t)
    \end{bmatrix}+\bm{u}(t),
\end{align}
with
\begin{equation}
    \bm{L_3}(t)=\left\{\begin{smallmatrix}
        -\frac{\gamma}{2} & -\mu(t) & -\mu(t) &0 \\
        \mu(t) & -\frac{\gamma}{2}+\ii[\omega_1(t)-\omega_2(t)] & 0&-\mu(t)\\
         \mu(t) & 0 &-\frac{\gamma}{2}-\ii[\omega_1(t)-\omega_2(t)]  &-\mu(t)\\
        0 & \mu(t) &\mu(t) &-\frac{\gamma}{2}  
    \end{smallmatrix}\right\},
\end{equation}
and 
\begin{equation}
    \bm{u}(t)=\frac{\gamma}{2}\begin{bmatrix}
        \bar{n}_1(t)\\
        0\\
        0\\
        \bar{n}_2(t)\\
    \end{bmatrix}.
\end{equation}
The solution of this equation is
\begin{align}\label{cdctH}
    \begin{bmatrix}
        (c_1^{\dagger}c_1)_{\rm H}(t)\\
        (c_1^{\dagger}c_2)_{\rm H}(t)\\
        (c_2^{\dagger}c_1)_{\rm H}(t)\\
        (c_2^{\dagger}c_2)_{\rm H}(t)
    \end{bmatrix}=\tilde{\bm{C}}(t,0)\begin{bmatrix}
        c_1^{\dagger}(0)c_1(0)\\
        c_1^{\dagger}(0)c_2(0)\\
        c_2^{\dagger}(0)c_1(0)\\
        c_2^{\dagger}(0)c_2(0)
    \end{bmatrix}+\tilde{\bm{y}}(t),
\end{align}
with $\tilde{\bm{y}}(t)=\int_0^t\tilde{\bm{C}}(t,s)\bm{u}(s)ds$ and $\tilde{\bm{C}}(t,s)=\mathcal{T}\exp\big[\int_s^t\bm{L_3}(s)ds\big]$. Once more, by \eqref{c(t)M(t)a},
\begin{equation}\label{cdcada}
    \begin{bmatrix}
        c_1^{\dagger}(t)c_1(t)\\
        c_1^{\dagger}(t)c_2(t)\\
        c_2^{\dagger}(t)c_1(t)\\
        c_2^{\dagger}(t)c_2(t)
    \end{bmatrix}=\bm{G}(t)\begin{pmatrix}
        a^\dagger a \\
        a^\dagger d \\
        d^\dagger a \\
        d^\dagger d
        \end{pmatrix},
\end{equation}
where
\begin{equation}
    \bm{G}(t)=\left\{\begin{smallmatrix}
        [Y_{11}(t)]^2 & Y_{11}(t) Y_{12}(t) &  Y_{11}(t) Y_{12}(t) & [Y_{12}(t)]^2\\
        Y_{11}(t) Y_{21}(t) & Y_{11}(t) Y_{22}(t) & Y_{12}(t) Y_{21}(t) & Y_{12}(t) Y_{22}(t) \\
        Y_{11}(t) Y_{21}(t) & Y_{12}(t) Y_{21}(t) & Y_{11}(t) Y_{22}(t) & Y_{12}(t) Y_{22}(t) \\
        [Y_{21}(t)]^2 & Y_{21}(t) Y_{22}(t) & Y_{21}(t) Y_{22}(t) & [Y_{22}(t)]^2
    \end{smallmatrix}\right\}.
\end{equation}
Again, from the fact that $\bm{Y}(t)$ is symmetric and orthogonal, one finds that the same is true for $\bm{G}(t)$, $\bm{G}^2(t)=\mathds{1}$. Applying $\Lambda_t^\star$ on \eqref{cdcada},
\begin{equation}
    \begin{bmatrix}
        (c_1^{\dagger}c_1)_{\rm H}(t)\\
        (c_1^{\dagger}c_2)_{\rm H}(t)\\
        (c_2^{\dagger}c_1)_{\rm H}(t)\\
        (c_2^{\dagger}c_2)_{\rm H}(t)
    \end{bmatrix}=\bm{G}(t)\begin{bmatrix}
        (a^\dagger a)_{\rm H}(t) \\
        (a^\dagger d)_{\rm H}(t) \\
        (d^\dagger a)_{\rm H}(t) \\
        (d^\dagger d)_{\rm H}(t)
        \end{bmatrix},
\end{equation}
and introducing this equation in \eqref{cdctH} we obtain
\begin{equation}\label{adaHt}
        \begin{bmatrix}
        (a^\dagger a)_{\rm H}(t) \\
        (a^\dagger d)_{\rm H}(t) \\
        (d^\dagger a)_{\rm H}(t) \\
        (d^\dagger d)_{\rm H}(t)
        \end{bmatrix}=\bm{C}(t)\begin{pmatrix}
        a^\dagger a \\
        a^\dagger d \\
        d^\dagger a \\
        d^\dagger d
        \end{pmatrix}+\bm{y}(t).
\end{equation}
Here, $\bm{C}(t):=\bm{G}(t)\tilde{\bm{C}}(t,0)\bm{G}(0)$ and $\bm{y}(t):=\bm{G}(t)\tilde{\bm{y}}(t)$. A reasoning similar to that in \eqref{auxL2_1} and \eqref{auxL2_2} leads to the equations
\begin{equation}
    \frac{d\tilde{\bm{C}}(t,0)}{dt} = \bm{L_3}(t)\tilde{\bm{C}}(t,0),
\end{equation}
with the initial condition $\tilde{\bm{C}}(0,0) = \mathds{1}$, and
\begin{equation}
    \frac{d\tilde{\bm{y}}(t)}{dt} = \bm{L_3}(t)\tilde{\bm{y}}(t) + \bm{u}(t),
\end{equation}
with the initial condition $\tilde{\bm{y}}(0) = (0,0,0,0)^{\ts}$. These two equations can again be solved numerically with ease.

Thus, for an arbitrary initial state, taking into account \eqref{atHa}, \eqref{aatHa} and \eqref{adaHt}, we obtain
\begin{align}
    \braket{q}_t&=\sqrt{2}\,\mathrm{Re}\braket{a}_t=\sqrt{2}\,\mathrm{Re}[A_{11}(t)\braket{a}+A_{12}(t)\braket{d}],
\end{align}
\begin{align}
    \braket{p}_t&=\sqrt{2}\,\mathrm{Im}\braket{a}_t=\sqrt{2}\,\mathrm{Im}[A_{11}(t)\braket{a}+A_{12}(t)\braket{d}],
\end{align}
\begin{align}
    \braket{qp+pq}_t&=2\,\mathrm{Im}\braket{a^2}_t\nonumber\\&=2\,\mathrm{Im}[B_{11}(t)\braket{a^2}+B_{12}(t)\braket{ad}+B_{13}(t)\braket{d^2}],
\end{align}
\begin{align}
    \braket{q^2}_t &=\braket{a^{\dagger}a}_t+\mathrm{Re}\braket{a^2}_t+\frac{1}{2}\nonumber\\
    &=\mathrm{Re}[B_{11}(t)\braket{a^2}+B_{12}(t)\braket{ad}+B_{13}(t)\braket{d^2}]\nonumber\\
    &\qquad+C_{11}(t)\braket{a^{\dagger}a}+C_{12}(t)\braket{a^{\dagger}d}+C_{13}(t)\braket{d^{\dagger}a}\nonumber\\
    &\qquad\qquad+C_{14}(t)\braket{d^{\dagger}d}+y_1(t)+\frac{1}{2},
\end{align}
and
\begin{align}
    \braket{p^2}_t &=\braket{a^{\dagger}a}_t-\mathrm{Re}\braket{a^2}_t+\frac{1}{2}\nonumber\\
    &=-\mathrm{Re}[B_{11}(t)\braket{a^2}+B_{12}(t)\braket{ad}+B_{13}(t)\braket{d^2}]\nonumber\\
    &\qquad+C_{11}(t)\braket{a^{\dagger}a}+C_{12}(t)\braket{a^{\dagger}d}+C_{13}(t)\braket{d^{\dagger}a}\nonumber\\
    &\qquad\qquad+C_{14}(t)\braket{d^{\dagger}d}+y_1(t)+\frac{1}{2}.
\end{align}
This allows us to determine the covariance matrix by Eq. \eqref{covariance_time_dependent}. Once $\sigma(t)$ is obtained, the entropy is calculated by \eqref{vn_entropy_gauss} and \eqref{nu_general}.

To obtain the work, since $\dot{H}_{\rm S}(t)=\dot{\omega}(t)a^\dagger a$, the integration of \eqref{work} in the Heisenberg picture gives 
\begin{align}
    W(t)&=\int_0^{t}dt'\Tr\left[\dot{H}_{\rm S}(t')\rho_{\rm S}(t')\right]\nonumber \\
    &=\int_0^{t}dt'\dot{\omega}(t')\Tr\left\{\left[(a^{\dagger}a)_{\rm H}(t')\right]\rho_{\rm S}(0)\otimes\rho_{O,\beta}\right\}\nonumber \\
    &=\int_0^{t}dt'\dot{\omega}(t')\Big[C_{11}(t')\braket{a^{\dagger}a}+C_{12}(t')\braket{a^{\dagger}d}\nonumber\\
    &\qquad\qquad+C_{13}(t')\braket{d^{\dagger}a}+C_{14}(t')\braket{d^{\dagger}d}+y_1(t')\Big].
\end{align}
Hence, by \eqref{W-sharp},
\begin{multline}\label{app_work_sharp}
    W_{\rm S}^{\sharp}(t)=\int_0^{t}dt'\dot{\omega}(t')\Big[C_{11}(t')\braket{a^{\dagger}a}+C_{12}(t')\braket{a^{\dagger}d}\\
    \qquad+C_{13}(t')\braket{d^{\dagger}a}+C_{14}(t')\braket{d^{\dagger}d}+y_1(t')\Big]-\Delta F_{\rm R|S}^{\sharp}.
\end{multline}
For the last term we have in this case
\begin{multline}
    \Delta F_{\rm R|S}^{\sharp}=\Delta F_{\rm SR}-\Delta F_{\rm S}^{\sharp}\\
    =-\beta^{-1}\log\dfrac{Z_{\rm SR}(t)}{Z_{\rm SR}(0)}+\beta^{-1}\log\dfrac{Z_{\rm S}^{\sharp}(t)}{Z_{\rm S}^{\sharp}(0)}\\
    \simeq-\beta^{-1}\log\dfrac{Z_{\rm SO}(t)}{Z_{\rm SO}(0)}+\beta^{-1}\log\dfrac{Z_{\rm S}^{\sharp}(t)}{Z_{\rm S}^{\sharp}(0)}\\
    =\beta^{-1}\left\{\log\dfrac{Z_{\rm S}^{\sharp}(t)}{Z_{\rm S}(0)}-\log\dfrac{\left[1-e^{-\beta\omega_1(0)}\right]\left[1-e^{-\beta\omega_2(0)}\right]}{\left[1-e^{-\beta\omega_1(t)}\right]\left[1-e^{-\beta\omega_2(t)}\right]}\right\}.
\end{multline}
For $Z_{\rm S}^{\sharp}(t)$ in this expression, analogy with the usual equilibrium state allows us to use \eqref{Zhash}, with
\begin{multline}
    S_{\rm vN}(t,\beta)=\frac{\nu(t)+1}{2}\log \left[\frac{\nu(t) + 1}{2}\right]\\
    -\frac{\nu(t) - 1}{2}\log \left[\frac{\nu(t) - 1}{2}\right],
\end{multline}
instead of $S_{\rm vN}(\beta)$ inside the integral, where $\nu(t)$ is the symplectic eigenvalue of $\rho_{\rm S,\eq}(t)=\Tr_{\rm R}[e^{-\beta H(t)}]/Z_{\rm SR}$. Following a similar procedure as in Sec. \ref{Sec:VA}, we obtain
\begin{equation}\label{NEnu}
    \nu(t)=\dfrac{2Y_{11}^2(t)}{e^{\beta\omega_1(t)}-1}+\dfrac{2Y_{12}^2(t)}{e^{\beta\omega_2(t)}-1}+1.
\end{equation}

To calculate the nonequilibrium Hamiltonian of mean force, we first note that since $\rho_{\rm S,eq}(t)$ has the single symplectic eigenvalue \eqref{NEnu}, it may be written as a thermal state $\rho_{\rm S,eq}(t)=e^{-\beta H_{\phi}(t)}/Z_{\phi}(t)$, with $H_{\phi}(t)=\phi(t)a^{\dagger}a$, $Z_{\phi}(t)=[1-e^{-\beta\phi(t)}]^{-1}$, and $\phi(t)$ defined by
\begin{equation}
    \dfrac{1}{e^{\beta\phi(t)}-1}=\dfrac{Y_{11}^2(t)}{e^{\beta\omega_1(t)}-1}+\dfrac{Y_{12}^2(t)}{e^{\beta\omega_2(t)}-1}.
\end{equation}
Hence, from \eqref{HashS} we obtain
\begin{equation}\label{HashHzetat}
    H_{\rm S}^{\sharp}(t)=\phi(t)a^{\dagger}a-\beta^{-1}\log \frac{Z_{\rm S}^{\sharp}(t)}{Z_\phi(t)}.
\end{equation}
The nonequilibrium internal energy is therefore given by
\begin{multline}  
\tilde{E}^{\sharp}_{\rm S}(t,\beta)=\Tr\big[\rho_{\rm S}(t)H_{\rm S}^{\sharp}(t)\big]\\
=\phi(t)\braket{a^\dagger a}_t
-\beta^{-1}\log \frac{Z_{\rm S}^{\sharp}(t)}{Z_\phi(t)} \\
=\phi(t)\Big[C_{11}(t)\braket{a^{\dagger}a}+C_{12}(t)\braket{a^{\dagger}d}+C_{13}(t)\braket{d^{\dagger}a}\\
    \quad +C_{14}(t)\braket{d^{\dagger}d}+y_1(t)\Big]-\beta^{-1}\log \frac{Z_{\rm S}^{\sharp}(t)}{Z_\phi(t)}.
\end{multline}
Thus, we can evaluate the thermodynamic quantities for arbitrary initial state of the system SO from the initial expected values $\braket{a}$, $\braket{d}$, $\braket{a^2}$, $\braket{ad}$, $\braket{d^2}$, $\braket{a^\dagger a}$, etc. In the case considered in Sec. \ref{sec:example_time_dep} of an initial equilibrium state $\rho_{\rm SO}(0)=\rho_{\rm SO,\beta}$ with a driving satisfying $\omega(0)=\omega_0$, these are given by
\begin{equation}
    \braket{a}=\braket{d}=\braket{a^2}=\braket{ad}=\braket{d^2}=0,
\end{equation}
\begin{equation}
    \braket{a^{\dagger}a}=\braket{d^{\dagger}d}=\dfrac{1}{2}\left[\dfrac{1}{e^{\beta\omega_1(0)}-1}+\dfrac{1}{e^{\beta\omega_2(0)}-1}\right],
\end{equation}
\begin{equation}
    \braket{a^{\dagger}d}=\braket{d^{\dagger}a}=\dfrac{1}{2}\left[\dfrac{1}{e^{\beta\omega_1(0)}-1}-\dfrac{1}{e^{\beta\omega_2(0)}-1}\right].
\end{equation}
Finally, we note that the expressions in this section can also be used to describe the case of an initial product state $\rho_{\rm SO}(0)=\rho_{\rm S}\otimes\rho_{\rm O, \beta}$ in the case of a sudden quench at $t=0$ followed by a driving for $t>0$. One just needs to keep in mind that for the system work $W^{\sharp}_{\rm S}(t>0)$ the additional term due to the quench \eqref{Wquench} must be added to \eqref{app_work_sharp}, and that for the internal energy, $H^{\sharp}_{\rm S}(0)=H_{\rm S}$.

\section{Relation between the entropy production in the intrinsic Hamiltonian of mean force and in reaction coordinate thermodynamics}\label{app:E}

Consider a system $\mathrm{S}$ coupled to a thermal reservoir $\mathrm{R}$ under general thermodynamic initial conditions, and assume that the reservoir $\mathrm{R} = \mathrm{RC} + \mathrm{R}'$ comprises a set of reaction coordinates $\mathrm{RC}$ weakly coupled to the remaining bath degrees of freedom $\mathrm{R}'$. The external driving only affects the $\mathrm{S+RC}$ degrees of freedom, i.e., $H_{\rm R'}$ is time-independent. We establish that
\begin{equation}\label{S>SRC}
    \Sigma_{\rm S}(t) \geq \Sigma_{\rm S+RC}(t).
\end{equation}
Namely, the entropy production calculated using the intrinsic Hamiltonian of mean force for the system, $\Sigma_{\rm S}$, is bounded from below by the entropy production from reaction coordinate thermodynamics, $\Sigma_{\rm S+RC}$—which incorporates the reaction coordinates (or pseudomodes) of the environment within the extended system $\mathrm{S+RC}$. This result directly reduces to Eq.~\eqref{S>SO} in the example of Sec.~\ref{sec:example}, where the single environmental discrete oscillator mode $\mathrm{O}$ serves as the reaction coordinate.

To prove Eq.~\eqref{S>SRC}, we express the entropy production $\Sigma_{\rm S}(t)$ in terms of the total work, and variations of the equilibrium free energy and relative entropy. Using the notation of Sec.~\ref{Sec:IVB}, we have
\begin{align}\label{appE_1}
    T\Sigma_{\rm S}(t) &= W_{\rm S}^\sharp(t) - \Delta \tilde{F}_{\rm S}^\sharp(t) \nonumber \\
    &= W(t) - \Delta F_{\rm R|S}^\sharp(t) - \Delta F_{\rm S}^\sharp(t) - \beta^{-1}\Delta D_{\rm S}(t).
\end{align}
For the equilibrium free-energy contribution, we find
\begin{align}\label{appE_2}
    \Delta F_{\rm S}^\sharp(t) + \Delta F_{\rm R|S}^\sharp(t) &= \Delta F_{\rm SR}(t) = -\beta^{-1}\log \frac{Z_{\rm SR}(t)}{Z_{\rm SR}(0)} \nonumber \\
    &= -\beta^{-1}\log \frac{Z_{\rm S+RC}(t)}{Z_{\rm S+RC}(0)} = \Delta F_{\rm S+RC}(t),
\end{align}
which holds because $\mathrm{S+RC}$ is weakly coupled to $\mathrm{R}'$, ensuring that the reservoir partition function $Z_{\rm R'}$ remains constant under driving and cancels in the ratio.

Next, consider the relative entropy variation:
\begin{align}\label{appE_3}
    \Delta D_{\rm S}(t) &= D[\rho_{\rm S}(t) \| \rho_{\rm S,eq}(t)] - D[\rho_{\rm S}(0) \| \rho_{\rm S,eq}(0)] \nonumber \\
    &= D[\rho_{\rm S}(t) \| \rho_{\rm S,eq}(t)] - D[\rho_{\rm S+RC}(0) \| \rho_{\rm S+RC,\beta}(0)] \nonumber \\
    &\leq \Delta D_{\rm S+RC}(t).
\end{align}
Here, the second equality follows from the initial thermodynamic conditions: since $\mathrm{S+RC}$ is weakly coupled to $\mathrm{R}'$, the initial global states factorize as 
\begin{align}
    &\rho_{\rm SR,\beta}(0) = \rho_{\rm S+RC,\beta}(0) \otimes \rho_{\rm R',\beta}(0), \nonumber \\
    &\rho_{\rm SR}(0) = \rho_{\rm S+RC}(0) \otimes \rho_{\rm R',\beta}(0), 
\end{align}
simplifying condition~\eqref{Petzequality} to 
\begin{equation}
    D[\rho_{\rm S+RC}(0) \| \rho_{\rm S+RC,\beta}(0)] = D[\rho_{\rm S}(0) \| \rho_{\rm S,eq}(0)].
\end{equation}
The inequality in the final step in \eqref{appE_3} then follows directly from the monotonicity of the quantum relative entropy under the partial trace, $D[\rho_{\rm S}(t) \| \rho_{\rm S,eq}(t)] \le D[\rho_{\rm S+RC}(t) \| \rho_{\rm S+RC,\beta}(t)]$.

Substituting Eqs.~\eqref{appE_2} and \eqref{appE_3} into Eq.~\eqref{appE_1}, and noting that the nonequilibrium free energy of the extended system reads $\tilde{F}_{\rm S+RC}(t) = F_{\rm S+RC}(t) + \beta^{-1} D_{\rm S+RC}(t)$, we arrive directly at Eq.~\eqref{S>SRC}.

\bibliography{strong_coupling.bib}

\end{document}